\documentclass[11pt, a4paper]{article}
\usepackage[mathscr]{euscript}

\PassOptionsToPackage{unicode}{hyperref}
\usepackage{jheppub}

\let\U\relax
\let\C\relax

\usepackage{tikz}
\usepackage{tikz-3dplot}

\usepackage{relsize}

\usetikzlibrary{calc, positioning, arrows.meta,
  decorations.pathreplacing}

\tikzset{font=\smaller, baseline={([yshift=-0.75ex]current bounding
    box.center)}, zerosep/.style={inner sep=0pt, outer sep=0pt,
    minimum size=0pt}, node distance=8pt, align at
  top/.style={baseline=(current bounding box.north)}, align at
  bottom/.style={baseline=(current bounding box.south)}, }

\usepackage{upgreek}

\newcommand{\gf}{\mathfrak{g}}

\newcommand{\ket}[1]{|#1\rangle}
\newcommand{\bra}[1]{\langle #1|}

\newcommand{\diag}{\mathop{\mathrm{diag}}\nolimits}

\newcommand{\Tr}{\mathop{\mathrm{tr}}\nolimits}
\newcommand{\Str}{\mathop{\mathrm{str}}\nolimits}
\newcommand{\End}{\mathop{\mathrm{End}}\nolimits}

\newcommand{\SO}{\mathrm{SO}}

\newcommand{\GL}{\mathrm{GL}}
\newcommand{\glf}{\mathfrak{gl}}
\newcommand{\slf}{\mathfrak{sl}}

\newcommand{\U}{\mathrm{U}}

\newcommand{\iso}{\cong}

\newcommand{\Z}{\mathbb{Z}}

\newcommand{\R}{\mathbb{R}}
\newcommand{\C}{\mathbb{C}}

\newcommand{\T}{\mathbb{T}}

\let\nc\newcommand
\let\renc\renewcommand
\nc{\wbar}{\overline}
\let\td\tilde
\let\wtd\widetilde
\let\wht\widehat
\let\mcl\mathcal

\nc{\ab}{{\bar{a}}} \nc{\at}{\tilde{a}} \nc{\ah}{\hat{a}}
\nc{\bb}{{\bar{b}}} \nc{\bt}{\tilde{b}} \nc{\bh}{\hat{b}}
\nc{\cb}{{\bar{c}}} \nc{\ct}{\tilde{c}} 
\nc{\db}{{\bar{d}}} \nc{\dt}{\tilde{d}} \renc{\dh}{\hat{d}}
\nc{\eb}{{\bar{e}}} \nc{\et}{\tilde{e}} \nc{\eh}{\hat{e}}
\nc{\fb}{{\bar{f}}} \nc{\ft}{\tilde{f}} \nc{\fh}{\hat{f}}
\nc{\gb}{{\bar{g}}} \nc{\gt}{\tilde{g}} \nc{\gh}{\hat{g}}
\nc{\hb}{{\bar{h}}} \nc{\hh}{\hat{h}} 
\nc{\ib}{{\bar{\imath}}} \nc{\ih}{\hat{\imath}} 
\nc{\jb}{{\bar{\jmath}}} \nc{\jt}{\tilde{\jmath}} \nc{\jh}{\hat{\jmath}}
\nc{\kb}{{\bar{k}}} \nc{\kt}{\tilde{k}} \nc{\kh}{\hat{k}}
\nc{\lb}{{\bar{l}}} \nc{\lt}{\tilde{l}} \nc{\lh}{\hat{l}}
\nc{\mb}{{\bar{m}}} \nc{\mt}{\tilde{m}} \nc{\mh}{\hat{m}}
\nc{\nb}{{\bar{n}}} \nc{\nt}{\tilde{n}} \nc{\nh}{\hat{n}}
\nc{\ob}{{\bar{o}}} \nc{\ot}{\tilde{o}} \nc{\oh}{\hat{o}}
\nc{\pb}{{\bar{p}}} \nc{\pt}{\tilde{p}} \nc{\ph}{\hat{p}}
\nc{\qb}{{\bar{q}}} \nc{\qt}{\tilde{q}} \nc{\qh}{\hat{q}}
\nc{\rb}{{\bar{r}}} \nc{\rt}{\tilde{r}} \nc{\rh}{{\hat{r}}}
\renc{\sb}{{\bar{s}}} \nc{\st}{\tilde{s}} \nc{\sh}{\hat{s}}
\nc{\tb}{{\bar{t}}} \renc{\th}{\hat{t}} 
\nc{\ub}{{\bar{u}}} \nc{\ut}{\tilde{u}} \nc{\uh}{\hat{u}}
\nc{\vb}{{\bar{v}}} \nc{\vt}{\tilde{v}} \nc{\vh}{\hat{v}}
\nc{\wb}{{\bar{w}}} \nc{\wt}{\tilde{w}} \nc{\wh}{\hat{w}}
\nc{\xb}{{\bar{x}}} \nc{\xt}{\tilde{x}} \nc{\xh}{\hat{x}}
\nc{\yb}{{\bar{y}}} \nc{\yt}{\tilde{y}} \nc{\yh}{\hat{y}}
\nc{\zb}{{\bar{z}}} \nc{\zt}{\tilde{z}} \nc{\zh}{\hat{z}}

\nc{\Ab}{{\wbar{A}}} \nc{\At}{{\wtd{A}}} \nc{\Ah}{{\wht{A}}}
\nc{\Bb}{{\wbar{B}}} \nc{\Bt}{{\wtd{B}}} \nc{\Bh}{{\wht{B}}}
\nc{\Cb}{{\wbar{C}}} \nc{\Ct}{{\wtd{C}}} \nc{\Ch}{{\wht{C}}}
\nc{\Db}{{\wbar{D}}} \nc{\Dt}{{\wtd{D}}} \nc{\Dh}{{\wht{D}}}
\nc{\Eb}{{\wbar{E}}} \nc{\Et}{{\wtd{E}}} \nc{\Eh}{{\wht{E}}}
\nc{\Fb}{{\wbar{F}}} \nc{\Ft}{{\wtd{F}}} \nc{\Fh}{{\wht{F}}}
\nc{\Gb}{{\wbar{G}}} \nc{\Gt}{{\wtd{G}}} \nc{\Gh}{{\wht{G}}}
\nc{\Hb}{{\wbar{H}}} \nc{\Ht}{{\wtd{H}}} \nc{\Hh}{{\wht{H}}}
\nc{\Ib}{{\bar{I}}} \nc{\It}{{\wtd{I}}} \nc{\Ih}{{\wht{I}}}
\nc{\Jb}{{\bar{J}}} \nc{\Jt}{{\wtd{J}}} \nc{\Jh}{{\wht{J}}}
\nc{\Kb}{{\wbar{K}}} \nc{\Kt}{{\wtd{K}}} \nc{\Kh}{{\wht{K}}}
\nc{\Lb}{{\wbar{L}}} \nc{\Lt}{{\wtd{L}}} \nc{\Lh}{{\wht{L}}}
\nc{\Mb}{{\wbar{M}}} \nc{\Mt}{{\wtd{M}}} \nc{\Mh}{{\wht{M}}}
\nc{\Nb}{{\wbar{N}}} \nc{\Nt}{{\wtd{N}}} \nc{\Nh}{{\wht{N}}}
\nc{\Ob}{{\wbar{O}}} \nc{\Ot}{{\wtd{O}}} \nc{\Oh}{{\wht{O}}}
\nc{\Pb}{{\wbar{P}}} \nc{\Pt}{{\wtd{P}}} \nc{\Ph}{{\wht{P}}}
\nc{\Qb}{{\wbar{Q}}} \nc{\Qt}{{\wtd{Q}}} \nc{\Qh}{{\wht{Q}}}
\nc{\Rb}{{\wbar{R}}} \nc{\Rt}{{\wtd{R}}} \nc{\Rh}{{\wht{R}}}
\nc{\Sb}{{\wbar{S}}} \nc{\St}{{\wtd{S}}} \nc{\Sh}{{\wht{S}}}
\nc{\Tb}{{\wbar{T}}} \nc{\Tt}{{\wtd{T}}} \nc{\Th}{{\wht{T}}}
\nc{\Ub}{{\wbar{U}}} \nc{\Ut}{{\wtd{U}}} \nc{\Uh}{{\wht{U}}}
\nc{\Vb}{{\wbar{V}}} \nc{\Vt}{{\wtd{V}}} \nc{\Vh}{{\wht{V}}}
\nc{\Wb}{{\wbar{W}}} \nc{\Wt}{{\wtd{W}}} \nc{\Wh}{{\wht{W}}}
\nc{\Xb}{{\wbar{X}}} \nc{\Xt}{{\wtd{X}}} \nc{\Xh}{{\wht{X}}}
\nc{\Yb}{{\wbar{Y}}} \nc{\Yt}{{\wtd{Y}}} \nc{\Yh}{{\wht{Y}}}
\nc{\Zb}{{\wbar{Z}}} \nc{\Zt}{{\wtd{Z}}} \nc{\Zh}{{\wht{Z}}}

\nc{\CA}{{\mcl{A}}} \nc{\CAb}{{\wbar{\CA}}} \nc{\CAt}{{\wtd{\CA}}} \nc{\CAh}{{\wht{\CA}}}
\nc{\CB}{{\mcl{B}}} \nc{\CBb}{{\wbar{\CB}}} \nc{\CBt}{{\wtd{\CB}}} \nc{\CBh}{{\wht{\CB}}}
\nc{\CC}{{\mcl{C}}} \nc{\CCb}{{\wbar{\CC}}} \nc{\CCt}{{\wtd{\CC}}} \nc{\CCh}{{\wht{\CC}}}
\nc{\cD}{{\mcl{D}}} \nc{\cDb}{{\wbar{\cD}}} \nc{\cDt}{{\wtd{\cC}}} \nc{\cDh}{{\wht{\cD}}}
\nc{\CE}{{\mcl{E}}} \nc{\CEb}{{\wbar{\CE}}} \nc{\CEt}{{\wtd{\CE}}} \nc{\CEh}{{\wht{\CE}}}
\nc{\CF}{{\mcl{F}}} \nc{\CFb}{{\wbar{\CF}}} \nc{\CFt}{{\wtd{\CF}}} \nc{\CFh}{{\wht{\CF}}}
\nc{\CG}{{\mcl{G}}} \nc{\CGb}{{\wbar{\CG}}} \nc{\CGt}{{\wtd{\CG}}} \nc{\CGh}{{\wht{\CG}}}
\nc{\CH}{{\mcl{H}}} \nc{\CHb}{{\wbar{\CH}}} \nc{\CHt}{{\wtd{\CH}}} \nc{\CHh}{{\wht{\CH}}}
\nc{\CI}{{\mcl{I}}} \nc{\CIb}{{\wbar{\CI}}} \nc{\CIt}{{\wtd{\CI}}} \nc{\CIh}{{\wht{\CI}}}
\nc{\CJ}{{\mcl{J}}} \nc{\CJb}{{\wbar{\CJ}}} \nc{\CJt}{{\wtd{\CJ}}} \nc{\CJh}{{\wht{\CJ}}}
\nc{\CK}{{\mcl{K}}} \nc{\CKb}{{\wbar{\CK}}} \nc{\CKt}{{\wtd{\CK}}} \nc{\CKh}{{\wht{\CK}}}
\nc{\CL}{{\mcl{L}}} \nc{\CLb}{{\wbar{\CL}}} \nc{\CLt}{{\wtd{\CL}}} \nc{\CLh}{{\wht{\CL}}}
\nc{\CM}{{\mcl{M}}} \nc{\CMb}{{\wbar{\CM}}} \nc{\CMt}{{\wtd{\CM}}} \nc{\CMh}{{\wht{\CM}}}
\nc{\CN}{{\mcl{N}}} \nc{\CNb}{{\wbar{\CN}}} \nc{\CNt}{{\wtd{\CN}}} \nc{\CNh}{{\wht{\CN}}}
\nc{\CO}{{\mcl{O}}} \nc{\COb}{{\wbar{\CO}}} \nc{\COt}{{\wtd{\CO}}} \nc{\COh}{{\wht{\CO}}}
\nc{\CP}{{\mcl{P}}} \nc{\CPb}{{\wbar{\CP}}} \nc{\CPt}{{\wtd{\CP}}} \nc{\CPh}{{\wht{\CP}}}
\nc{\CQ}{{\mcl{Q}}} \nc{\CQb}{{\wbar{\CQ}}} \nc{\CQt}{{\wtd{\CQ}}} \nc{\CQh}{{\wht{\CQ}}}
\nc{\CR}{{\mcl{R}}} \nc{\CRb}{{\wbar{\CR}}} \nc{\CRt}{{\wtd{\CR}}} \nc{\CRh}{{\wht{\CR}}}
\nc{\CS}{{\mcl{S}}} \nc{\CSb}{{\wbar{\CS}}} \nc{\CSt}{{\wtd{\CS}}} \nc{\CSh}{{\wht{\CS}}}
\nc{\CT}{{\mcl{T}}} \nc{\CTb}{{\wbar{\CT}}} \nc{\CTt}{{\wtd{\CT}}} \nc{\CTh}{{\wht{\CT}}}
\nc{\CU}{{\mcl{U}}} \nc{\CUb}{{\wbar{\CU}}} \nc{\CUt}{{\wtd{\CU}}} \nc{\CUh}{{\wht{\CU}}}
\nc{\CV}{{\mcl{V}}} \nc{\CVb}{{\wbar{\CV}}} \nc{\CVt}{{\wtd{\CV}}} \nc{\CVh}{{\wht{\CV}}}
\nc{\CW}{{\mcl{W}}} \nc{\CWb}{{\wbar{\CW}}} \nc{\CWt}{{\wtd{\CW}}} \nc{\CWh}{{\wht{\CW}}}
\nc{\CX}{{\mcl{X}}} \nc{\CXb}{{\wbar{\CX}}} \nc{\CXt}{{\wtd{\CX}}} \nc{\CXh}{{\wht{\CX}}}
\nc{\CY}{{\mcl{Y}}} \nc{\CYb}{{\wbar{\CY}}} \nc{\CYt}{{\wtd{\CY}}} \nc{\CYh}{{\wht{\CY}}}
\nc{\CZ}{{\mcl{Z}}} \nc{\CZb}{{\wbar{\CZ}}} \nc{\CZt}{{\wtd{\CZ}}} \nc{\CZh}{{\wht{\CZ}}}

\let\eps\epsilon
\let\ups\upsilon
\let\veps\varepsilon
\let\vtht\vartheta

\let\vsgm\varsigma
\let\vphi\varphi
\let\vrho\varrho

\nc{\alphab}{{\bar{\alpha}}} \nc{\alphat}{{\td{\alpha}}} \nc{\alphah}{{\hat{\alpha}}}
\nc{\betab}{{\bar{\beta}}}   \nc{\betat}{{\td{\beta}}}   \nc{\betah}{{\hat{\beta}}} 
\nc{\gammab}{{\bar{\gamma}}} \nc{\gammat}{{\td{\gamma}}} \nc{\gammah}{{\hat{\gamma}}} 
\nc{\deltab}{{\bar{\delta}}} \nc{\deltat}{{\td{\delta}}} \nc{\deltah}{{\hat{\delta}}} 
\nc{\epsilonb}{{\bar{\eps}}} \nc{\epsilont}{{\td{\eps}}} \nc{\epsilonh}{{\hat{\eps}}} 
\nc{\vepsb}{{\bar{\veps}}}   \nc{\vepst}{{\td{\veps}}}   \nc{\vepsh}{{\hat{\veps}}} 
\nc{\zetab}{{\bar{\zeta}}}   \nc{\zetat}{{\td{\zeta}}}   \nc{\zetah}{{\hat{\zeta}}} 
\nc{\etab}{{\bar{\eta}}}     \nc{\etat}{{\td{\eta}}}     \nc{\etah}{{\hat{\eta}}} 
\nc{\thetab}{{\bar{\theta}}} \nc{\thetat}{{\td{\theta}}} \nc{\thetah}{{\hat{\theta}}} 
\nc{\vthetab}{{\bar{\vtht}}} \nc{\vthetat}{{\td{\vtht}}} \nc{\vthetah}{{\hat{\vtht}}} 
\nc{\lambdab}{{\bar{\lambda}}} \nc{\lambdat}{{\td{\lambda}}} \nc{\lambdah}{{\hat{\lambda}}} 
\nc{\iotab}{{\bar{\iota}}}   \nc{\iotat}{{\td{\iota}}}   \nc{\iotah}{{\hat{\iota}}} 
\nc{\kappab}{{\bar{\kappa}}} \nc{\kappat}{{\td{\kappa}}} \nc{\kappah}{{\hat{\kappa}}} 
\nc{\lmdb}{{\bar{\lmd}}}     \nc{\lmdt}{{\td{\lmd}}}     \nc{\lmdh}{{\hat{\lmd}}} 
\nc{\mub}{{\bar{\mu}}}       \nc{\mut}{{\td{\mu}}}       \nc{\muh}{{\hat{\mu}}} 
\nc{\nub}{{\bar{\nu}}}       \nc{\nut}{{\td{\nu}}}       \nc{\nuh}{{\hat{\nu}}} 
\nc{\xib}{{\bar{\xi}}}       \nc{\xit}{{\td{\xi}}}       \nc{\xih}{{\hat{\xi}}} 
\nc{\pib}{{\bar{\pi}}}       \nc{\pit}{{\td{\pi}}}       \nc{\pih}{{\hat{\pi}}} 
\nc{\vpib}{{\bar{\vpi}}}     \nc{\vpit}{{\td{\vpi}}}     \nc{\vpih}{{\hat{\vpi}}} 
\nc{\rhob}{{\bar{\rho}}}     \nc{\rhot}{{\td{\rho}}}     \nc{\rhoh}{{\hat{\rho}}} 
\nc{\vrhob}{{\bar{\vrho}}}   \nc{\vrhot}{{\td{\vrho}}}   \nc{\vrhoh}{{\hat{\vrho}}} 
\nc{\sigmab}{{\bar{\sigma}}} \nc{\sigmat}{{\td{\sigma}}} \nc{\sigmah}{{\hat{\sigma}}} 
\nc{\vsigmab}{{\bar{\vsgm}}} \nc{\vsigmat}{{\td{\vsgm}}} \nc{\vsigmah}{{\hat{\vsgm}}} 
\nc{\taub}{{\bar{\tau}}}     \nc{\taut}{{\td{\tau}}}     \nc{\tauh}{{\hat{\tau}}} 
\nc{\upsb}{{\bar{\ups}}} \nc{\upst}{{\td{\ups}}} \nc{\upsh}{{\hat{\ups}}} 
\nc{\phib}{{\bar{\phi}}}     \nc{\phit}{{\td{\phi}}}     \nc{\phih}{{\hat{\phi}}} 
\nc{\varphib}{{\bar{\vphi}}}   \nc{\varphit}{{\td{\vphi}}}   \nc{\varphih}{{\hat{\vphi}}} 
\nc{\chib}{{\bar{\chi}}}     \nc{\chit}{{\td{\chi}}}     \nc{\chih}{{\hat{\chi}}} 
\nc{\psib}{{\bar{\psi}}}     \nc{\psit}{{\td{\psi}}}     \nc{\psih}{{\hat{\psi}}} 
\nc{\omegab}{{\bar{\omega}}} \nc{\omegat}{{\td{\omega}}} \nc{\omegah}{{\hat{\omega}}} 

\nc{\Gammab}{{\wbar{\Gamma}}}     \nc{\Gammat}{{\wtd{\Gamma}}}     \nc{\Gammah}{{\wht{\Gamma}}}
\nc{\Deltab}{{\wbar{\Delta}}}     \nc{\Deltat}{{\wtd{\Delta}}}     \nc{\Deltah}{{\wht{\Delta}}}
\nc{\Thetab}{{\wbar{\Theta}}}     \nc{\Thetat}{{\wtd{\Theta}}}     \nc{\Thetah}{{\wht{\Theta}}}
\nc{\Lambdab}{{\wbar{\Lambda}}}   \nc{\Lambdat}{{\wtd{\Lambda}}}   \nc{\Lambdah}{{\wht{\Lambda}}}
\nc{\Xib}{{\wbar{\Xi}}}           \nc{\Xit}{{\wtd{\Xi}}}           \nc{\Xih}{{\wht{\Xi}}}
\nc{\Pib}{{\wbar{\Pi}}}           \nc{\Pit}{{\wtd{\Pi}}}           \nc{\Pih}{{\wht{\Pi}}}
\nc{\Sigmab}{{\wbar{\Sigma}}}     \nc{\Sigmat}{{\wtd{\Sigma}}}     \nc{\Sigmah}{{\wht{\Sigma}}}
\nc{\Upsilonb}{{\wbar{\Upsilon}}} \nc{\Upsilont}{{\wtd{\Upsilon}}} \nc{\Upsilonh}{{\wht{\Upsilon}}}
\nc{\Phib}{{\wbar{\Phi}}} \nc{\Phit}{{\wtd{\Phi}}} \nc{\Phih}{{\wht{\Phi}}}
\nc{\Psib}{{\wbar{\Psi}}}         \nc{\Psit}{{\wtd{\Psi}}}         \nc{\Psih}{{\wht{\Psi}}}
\nc{\Omegab}{{\wbar{\Omega}}}     \nc{\Omegat}{{\wtd{\Omega}}}     \nc{\Omegah}{{\wht{\Omega}}}

\newcommand{\rmd}{\mathrm{d}}
\newcommand{\epsb}{\epsilonb}

\newcommand{\iu}{\mathrm{i}}

\title{Integrable 3D lattice model in M-theory}

\author{Junya Yagi}

\emailAdd{junyagi@tsinghua.edu.cn}

\affiliation{Yau Mathematical Sciences Center, Tsinghua University, \\
  Beijing, 100084, China}

\abstract{It is argued that the supersymmetric index of a certain
  system of branes in M-theory is equal to the partition function of
  an integrable three-dimensional lattice model.  The local Boltzmann
  weights of the lattice model satisfy a generalization of
  Zamolodchikov's tetrahedron equation.  In a special case the model
  is described by a solution of the tetrahedron equation discovered by
  Kapranov and Voevodsky and by Bazhanov and Sergeev.}


\arxivnumber{2203.09706}

\renewcommand{\circle}{{\mathbb{S}}}

\newcommand{\ten}{{\natural}}
\newcommand{\point}{{\boldsymbol{\cdot}}}
\newcommand{\rline}{{-}}
\newcommand{\interval}{{\point\mspace{-4mu}\rline\mspace{-4mu}\point}}
\newcommand{\atzero}{{\circ}}

\newcommand{\Mtwo}{\mathrm{M2}}
\newcommand{\Mfive}{\mathrm{M5}}
\newcommand{\Fone}{\mathrm{F1}}
\newcommand{\Dthree}{\mathrm{D3}}
\newcommand{\Dfour}{\mathrm{D4}}
\newcommand{\Dfive}{\mathrm{D5}}

\newcommand{\BPS}{{\mathrm{BPS}}}
\newcommand{\Man}{M}
\newcommand{\TN}{\mathrm{TN}}
\newcommand{\cigar}{D}

\newcommand{\charge}[1]{{\mathsf{#1}}}
\newcommand{\config}[1]{{\mathbf{#1}}}

\newcommand{\TDL}{\mathscr{L}}
\newcommand{\TDM}{\mathscr{M}}
\newcommand{\TDN}{\mathscr{N}}
\newcommand{\TDR}{\mathscr{R}}
\newcommand{\TDS}{\mathscr{S}}
\newcommand{\TDT}{\mathscr{T}}

\begin{document}
\maketitle

\section{Introduction}
\label{sec:introduction}

Over the past decade a family of integrable two-dimensional (2D)
lattice models have been embedded into string
theory~\cite{Yamazaki:2012cp, Yagi:2015lha, Yagi:2016oum,
  Maruyoshi:2016caf, Yagi:2017hmj, Ashwinkumar:2018tmm,
  Costello:2018txb, Ishtiaque:2021jan}.  Using open strings ending on
a stack of D5-branes, one can construct the eight-vertex model and its
trigonometric and rational limits \cite{Costello:2018txb}.  Combined
with string dualities, the brane construction provides a unified
understanding~\cite{Costello:2018txb} of the appearances of these
models in supersymmetric quantum field theories (QFTs) in diverse
spacetime dimensions \cite{Nekrasov:2009rc, Dorey:2011pa, Chen:2011sj,
  Gaiotto:2012xa, Gadde:2013ftv, Gaiotto:2015usa, Maruyoshi:2016caf,
  Yagi:2017hmj, Maruyoshi:2020cwy, Bullimore:2015lsa,
  Braverman:2016pwk, Dedushenko:2020yzd, Nekrasov:2018gne}.

In this paper I propose an analogous construction in one dimension
higher: the supersymmetric index of a certain brane system in M-theory
computes the partition function of an integrable 3D lattice model.  In
a special case the local Boltzmann weights of the model are given by a
solution of Zamolodchikov's tetrahedron equation \cite{MR611994}, a 3D
analog of the Yang--Baxter equation.  More generally, the model is
built from eight types of local Boltzmann weights.  They satisfy 16
quartic relations, which have appeared in~\cite{MR2554447,
  Yoneyama:2020duw} and will be collectively referred to here as the
supertetrahedron equation.

The brane system in question consists of M5-branes that intersect a
$3$-torus $\T^3$ along $2$-tori.  The 3D lattice model emerges on the
periodic cubic lattice formed by these $2$-tori.  The excited states
of the system contributing to the supersymmetric index contains
M2-branes suspended between M5-branes along the edges of the lattice.
These M2-branes play the roles of spin variables for the lattice
model.

The masses of the M2-branes are determined by the positions of
M5-branes in the directions transverse to $\T^3$.  When the positions
are adjusted in such a way that all M2-branes become massless, the
supersymmetric index has a description in terms of a 3D topological
quantum field theory (TQFT) defined on $\T^3$.  The M5-branes are
represented by surface defects in the TQFT.  This description enables
the supersymmetric index to be decomposed into contributions localized
around the vertices of the lattice.  These local contributions are the
local Boltzmann weights of the lattice model.

For each M5-brane, there are two discrete choices for the directions
of its worldvolume.  The two choices lead to two kinds of surface
defect in the TQFT, which we distinguish by the signs $+$ and $-$.  An
edge of the lattice is where two surface defects intersect.  The
Hilbert space for the spin variable living there depends on the pair
of signs $(\sigma_1, \sigma_2)$ representing the types of the two
defects: the bosonic Fock space if $\sigma_1 = \sigma_2$ and the
fermionic Fock space if $\sigma_1 \neq \sigma_2$.  A local Boltzmann
weight arises at a vertex of the lattice where three surface defects
intersect, and is therefore labeled by a triplet of signs
$(\sigma_1,\sigma_2,\sigma_3)$.

The topological invariance, together with the fact that the TQFT has
hidden extra dimensions in the 11D spacetime, implies that the
correlation function is invariant under movement of surface
defects~\cite{Costello:2013zra, Costello:2013sla, Yagi:2015lha,
  Yagi:2016oum}.  A local version of this statement is an equivalence
between two different arrangements of four surface defects forming
tetrahedra.  In the language of the lattice model the equivalence
translates to an equation satisfied by local Boltzmann weights, and
there is one equation for each quadruple of signs
$(\sigma_1, \sigma_2, \sigma_3, \sigma_4)$ which specifies the signs
of the four defects.  The set of equations thus obtained is the
supertetrahedron equation.  For
$(\sigma_1, \sigma_2, \sigma_3, \sigma_4) = (+,+,+,+)$ and
$(-,-,-,-)$, the equations take the form of the ordinary tetrahedron
equation.

When the M5-branes are freed from the adjusted configuration and the
M2-branes become massive, the TQFT description is deformed by
continuous parameters corresponding to the M5-brane positions.  These
parameters control the twisting of the periodic boundary conditions of
the lattice model.  The supertetrahedron equation and the existence of
the twisting parameters together imply the integrability of the model.

The tetrahedron equation is a highly overdetermined system of
equations.  If a 3D lattice model (of the vertex type) has a single
type of spin variable that takes $N$ different values, then the local
Boltzmann weights are encapsulated in an $N^3 \times N^3$ matrix,
called the R-matrix of the model.  The tetrahedron equation, on the
other hand, has $N^6 \times N^6$ components.

For the lattice model constructed from branes in this paper,
constraints imposed on the local Boltzmann weights are even more
stringent.  The local Boltzmann weights are defined by eight
R-matrices $R^{\sigma_1\sigma_2\sigma_3}$, $\sigma_1$, $\sigma_2$,
$\sigma_3 = \pm$, satisfying intricate relations generalizing the
tetrahedron equation.  Although there are symmetries that reduce the
number of independent R-matrices to three, the supertetrahedron
equation may still appear too constraining to admit a solution.

Remarkably, a solution of the supertetrahedron equation is
known~\cite{MR2554447, Yoneyama:2020duw}.  For this solution,
$R^{+++}$ is the R-matrix discovered independently by Kapranov and
Voevodsky \cite{MR1278735} and by Bazhanov and Sergeev
\cite{Bazhanov:2005as}, whereas $R^{++-}$ was obtained in
\cite{Bazhanov:2005as} where it was called an L-operator.  The
remaining R-matrices were found by Yoneyama~~\cite{Yoneyama:2020duw}.
Furthermore, these R-matrices possess all the properties that we
expect for the R-matrices of our model to have, such as a charge
conservation rule and involutivity.  Therefore, I propose that the
lattice model constructed from branes is described by this solution of
the supertetrahedron equation.

Probably the most convincing evidence for this proposal is that the
behavior of the brane system under string dualities correctly
reproduces the behavior of the 3D lattice model defined by this
solution under reduction along one of the directions of the lattice.
By reduction to 10D and T-duality, the brane system under
consideration is mapped to a brane configuration studied in
\cite{Costello:2018txb, Ishtiaque:2021jan}, which constructs a 2D
lattice model defined by a trigonometric solution of the Yang--Baxter
equation whose symmetry algebra is a general linear Lie superalgebra
$\glf(L_+|L_-)$.  This is consistent with the results of
\cite{Bazhanov:2005as, MR2554447, Kuniba:2015sca}.

Turning the logic around, the brane construction of this paper
explains characteristic features of the above solution of the
supertetrahedron equation.  For example, the R-matrices are constant
and the spectral parameters, which are crucial for the integrability
of the model, enter the partition function as twisting parameters for
the periodic boundary conditions.  This is in contrast with familiar
solutions of the Yang--Baxter equation describing integrable 2D
lattice models, for which spectral parameters directly appear in the
R-matrices.  As we will see, this feature finds a natural explanation
in the brane construction.

The rest of the paper is organized as follows.  In
section~\ref{sec:branes} we introduce the brane system whose
supersymmetric index is argued to define an integrable 3D lattice
model.  Section~\ref{sec:TQFT} discusses this system from the point of
view of 3D TQFT.  We explain how the structure of a 3D TQFT arises
from the special case of the brane configuration, how the 3D TQFT
gives rise to a 3D lattice model in the presence of surface defects,
and how the existence of extra dimensions implies the tetrahedron
equation.  In section~\ref{sec:properties} we deduce key properties of
the lattice model based on the brane construction.  Finally, in
section~\ref{sec:identification} we identify a solution of the
supertetrahedron equation that has the properties deduced in the
previous section and which I propose describes this lattice model.  An
analysis of the supersymmetry preserved by the brane system is carried
out in appendix~\ref{sec:susy}.

\section{3D lattice model constructed from branes}
\label{sec:branes}

We will construct an integrable 3D lattice model in M-theory
formulated in the 11D spacetime
\begin{equation}
  \label{eq:spacetime}
  \R_0 \times \T^3_{123} \times \R^3_{456} \times \R^2_{78} \times \R^2_{9\ten} \,,
\end{equation}
endowed with the flat metric
\begin{equation}
  \eta =
  -(\rmd x^0)^2 + (\rmd x^1)^2 + (\rmd x^2)^2 + \dotsb + (\rmd x^9)^2
  + (\rmd x^\ten)^2 \,.
\end{equation}
Here $\R^n_{\mu_1 \mu_2 \dots \mu_n}$ is an $n$-dimensional Euclidean
space with coordinates $(x^{\mu_1}, x^{\mu_2}, \dotsc, x^{\mu_n})$ and
$\T^3_{123}$ is a $3$-torus with periodic coordinates
$(x^1, x^2, x^3)$, respectively.  We use the notation $\ten = 10$.  We
will write $\circle_\mu$ to denote a circle with periodic coordinate
$x^\mu$.

To this spacetime we introduce six types of M5-branes, which we call
\emph{type $i\sigma$}, $i = 1$, $2$, $3$, $\sigma = +$, $-$.
M5-branes of different types have worldvolumes extending in different
directions, as listed in table \ref{tab:M5-M2}.  We will say that an
M5-brane of type $i\sigma$ has \emph{sign $\sigma$} and refer to
M5-branes of either type $i+$ or type $i-$ simply as M5-branes of
\emph{type $i$}.  The orientations of the M5-branes are chosen in such
a way that the setup is symmetric under the cyclic permutation
$(1,4) \to (2,5) \to (3,6) \to (1,4)$ and the exchange
$(7,8) \leftrightarrow (9,\ten)$ of directions.

\begin{table}
  \centering
  \begin{tabular}{l|c|ccc|ccc|cc|cc}
    \hline \hline
    & 0 & 1 & 2 & 3 & 4 & 5 & 6 & 7 & 8 & 9 & $\ten$
    \\
    \hline
    $\Mfive_{1+}$
    & $\rline$
    & $\point$ & $\rline$ & $\rline$
    & $\rline$ & $\point$ & $\point$
    & $\rline$ & $\rline$
    & $\atzero$ & $\atzero$
    \\
    $\Mfive_{2+}$
    & $\rline$
    & $\rline$ & $\point$ & $\rline$
    & $\point$ & $\rline$ & $\point$
    & $\rline$ & $\rline$
    & $\atzero$ & $\atzero$
    \\
    $\Mfive_{3+}$
    & $\rline$
    & $\rline$ & $\rline$ & $\point$
    & $\point$ & $\point$ & $\rline$
    & $\rline$ & $\rline$
    & $\atzero$ & $\atzero$
    \\
    \hline
    $\Mfive_{1-}$
    & $\rline$
    & $\point$ & $\rline$ & $\rline$
    & $\rline$ & $\point$ & $\point$
    & $\atzero$ & $\atzero$
    & $\rline$ & $\rline$
    \\
    $\Mfive_{2-}$
    & $\rline$
    & $\rline$ & $\point$ & $\rline$
    & $\point$ & $\rline$ & $\point$
    & $\atzero$ & $\atzero$
    & $\rline$ & $\rline$
    \\
    $\Mfive_{3-}$
    & $\rline$
    & $\rline$ & $\rline$ & $\point$
    & $\point$ & $\point$ & $\rline$
    & $\atzero$ & $\atzero$
    & $\rline$ & $\rline$
    \\
    \hline
    $\Mtwo_1$
    & $\rline$
    & $\rline$ & $\point$ & $\point$
    & $\interval$ & $\point$ & $\point$
    & $\atzero$ & $\atzero$
    & $\atzero$ & $\atzero$
    \\
    $\Mtwo_2$
    & $\rline$
    & $\point$ & $\rline$ & $\point$
    & $\point$ & $\interval$ & $\point$
    & $\atzero$ & $\atzero$
    & $\atzero$ & $\atzero$
    \\
    $\Mtwo_3$
    & $\rline$
    & $\point$ & $\point$ & $\rline$
    & $\point$ & $\point$ & $\interval$
    & $\atzero$ & $\atzero$
    & $\atzero$ & $\atzero$
    \\
    \hline \hline
  \end{tabular}
  \caption{A configuration of M5-branes and M2-branes.  The symbol
    $\rline$ indicates that the brane extends in the corresponding
    direction, while a circle $\atzero$ or a dot $\point$ means that
    it is located at the origin or any point, respectively.  The
    symbol $\interval$ means an interval between two points; M2-branes
    are suspended between two M5-branes.}

  \label{tab:M5-M2}
\end{table}

This configuration of M5-branes is invariant under 1/8 of the
supersymmetry of M-theory.  The unbroken supersymmetry is generated by
two supercharges $Q_+$, $Q_-$, defined and analyzed in
appendix~\ref{sec:susy}.

These supercharges have a couple of properties that are important to
us.  One is that
\begin{equation}
  Q_+^2 + Q_-^2
  = \frac12 (H - Z^{(2)}_{14} - Z^{(2)}_{25} - Z^{(2)}_{36} - Y) \,,
\end{equation}
where $H = P^0$ is the Hamiltonian, $Z^{(2)}_{14}$, $Z^{(2)}_{25}$,
$Z^{(2)}_{36}$ are components of a $2$-form charge $Z^{(2)}$, and $Y$
is a linear combination of components of a $5$-form charge $Z^{(5)}$.
(The momentum $P$ and the charges $Z^{(2)}$, $Z^{(5)}$ commute with
the supercharges and with each other.)  Since $Q_\pm$ are hermitian,
the right-hand side is nonnegative.

Another property of $Q_\pm$ is the invariance under the action of the
antidiagonal subgroup of the rotation groups $\SO(2)_{78}$ of
$\R^2_{78}$ and $\SO(2)_{9\ten}$ of $\R^2_{9\ten}$.  In other words,
$Q_\pm$ commute with the difference $J_{78} - J_{9\ten}$ of the
generators $J_{78}$ of $\SO(2)_{78}$ and $J_{9\ten}$ of
$\SO(2)_{9\ten}$:
\begin{equation}
  [J_{78} - J_{9\ten}, Q_\pm] = 0 \,.
\end{equation}
To make use of this property, we place the M5-branes at either the
origin of $\R^2_{78}$ or the origin of $\R^2_{9\ten}$.  Then, the
M5-brane system is invariant under the action of the antidiagonal
subgroup of $\SO(2)_{78} \times \SO(2)_{9\ten}$, and its Hilbert space
$\CH$ (the subspace of the Hilbert space of M-theory consisting of
states that contain this configuration of M5-branes) is acted on by
$J_{78} - J_{9\ten}$ in addition to $Q_\pm$.

Now, let us define the supersymmetric index of the M5-brane system by
\begin{equation}
  \label{eq:Z}
  Z
  =
  \Tr_\CH\bigl((-1)^F e^{\iu\theta(J_{78} - J_{9\ten})} e^{-\beta(H - E_0)}\bigr) \,,
\end{equation}
where $E_0$ is the ground state energy.  The index $Z$ can be computed
by the Euclidean path integral in which the imaginary time $\iu x^0$
is periodic with period $\beta$.  In this language, we have chosen a
periodic boundary condition for fermions, which is responsible for the
appearance of the fermion parity operator $(-1)^F$.  The insertion of
$e^{\iu\theta(J_{78} - J_{9\ten})}$ means that the periodic boundary
condition is twisted: if the time direction is viewed as the interval
$[0,\beta]$ with the end points identified, the Hilbert space at
$\iu x^0 = \beta$ is glued back to the Hilbert space at $\iu x^0 = 0$
after $\R^2_{78}$ and $\R^2_{9\ten}$ are rotated by angles $+\theta$
and $-\theta$, respectively.

The Hilbert space $\CH$ of the M5-brane system is graded by
$J_{78} - J_{9\ten}$, which we normalize to take integer values: $\CH$
has the decomposition
\begin{equation}
  \CH = \bigoplus_{j=-\infty}^{\infty} \CH^j \,,
\end{equation}
where $\CH^j$ is the eigenspace of $J_{78} - J_{9\ten}$ with
eigenvalue $j$.  Hence, the partition function is a formal power
series
\begin{equation}
  Z = \sum_{j=-\infty}^\infty Z_j q^j
  \,,
  \quad
  q = e^{\iu\theta} \,,
\end{equation}
with the coefficients given by
\begin{equation}
  Z_j = \Tr_{\CH^j} \bigl((-1)^F e^{-\beta(H - E_0)}\bigr) \,.
\end{equation}

Eventually, we will identify $Z$ with the partition function of an
integrable 3D lattice model.  To see how such a model can arise from
$Z$, let us look at the configuration of M5-branes a little more
closely.  The M5-branes generically do not intersect with each other
in the 11D spacetime because any two of them can be separated in
$\R^3_{678}$.  Inside $\T^3_{123}$, however, they represent
intersecting $2$-tori that form a periodic cubic lattice.%
\footnote{Analogous cubic lattice configurations in M-theory were
  considered in~\cite{Duff:1997fd} where the authors were motivated by
  an observed structure of galaxy superclusters.}
Given a lattice made up by M5-branes, the operation of taking its
supersymmetric index produces a complex number.  In this sense, the
brane system defines a physical model associated with a 3D lattice.

Although this is certainly what we want, it is a rather weak
conclusion.  Why should the model thus defined by the M5-branes be a
lattice model in the usual sense of the term in statistical mechanics,
that is, a model describing interactions of ``spin variables'' located
at lattice points?

The answer is because only special states contribute to $Z$.  We can
write $Z$ as the sum of a trace over $\ker Q_+ = \ker Q_+^2$ and a
trace over the orthogonal complement $(\ker Q_+)^\perp$ of $\ker Q_+$.
In the latter subspace $Q_+$ is invertible.  (In the eigenspace of
$Q_+^2$ with eigenvalue $q_+^2 > 0$, we have $Q_+^{-1} = Q_+/q_+^2$.)
The action of $Q_+$ gives a one-to-one correspondence between the
bosonic states and the fermionic states in $(\ker Q_+)^\perp$, and
their contributions to $Z$ cancel.  As a result, only the states in
$\ker Q_+$ contribute to $Z$.  The same argument applies to $Q_-$
(which leaves $\ker Q_+$ invariant).  Hence, $Z$ receives
contributions only from $\ker Q_+ \cap \ker Q_-$, which is the space
of states whose energy satisfies the
\emph{Bogomol'nyi--Prasad--Sommerfield (BPS) condition}
\begin{equation}
  H = Z^{(2)}_{14} + Z^{(2)}_{25} + Z^{(2)}_{36} + Y
  \,.
\end{equation}
%

The relevant BPS states are represented by configurations of M2-branes
and M5-branes added to the system on top of those M5-branes that are
already present.  BPS M2-branes contribute to $Z^{(2)}_{14}$,
$Z^{(2)}_{25}$, $Z^{(2)}_{36}$, whereas BPS M5-branes contribute to
$Y$.

We can introduce, for example, an M5-brane along $\R^{5,1}_{045678}$.
However, this M5-brane has an infinite spatial volume, so adding it to
the system increases the energy by an infinite amount and BPS states
containing it do not contribute to $Z$.  By the same token, any BPS
states with additional M5-branes make no contributions to $Z$ because
those M5-branes necessarily have infinite spatial volumes.%
\footnote{From the expression \eqref{eq:Y} of $Y$ we see that there
  may also be BPS states that have nonzero $Z^{(5)}_{01245}$,
  $Z^{(5)}_{01346}$, $Z^{(5)}_{02356}$ or $Z^{(5)}_{0789\ten}$.  Such
  a BPS state corresponds to a spacetime in which
  $\T^2_{12} \times \R^2_{45}$, $\T^2_{23} \times \R^2_{56}$,
  $\T^2_{13} \times \R^2_{46}$ or $\R^4_{789\ten}$ is replaced by a
  Taub--NUT space.  Since a Taub--NUT geometry becomes an infinite
  D6-brane in type IIA string theory, it does not contribute to $Z$.
  In section \ref{sec:TQFT} we will modify the system so that it has a
  Taub--NUT space in place of $\R^4_{789\ten}$.}
For this reason, on the states relevant to $Z$ the value of $Y$ is
fixed to the value for the initial brane configuration, which we have
called $E_0$.  In the definition \eqref{eq:Z} of $Z$, we subtracted
$E_0$ from $H$ to offset the ground state energy so that $Z$ can be
nonvanishing.

In contrast, BPS M2-branes can have finite spatial area.  There are
three types of BPS M2-branes, which we call \emph{type $i$}, $i = 1$,
$2$, $3$.  An M2-brane of type $i$ extends in the directions of
$\R_0 \times \circle_i \times \R_{i+3}$.  For example, the spatial
area of an M2-brane of type $1$ is finite if it stretches between two
M5-branes, one of type $2$ and one of type $3$, both of which are
localized at points in $\R_4$.  Therefore, the contributions to $Z$
come from those BPS states that only have finite-area M2-branes
excited.  Writing $\CH_\BPS$ for the space of such BPS states, we have
\begin{equation}
  Z_j
  = \Tr_{\CH_\BPS^j} \bigl((-1)^F
    e^{-\beta(Z^{(2)}_{14} + Z^{(2)}_{25} + Z^{(2)}_{36})}\bigr) \,,
\end{equation}
where
\begin{equation}
  \CH_\BPS^j = \CH_\BPS \cap \CH^j \,.
\end{equation}

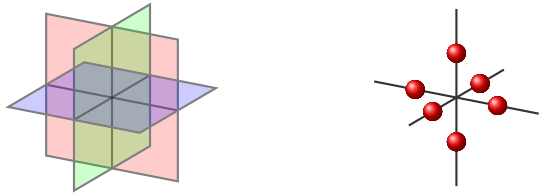
\begin{figure}
  \centering
  \tdplotsetmaincoords{70}{120}
  \begin{tikzpicture}[tdplot_main_coords, scale=0.5]
    \draw[thick, opacity=0.5] (0,-2,0) -- (0,2,0);
    \draw[thick, opacity=0.5] (2,0,0) -- (-2,0,0);
    \draw[thick, opacity=0.5] (0,0,-2) -- (0,0,2);
    
    \draw[fill=red, opacity=0.2] (0,-2,-2) -- (0,2,-2)-- (0,2,2) -- (0,-2,2)
    -- cycle;
    \draw[black!50, thick] (0,-2,-2) -- (0,2,-2) -- (0,2,2) -- (0,-2,2) -- cycle;
    
    \draw[fill=green, opacity=0.2] (-2,0,-2) -- (-2,0,2) -- (2,0,2) -- (2,0,-2)
    -- cycle;
    \draw[black!50, thick] (-2,0,-2) -- (-2,0,2) -- (2,0,2) -- (2,0,-2) -- cycle;

    \draw[fill=blue, opacity=0.2] (-2,-2,0) -- (2,-2,0) -- (2,2,0) -- (-2,2,0)
    -- cycle;
    \draw[black!50, thick] (-2,-2,0) -- (2,-2,0) -- (2,2,0) -- (-2,2,0) -- cycle;
  \end{tikzpicture}
  \qquad\qquad
  \tdplotsetmaincoords{70}{120}
  \begin{tikzpicture}[tdplot_main_coords, scale=0.5]
    \draw[black!80, thick] (0,-2.5,0) -- (0,2.5,0);
    \draw[black!80, thick] (2.5,0,0) -- (-2.5,0,0);
    \draw[black!80, thick] (0,0,-2.5) -- (0,0,2.5);
  
    \tdplottransformmainscreen{0}{0}{-1.25}
    \shade[tdplot_screen_coords, ball color=red, opacity=0.9] (\tdplotresx,\tdplotresy) circle (0.25);
    \tdplottransformmainscreen{0}{-1.25}{0}
    \shade[tdplot_screen_coords, ball color=red, opacity=0.9] (\tdplotresx,\tdplotresy) circle (0.25);
    \tdplottransformmainscreen{-1.25}{0}{0}
    \shade[tdplot_screen_coords, ball color=red, opacity=0.9] (\tdplotresx,\tdplotresy) circle (0.25);
    \tdplottransformmainscreen{0}{0}{1.25}
    \shade[tdplot_screen_coords, ball color=red, opacity=0.9] (\tdplotresx,\tdplotresy) circle (0.25);
    \tdplottransformmainscreen{0}{1.25}{0}
    \shade[tdplot_screen_coords, ball color=red, opacity=0.9] (\tdplotresx,\tdplotresy) circle (0.25);
    \tdplottransformmainscreen{1.25}{0}{0}
    \shade[tdplot_screen_coords, ball color=red, opacity=0.9] (\tdplotresx,\tdplotresy) circle (0.25);
  \end{tikzpicture}

  \caption{A neighborhood of a vertex where three M5-branes intersect.
    The spin variables on the six edges, represented here by six
    balls, are the numbers of M2-branes suspended between M5-branes
    along the edges.}

\label{fig:lattice}
\end{figure}

Let us focus on a single vertex of the lattice and ask what sort of
configurations of M2-branes are possible around it.  Six edges meet at
the vertex, two from the intersection of M5-branes of type $1$ and
type $2$, two from type $2$ and type $3$, and two from type $3$ and
type $1$; see figure~\ref{fig:lattice}.  Take the pair of edges where
the M5-branes of type $2$ and type $3$ intersect.  Along these edges
we can suspend M2-branes of type $1$.  The number of suspended
M2-branes may be different for the two edges.  Similarly, we can
suspend some numbers of M2-branes of type $2$ between the M5-branes of
type $3$ and type $1$ along the pair of edges coming from these
M5-branes, and some numbers of M2-branes of type $3$ between the
M5-branes of type $1$ and type $2$ along the remaining pair of edges.

Thus we can assign an integer to each edge of the lattice, namely the
number of M2-branes suspended between M5-branes along that edge.  We
can interpret this integer as a spin variable of the lattice model.
Near a vertex the M2-branes along the six edges combine in some
manner, giving rise to interactions among the spin variables.

The picture that emerges from the above consideration is that the
supersymmetric index of the brane system defines what is known as a
\emph{vertex model} in statistical mechanics: spin variables live on
the edges and interact at the vertices of a lattice.

We have understood, more or less, how a lattice model arises from the
brane system.  An important question remains to be answered,
however: why should this model be integrable?

\section{TQFT origin of the model}
\label{sec:TQFT}

Before trying to answer the question of integrability, we should first
understand the origin of the lattice model better.  To this end let us
consider the special case of the above brane system in which the
positions of the M5-branes are adjusted so that all of them intersect
the origin of $\R^3_{456}$.  This is not so severe a restriction
because, as it turns out, the general case merely differs by twisting
of the periodic boundary conditions of the lattice model.  We will see
that the supersymmetric index of the brane system in this special case
has the structure of a 3D TQFT, and this structure is what underlies
the emergence of the lattice model.  The line of reasoning in this
section in large part follows similar arguments given in
\cite{Yagi:2015lha, Yagi:2016oum}, which adapt ideas of
Costello~\cite{Costello:2013zra, Costello:2013sla} to brane
constructions of gauge theories.

\subsection{From branes to 3D TQFT}
\label{sec:branes-TQFT}

While we restrict ourselves to the special case of the brane system as
just mentioned, we also make the following two generalizations to the
setup.

First, we replace $\T^3_{123} \times \R^3_{456}$ with the cotangent
bundle $T^*\Man$ of a closed oriented Riemannian 3-manifold
$\Man$.%
\footnote{\label{fn:X} More precisely, we isometrically embed $\Man$ into a
  Calabi--Yau threefold $X$ as a special Lagrangian submanifold, which
  is always possible \cite{MR1795105}.  A neighborhood of
  $\Man \subset X$ can be identified with a neighborhood of the zero
  section of $T^*\Man$.  For the following discussion it is sufficient
  to consider this neighborhood.}
The total space of $T^*\Man$ is the phase space of a particle moving
in $\Man$, locally parametrized by coordinates $(x^1, x^2, x^3)$ on
$\Man$ and their conjugate momenta $(p_1, p_2, p_3)$.  The zero
section of $T^*\Man$, where $p_1 = p_2 = p_3 = 0$, may be identified
with $\Man$.  In the original setup, $\T^3_{123} \times \R^3_{456}$
can be regarded as $T^*\T^3_{123}$ by the identification
$p_i = x^{i+3}$, $i = 1$, $2$, $3$.  There, the M5-branes wrap
submanifolds parametrized by the coordinates $(x^2, x^3, p_1)$,
$(x^3, x^1, p_2)$ or $(x^1, x^2, p_3)$, and we have specialized the
positions of the M5-branes in $\R^3_{456}$ so that these submanifolds
intersect within the zero section.  In the generalized setup, we
require each M5-brane to wrap the conormal bundle $N^*\Sigma$ of some
oriented surface $\Sigma \subset \Man$.  This is a subbundle of
$T^*\Man|_\Sigma$ consisting of all cotangent vectors that annihilate
the tangent vectors of $\Sigma$.  For two surfaces $\Sigma_1$,
$\Sigma_2 \subset \Man$ intersecting along a curve, their conormal
bundles $N^*\Sigma_1$, $N^*\Sigma_2$ intersect solely in the zero
section $\Man \subset T^*\Man$ along $\Sigma_1 \cap \Sigma_2$; if
$(x,p) \in N^*\Sigma_1 \cap N^*\Sigma_2$, then $p(v_1) = p(v_2) = 0$
for all $v_1 \in T_x\Sigma_1$ and $v_2 \in T_x\Sigma_2$, which implies
$p = 0$ since $T_x\Sigma_1 + T_x\Sigma_2 = T_xM$.

Second, we replace $\R^4_{789\ten}$ with a Taub--NUT space $\TN$.
This space is a hyperk\"ahler manifold and can be presented as a
circle fibration over $\R^3$. The circle fiber shrinks to a point at
the origin of $\R^3$.  Along a half-line emanating from the origin,
the radius of the fiber increases and approaches an asymptotic value
$R$.  The fibers over the half-line thus make a cigar shape
homeomorphic to $\R^2$.  We choose a line through the origin of the
base $\R^3$ and view it as the union of two half-lines $\R^+$ and
$\R^-$ meeting at the origin.  The fibers over $\R^+$ and $\R^-$ form
two cigars, $\cigar^+$ and $\cigar^-$, touching each other at the
tips.  We wrap each M5-brane on either $\cigar^+$ or $\cigar^-$.

In summary, in the modified brane system the spacetime is
\begin{equation}
  \R_0 \times T^*M \times \TN \,,
\end{equation}
and each M5-brane is supported on a submanifold of the form
\begin{equation}
  \R_0 \times N^*\Sigma \times \cigar^{\sigma[\Sigma]} \,.
\end{equation}
The worldvolume of an M5-brane is specified by a \emph{signed surface}
$(\Sigma,\sigma[\Sigma])$, a pair of an oriented surface
$\Sigma \subset M$ and a sign $\sigma[\Sigma] \in \{+, -\}$.  If we
choose $M = \T^3$ and take the limit $R \to \infty$ (in which $\TN$
reduces to $\R^4$), the modified spacetime becomes the one considered
in section~\ref{sec:branes}; the M5-branes are allowed to wrap more
general surfaces in $\T^3$ but required to intersect the origin of
$\R^3_{456}$.  As in the original brane system, the modified system
preserves two supercharges.

Let us discuss implications of the modifications we have made to the
brane system one by one.

By forcing the M5-branes to intersect $\R^3_{456}$ at the origin, we
have reduced the spatial areas of BPS M2-branes suspended between
M5-branes to zero.  Thus we have made all M2-branes massless.  Then,
the states in $\CH_\BPS$ all have the same energy $H = E_0$, and the
index $Z$ reduces to
\begin{equation}
  \label{eq:Z0}
  \mathring{Z}
  =
  \sum_{j=-\infty}^\infty
  \Tr_{\CH_\BPS^j} (-1)^F q^j \,.
\end{equation}
The integer coefficient of $q^j$ in $\mathring{Z}$ is known as the
\emph{Witten index} of $\CH^j$.

What happens to the M2-branes when we replace
$\T^3_{123} \times \R^3_{456}$ with $T^*\Man$?  Suppose that two
M5-branes wrap surfaces $\Sigma_1$, $\Sigma_2 \subset \Man$ that
intersect along a curve.  The cotangent bundle
$T^*(\Sigma_1 \cap \Sigma_2)$ of this curve may be considered as a
submanifold of $T^*\Man$ via the sequence of maps
$T^*(\Sigma_1 \cap \Sigma_2) \iso T(\Sigma_1 \cap \Sigma_2)
\hookrightarrow TM \iso T^*M$, where the isomorphisms between the
tangent and cotangent bundles are induced by the canonical symplectic
forms.  (The canonical symplectic form of $T^*M$ is
$\sum_{i=1}^3 \rmd x^i \wedge \rmd p_i$.)  An M2-brane suspended
between these two M5-branes is BPS if its spatial extent lies in
$T^*(\Sigma_1 \cap \Sigma_2)$.  Since both M5-branes intersect
$T^*(\Sigma_1 \cap \Sigma_2)$ only along the curve
$\Sigma_1 \cap \Sigma_2 \subset M \subset T^*M$, such an M2-brane has
a zero spatial area.  Hence, the index is still a series of the
form~\eqref{eq:Z0}.

The coefficients being Witten indices, $\mathring{Z}$ is invariant
under continuous deformations of parameters of the system: although
some states may enter or leave $\CH_\BPS^j$ under deformations, due to
supersymmetry such changes occur in boson--fermion pairs and their
contributions to the Witten index of $\CH^j$ cancel.  In particular,
$\mathring{Z}$ is protected against deformations of the metric of $M$
and the shapes of surfaces in $M$ on which the M5-branes are
supported.  Since the M5-brane worldvolumes are completely specified
by these surfaces and the choice of their signs, we may think of
$\mathring{Z}$ as a topological invariant of the configuration of
signed surfaces in $M$.

Then we replaced $\R^4_{789\ten}$ with the Taub--NUT space $\TN$.  The
spacetime now contains a spatial circle around which all M5-branes
wrap, the circle fiber of $\TN$.  If we reduce M-theory to type IIA
string theory on this circle, the spacetime becomes
$\R_0 \times T^*M \times \R^3$ and a D6-brane with worldvolume
$\R_0 \times T^*M \times \{0\}$ appears.  An M5-brane with worldvolume
$\R_0 \times N^*\Sigma \times D^{\sigma[\Sigma]}$ is converted to a
D4-brane with worldvolume
$\R_0 \times N^*\Sigma \times \R^{\sigma[\Sigma]}$, ending on the
D6-brane.  The resulting D4--D6 brane system is summarized in
table~\ref{tab:D4-D6}, where $D^\pm$ are chosen to be the cigars along
$\R^\pm_7$ in the base $\R^3_{789}$ of $\TN$.

\begin{table}
  \centering
  \begin{tabular}{l|c|ccc|ccc|ccc}
    \hline \hline
    & 0 & 1 & 2 & 3 & 4 & 5 & 6 & 7 & 8 & 9
    \\
    \hline
    $\mathrm{D6}$
    & $-$
    & $-$ & $-$ & $-$
    & $-$ & $-$ & $-$
    & $\circ$ & $\circ$ & $\circ$
    \\
    \hline
    $\mathrm{D4}_{1+}$
    & $-$
    & $\point$ & $-$ & $-$
    & $-$ & $\circ$ & $\circ$
    & $\circ\!-$ & $\circ$ & $\circ$
    \\
    $\mathrm{D4}_{2+}$
    & $-$
    & $-$ & $\point$ & $-$
    & $\circ$ & $-$ & $\circ$
    & $\circ\!-$ & $\circ$ & $\circ$
    \\
    $\mathrm{D4}_{3+}$
    & $-$
    & $-$ & $-$ & $\point$
    & $\circ$ & $\circ$ & $-$
    & $\circ\!-$ & $\circ$ & $\circ$
    \\
    \hline
    $\mathrm{D4}_{1-}$
    & $-$
    & $\point$ & $-$ & $-$
    & $-$ & $\circ$ & $\circ$
    & $-\!\circ$ & $\circ$ & $\circ$
    \\
    $\mathrm{D4}_{2-}$
    & $-$
    & $-$ & $\point$ & $-$
    & $\circ$ & $-$ & $\circ$
    & $-\!\circ$ & $\circ$ & $\circ$
    \\
    $\mathrm{D4}_{3-}$
    & $-$
    & $-$ & $-$ & $\point$
    & $\circ$ & $\circ$ & $-$
    & $-\!\circ$ & $\circ$ & $\circ$
    \\
    \hline \hline
  \end{tabular}
  \caption{The D4--D6 brane configuration defining a 3D TQFT.  The
    symbols $\circ\!-$ and $-\!\circ$ mean that the brane extends
    along the positive and negative half-lines, respectively.}

  \label{tab:D4-D6}
\end{table}

The low-energy physics of a D6-brane is described by a 7D QFT, and the
D4-branes create 4D defects in this theory.  This QFT description is
sufficient for us as we are interested in $\mathring{Z}$, which
receives contributions only from zero-energy states.  To compute
$\mathring{Z}$, we perform a Wick rotation to Euclidean signature,
place the theory on $\circle_0 \times T^*M$, and twist the periodic
boundary condition in the time direction by the $\U(1)$ symmetry
originating from the rotation symmetry of the circle fibers of $\TN$.
In the presence of the codimension-two defects created by the
D4-branes, the path integral yields $\mathring{Z}$.

The 7D theory on $\circle_0 \times T^*M$ may be regarded as a 3D
theory on $M$, with fields taking values in infinite-dimensional
spaces.  For example, the theory has three real scalar fields
parametrizing the positions of the D6-brane in $\R^3_{789}$, and they
can be thought of as sections of a fiber bundle over $M$ whose fiber
at $x \in M$ is the space of maps from $\circle_0 \times T^*_xM$ to
$\R$.  From this point of view, $\mathring{Z}$ is captured by a 3D QFT
which has two types of surface defects distinguished by signs.

Combining what we have deduced, we reach the following conclusion: the
reduced index $\mathring{Z}$ defines a 3D TQFT that assigns to a
configuration of signed surfaces in a 3-manifold $M$ a formal power
series in $q$ and $q^{-1}$ with integer coefficients.

\subsection{From 3D TQFT to 3D lattice model}
\label{sec:TQFT-lattice}

With the understanding of $\mathring{Z}$ as a 3D TQFT with surface
defects, we can say more about its structures using properties of
TQFT.  We argue that for $M = \T^3$ and surface defects making a cubic
lattice, $\mathring{Z}$ is the partition function of a 3D lattice
model.

Suppose that a 3D TQFT is placed on a $3$-manifold $M$.  For
simplicity let us assume that the theory has only one type of surface
defects.  Given a configuration of surface defects, we wish to compute
their correlation function.

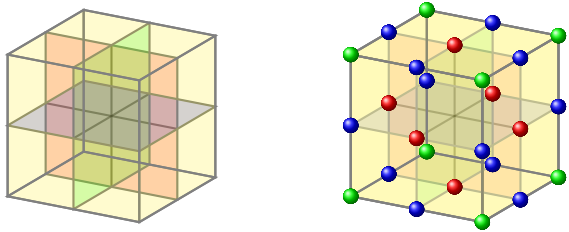
\begin{figure}
  \centering
  \tdplotsetmaincoords{70}{120}
  \begin{tikzpicture}[tdplot_main_coords, scale=0.5]
    \draw[thick, opacity=0.5] (0,-2,0) -- (0,2,0);
    \draw[thick, opacity=0.5] (2,0,0) -- (-2,0,0);
    \draw[thick, opacity=0.5] (0,0,-2) -- (0,0,2);

    \draw[fill=red, opacity=0.2] (0,-2,-2) -- (0,2,-2)-- (0,2,2) -- (0,-2,2)
    -- cycle;
    \draw[black!50, thick] (0,-2,-2) -- (0,2,-2) -- (0,2,2) -- (0,-2,2) -- cycle;

    \draw[fill=green, opacity=0.2] (-2,0,-2) -- (-2,0,2) -- (2,0,2) -- (2,0,-2)
    -- cycle;
    \draw[black!50, thick] (-2,0,-2) -- (-2,0,2) -- (2,0,2) -- (2,0,-2) -- cycle;

    \draw[fill=blue, opacity=0.2] (-2,-2,0) -- (2,-2,0) -- (2,2,0) -- (-2,2,0)
    -- cycle;
    \draw[black!50, thick] (-2,-2,0) -- (2,-2,0) -- (2,2,0) -- (-2,2,0) -- cycle;

    \draw[fill=yellow, opacity=0.1] (-2,-2,-2) -- (-2,2,-2)-- (-2,2,2) -- (-2,-2,2) -- cycle;
    \draw[black!50, thick] (-2,-2,-2) -- (-2,2,-2) -- (-2,2,2) -- (-2,-2,2) -- cycle;

    \draw[fill=yellow, opacity=0.1] (2,-2,-2) -- (2,2,-2)-- (2,2,2) -- (2,-2,2)
    -- cycle;
    \draw[black!50, thick] (2,-2,-2) -- (2,2,-2) -- (2,2,2) -- (2,-2,2) -- cycle;

    \draw[fill=yellow, opacity=0.1] (-2,-2,-2) -- (-2,-2,2) -- (2,-2,2) -- (2,-2,-2) -- cycle;
    \draw[black!50, thick] (-2,-2,-2) -- (-2,-2,2) -- (2,-2,2) -- (2,-2,-2) -- cycle;

    \draw[fill=yellow, opacity=0.1] (-2,2,-2) -- (-2,2,2) -- (2,2,2) -- (2,2,-2) -- cycle;
    \draw[black!50, thick] (-2,2,-2) -- (-2,2,2) -- (2,2,2) -- (2,2,-2) -- cycle;

    \draw[fill=yellow, opacity=0.1] (-2,-2,-2) -- (2,-2,-2) -- (2,2,-2) -- (-2,2,-2)
     -- cycle;
    \draw[black!50, thick] (-2,-2,-2) -- (2,-2,-2) -- (2,2,-2) -- (-2,2,-2) -- cycle;

    \draw[fill=yellow, opacity=0.1] (-2,-2,2) -- (2,-2,2) -- (2,2,2) -- (-2,2,2)
    -- cycle;
    \draw[black!50, thick] (-2,-2,2) -- (2,-2,2) -- (2,2,2) -- (-2,2,2) -- cycle;
  \end{tikzpicture}
  \qquad\qquad
  \begin{tikzpicture}[tdplot_main_coords, scale=0.5]
    \draw[thick, opacity=0.3] (0,-2,0) -- (0,2,0);
    \draw[thick, opacity=0.3] (2,0,0) -- (-2,0,0);
    \draw[thick, opacity=0.3] (0,0,-2) -- (0,0,2);

    \draw[fill=yellow, opacity=0.1] (2,-2,-2) -- (2,2,-2)-- (2,2,2) -- (2,-2,2)
    -- cycle;
    \draw[black!50, thick] (2,-2,-2) -- (2,2,-2) -- (2,2,2) -- (2,-2,2) -- cycle;
    
    \draw[fill=yellow, opacity=0.1] (-2,2,-2) -- (-2,2,2) -- (2,2,2) -- (2,2,-2) -- cycle;
    \draw[black!50, thick] (-2,2,-2) -- (-2,2,2) -- (2,2,2) -- (2,2,-2) -- cycle;
    
    \draw[fill=yellow, opacity=0.1] (-2,-2,2) -- (2,-2,2) -- (2,2,2) -- (-2,2,2)
    -- cycle;
    \draw[black!50, thick] (-2,-2,2) -- (2,-2,2) -- (2,2,2) -- (-2,2,2) -- cycle;
    
    \draw[fill=red, opacity=0.1] (0,-2,-2) -- (0,2,-2)-- (0,2,2) -- (0,-2,2)
    -- cycle;
    \draw[black!30, thick] (0,-2,-2) -- (0,2,-2) -- (0,2,2) -- (0,-2,2) -- cycle;
    
    \draw[fill=green, opacity=0.1] (-2,0,-2) -- (-2,0,2) -- (2,0,2) -- (2,0,-2)
    -- cycle;
    \draw[black!30, thick] (-2,0,-2) -- (-2,0,2) -- (2,0,2) -- (2,0,-2) -- cycle;

    \draw[fill=blue, opacity=0.1] (-2,-2,0) -- (2,-2,0) -- (2,2,0) -- (-2,2,0)
    -- cycle;
    \draw[black!30, thick] (-2,-2,0) -- (2,-2,0) -- (2,2,0) -- (-2,2,0) -- cycle;

    \draw[fill=yellow, opacity=0.1] (-2,-2,-2) -- (-2,2,-2)-- (-2,2,2) -- (-2,-2,2) -- cycle;
    \draw[black!50, thick] (-2,-2,-2) -- (-2,2,-2) -- (-2,2,2) -- (-2,-2,2) -- cycle;
    
    \draw[fill=yellow, opacity=0.1] (2,-2,-2) -- (2,2,-2)-- (2,2,2) -- (2,-2,2)
    -- cycle;
    \draw[black!50, thick] (2,-2,-2) -- (2,2,-2) -- (2,2,2) -- (2,-2,2) -- cycle;
    
    \draw[fill=yellow, opacity=0.1] (-2,-2,-2) -- (-2,-2,2) -- (2,-2,2) -- (2,-2,-2) -- cycle;
    \draw[black!50, thick] (-2,-2,-2) -- (-2,-2,2) -- (2,-2,2) -- (2,-2,-2) -- cycle;
    
    \draw[fill=yellow, opacity=0.1] (-2,2,-2) -- (-2,2,2) -- (2,2,2) -- (2,2,-2) -- cycle;
    \draw[black!50, thick] (-2,2,-2) -- (-2,2,2) -- (2,2,2) -- (2,2,-2) -- cycle;
    
    \draw[fill=yellow, opacity=0.1] (-2,-2,-2) -- (2,-2,-2) -- (2,2,-2) -- (-2,2,-2)
    -- cycle;
    \draw[black!50, thick] (-2,-2,-2) -- (2,-2,-2) -- (2,2,-2) -- (-2,2,-2) -- cycle;
    
    \draw[fill=yellow, opacity=0.1] (-2,-2,2) -- (2,-2,2) -- (2,2,2) -- (-2,2,2)
    -- cycle;
    \draw[black!50, thick] (-2,-2,2) -- (2,-2,2) -- (2,2,2) -- (-2,2,2) -- cycle;
    
    \tdplottransformmainscreen{-2}{-2}{-2}
    \shade[tdplot_screen_coords, ball color=green, opacity=0.9] (\tdplotresx,\tdplotresy) circle (0.2);
    \tdplottransformmainscreen{2}{-2}{-2}
    \shade[tdplot_screen_coords, ball color=green, opacity=0.9] (\tdplotresx,\tdplotresy) circle (0.2);
    \tdplottransformmainscreen{-2}{2}{-2}
    \shade[tdplot_screen_coords, ball color=green, opacity=0.9] (\tdplotresx,\tdplotresy) circle (0.2);
    \tdplottransformmainscreen{-2}{-2}{2}
    \shade[tdplot_screen_coords, ball color=green, opacity=0.9] (\tdplotresx,\tdplotresy) circle (0.2);
    \tdplottransformmainscreen{2}{2}{2}
    \shade[tdplot_screen_coords, ball color=green, opacity=0.9] (\tdplotresx,\tdplotresy) circle (0.2);
    \tdplottransformmainscreen{-2}{2}{2}
    \shade[tdplot_screen_coords, ball color=green, opacity=0.9] (\tdplotresx,\tdplotresy) circle (0.2);
    \tdplottransformmainscreen{2}{-2}{2}
    \shade[tdplot_screen_coords, ball color=green, opacity=0.9] (\tdplotresx,\tdplotresy) circle (0.2);
    \tdplottransformmainscreen{2}{2}{-2}
    \shade[tdplot_screen_coords, ball color=green, opacity=0.9] (\tdplotresx,\tdplotresy) circle (0.2);
    
    \tdplottransformmainscreen{0}{-2}{-2}
    \shade[tdplot_screen_coords, ball color=blue, opacity=0.9] (\tdplotresx,\tdplotresy) circle (0.2);
    \tdplottransformmainscreen{-2}{0}{-2}
    \shade[tdplot_screen_coords, ball color=blue, opacity=0.9] (\tdplotresx,\tdplotresy) circle (0.2);
    \tdplottransformmainscreen{-2}{-2}{0}
    \shade[tdplot_screen_coords, ball color=blue, opacity=0.9] (\tdplotresx,\tdplotresy) circle (0.2);
    \tdplottransformmainscreen{0}{2}{-2}
    \shade[tdplot_screen_coords, ball color=blue, opacity=0.9] (\tdplotresx,\tdplotresy) circle (0.2);
    \tdplottransformmainscreen{2}{0}{-2}
    \shade[tdplot_screen_coords, ball color=blue, opacity=0.9] (\tdplotresx,\tdplotresy) circle (0.2);
    \tdplottransformmainscreen{2}{-2}{0}
    \shade[tdplot_screen_coords, ball color=blue, opacity=0.9] (\tdplotresx,\tdplotresy) circle (0.2);
    \tdplottransformmainscreen{0}{-2}{2}
    \shade[tdplot_screen_coords, ball color=blue, opacity=0.9] (\tdplotresx,\tdplotresy) circle (0.2);
    \tdplottransformmainscreen{-2}{0}{2}
    \shade[tdplot_screen_coords, ball color=blue, opacity=0.9] (\tdplotresx,\tdplotresy) circle (0.2);
    \tdplottransformmainscreen{-2}{2}{0}
    \shade[tdplot_screen_coords, ball color=blue, opacity=0.9] (\tdplotresx,\tdplotresy) circle (0.2);
    \tdplottransformmainscreen{0}{2}{2}
    \shade[tdplot_screen_coords, ball color=blue, opacity=0.9] (\tdplotresx,\tdplotresy) circle (0.2);
    \tdplottransformmainscreen{2}{0}{2}
    \shade[tdplot_screen_coords, ball color=blue, opacity=0.9] (\tdplotresx,\tdplotresy) circle (0.2);
    \tdplottransformmainscreen{2}{2}{0}
    \shade[tdplot_screen_coords, ball color=blue, opacity=0.9] (\tdplotresx,\tdplotresy) circle (0.2);
    
    \tdplottransformmainscreen{0}{0}{-2}
    \shade[tdplot_screen_coords, ball color=red, opacity=0.9] (\tdplotresx,\tdplotresy) circle (0.2);
    \tdplottransformmainscreen{0}{-2}{0}
    \shade[tdplot_screen_coords, ball color=red, opacity=0.9] (\tdplotresx,\tdplotresy) circle (0.2);
    \tdplottransformmainscreen{-2}{0}{0}
    \shade[tdplot_screen_coords, ball color=red, opacity=0.9] (\tdplotresx,\tdplotresy) circle (0.2);
    \tdplottransformmainscreen{0}{0}{2}
    \shade[tdplot_screen_coords, ball color=red, opacity=0.9] (\tdplotresx,\tdplotresy) circle (0.2);
    \tdplottransformmainscreen{0}{2}{0}
    \shade[tdplot_screen_coords, ball color=red, opacity=0.9] (\tdplotresx,\tdplotresy) circle (0.2);
    \tdplottransformmainscreen{2}{0}{0}
    \shade[tdplot_screen_coords, ball color=red, opacity=0.9] (\tdplotresx,\tdplotresy) circle (0.2);
  \end{tikzpicture}
  
  \caption{Left: a hexahedron containing three intersecting surfaces.
    Right: the boundary conditions on the faces, edges and corners
    represented by balls; they are spin variables of a lattice model.}

  \label{fig:hexahedron}
\end{figure}

Our strategy is to use the cutting-and-gluing property of QFT.  We
dice $M$ up into hexahedra $H_\alpha$, $\alpha = 1$, $\dotsc$, $N$, in
such a way that each hexahedron contains an intersection of three
surface defects.  This is possible if the configuration of surfaces is
generic, which we assume is the case.  Since we are considering a
TQFT, detailed shapes of the hexahedra are irrelevant.  An example of
a hexahedron is illustrated in figure \ref{fig:hexahedron}.

For each hexahedron, we choose boundary conditions on its faces, edges
and vertices.  To a hexahedron $H_\alpha$ equipped with a choice
$b_\alpha$ of boundary conditions, the TQFT assigns a value
$W_\alpha(b_\alpha)$ which is an element of a commutative ring
$\Lambda$.  Once boundary conditions are chosen for all hexahedra, we
have a set of values $\{W_1(b_1), \dotsc, W_N(b_N)\}$.  As we vary the
boundary conditions, these values also vary, giving a function
$W = (W_1, \dotsc, W_N)$ from the set of boundary conditions to
$\Lambda^N$.  According to the gluing property of QFT, the value
assigned to the entire configuration of surface defects in $M$ is
reconstructed from this function as
\begin{equation}
  \sum_{(b_1, \dotsc, b_N) \in \config{B}} \prod_\alpha W_\alpha(b_\alpha) \,,
\end{equation}
where the sum is taken over the set $\config{B}$ of consistent boundary
conditions.  The consistency means that glued faces, edges or vertices
have matching boundary conditions.

This procedure of reconstructing the correlation function of surface
defects from hexahedra parallels how the partition function of a
classical spin model is defined.  Spin variables reside on the faces,
edges and vertices of the hexahedra and take values in appropriate
sets of boundary conditions.  To a hexahedron $H_\alpha$ with a
specified configuration $b_\alpha$ of spin variables, the model
assigns a value $W_\alpha(b_\alpha)$, called the \emph{local Boltzmann
  weight} for the spin configuration.  The partition function of the
model is defined to be the product of the local Boltzmann weights for
all hexahedra, summed over the set $\config{B}$ of all spin
configurations.

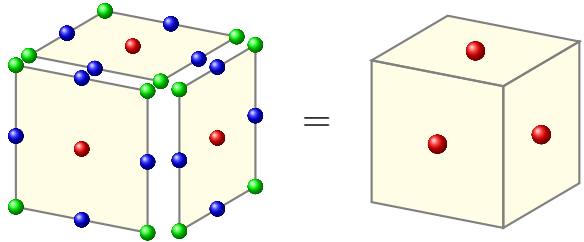
\begin{figure}
  \centering
  \tdplotsetmaincoords{70}{120}
  \begin{tikzpicture}[tdplot_main_coords, scale=0.5]

    \tdplottransformmainscreen{0}{0.6}{0}
    \begin{scope}[shift={(\tdplotresx,\tdplotresy)}]
      \draw[fill=yellow, opacity=0.1] (2,-2,-2) -- (2,2,-2)-- (2,2,2) -- (2,-2,2) -- cycle;
      \draw[black!50, thick] (2,-2,-2) -- (2,2,-2) -- (2,2,2) -- (2,-2,2) -- cycle;
      \tdplottransformmainscreen{2}{-2}{-2}
      \shade[tdplot_screen_coords, ball color=green, opacity=0.9] (\tdplotresx,\tdplotresy) circle (0.2);

      \tdplottransformmainscreen{2}{2}{2}
      \shade[tdplot_screen_coords, ball color=green, opacity=0.9] (\tdplotresx,\tdplotresy) circle (0.2);
      
      \tdplottransformmainscreen{2}{-2}{2}
      \shade[tdplot_screen_coords, ball color=green, opacity=0.9] (\tdplotresx,\tdplotresy) circle (0.2);

      \tdplottransformmainscreen{2}{2}{-2}
      \shade[tdplot_screen_coords, ball color=green, opacity=0.9] (\tdplotresx,\tdplotresy) circle (0.2);

      \tdplottransformmainscreen{2}{0}{-2}
      \shade[tdplot_screen_coords, ball color=blue, opacity=0.9] (\tdplotresx,\tdplotresy) circle (0.2);

      \tdplottransformmainscreen{2}{-2}{0}
      \shade[tdplot_screen_coords, ball color=blue, opacity=0.9] (\tdplotresx,\tdplotresy) circle (0.2);

      \tdplottransformmainscreen{2}{0}{2}
      \shade[tdplot_screen_coords, ball color=blue, opacity=0.9] (\tdplotresx,\tdplotresy) circle (0.2);

      \tdplottransformmainscreen{2}{2}{0}
      \shade[tdplot_screen_coords, ball color=blue, opacity=0.9] (\tdplotresx,\tdplotresy) circle (0.2);

      \tdplottransformmainscreen{2}{0}{0}
      \shade[tdplot_screen_coords, ball color=red, opacity=0.9] (\tdplotresx,\tdplotresy) circle (0.2);
    \end{scope}
    
    \tdplottransformmainscreen{0}{0}{0.6}
    \begin{scope}[shift={(\tdplotresx,\tdplotresy)}]
      \draw[fill=yellow, opacity=0.1] (-2,2,-2) -- (-2,2,2) -- (2,2,2) -- (2,2,-2) -- cycle;
      \draw[black!50, thick] (-2,2,-2) -- (-2,2,2) -- (2,2,2) -- (2,2,-2) -- cycle;

      \tdplottransformmainscreen{-2}{2}{-2}
      \shade[tdplot_screen_coords, ball color=green, opacity=0.9] (\tdplotresx,\tdplotresy) circle (0.2);
      
      \tdplottransformmainscreen{2}{2}{2}
      \shade[tdplot_screen_coords, ball color=green, opacity=0.9] (\tdplotresx,\tdplotresy) circle (0.2);
    
      \tdplottransformmainscreen{-2}{2}{2}
      \shade[tdplot_screen_coords, ball color=green, opacity=0.9] (\tdplotresx,\tdplotresy) circle (0.2);

      \tdplottransformmainscreen{2}{2}{-2}
      \shade[tdplot_screen_coords, ball color=green, opacity=0.9] (\tdplotresx,\tdplotresy) circle (0.2);

      \tdplottransformmainscreen{-2}{2}{0}
      \shade[tdplot_screen_coords, ball color=blue, opacity=0.9] (\tdplotresx,\tdplotresy) circle (0.2);
    
      \tdplottransformmainscreen{0}{2}{2}
      \shade[tdplot_screen_coords, ball color=blue, opacity=0.9] (\tdplotresx,\tdplotresy) circle (0.2);

      \tdplottransformmainscreen{2}{2}{0}
      \shade[tdplot_screen_coords, ball color=blue, opacity=0.9] (\tdplotresx,\tdplotresy) circle (0.2);
    
      \tdplottransformmainscreen{0}{2}{-2}
      \shade[tdplot_screen_coords, ball color=blue, opacity=0.9] (\tdplotresx,\tdplotresy) circle (0.2);

        \tdplottransformmainscreen{0}{2}{0}
        \shade[tdplot_screen_coords, ball color=red, opacity=0.9] (\tdplotresx,\tdplotresy) circle (0.2);
      \end{scope}
 
    \tdplottransformmainscreen{0.6}{0}{0}
    \begin{scope}[shift={(\tdplotresx,\tdplotresy)}]
      \draw[fill=yellow, opacity=0.1] (-2,-2,2) -- (2,-2,2) -- (2,2,2) -- (-2,2,2) -- cycle;
      \draw[black!50, thick] (-2,-2,2) -- (2,-2,2) -- (2,2,2) -- (-2,2,2) -- cycle;
      
      \tdplottransformmainscreen{-2}{-2}{2}
      \shade[tdplot_screen_coords, ball color=green, opacity=0.9] (\tdplotresx,\tdplotresy) circle (0.2);
   
      \tdplottransformmainscreen{2}{2}{2}
      \shade[tdplot_screen_coords, ball color=green, opacity=0.9] (\tdplotresx,\tdplotresy) circle (0.2);
      
      \tdplottransformmainscreen{-2}{2}{2}
      \shade[tdplot_screen_coords, ball color=green, opacity=0.9] (\tdplotresx,\tdplotresy) circle (0.2);
      
      \tdplottransformmainscreen{2}{-2}{2}
      \shade[tdplot_screen_coords, ball color=green, opacity=0.9] (\tdplotresx,\tdplotresy) circle (0.2);
      
      \tdplottransformmainscreen{0}{-2}{2}
      \shade[tdplot_screen_coords, ball color=blue, opacity=0.9] (\tdplotresx,\tdplotresy) circle (0.2);

      \tdplottransformmainscreen{-2}{0}{2}
      \shade[tdplot_screen_coords, ball color=blue, opacity=0.9] (\tdplotresx,\tdplotresy) circle (0.2);
      
      \tdplottransformmainscreen{0}{2}{2}
      \shade[tdplot_screen_coords, ball color=blue, opacity=0.9] (\tdplotresx,\tdplotresy) circle (0.2);
      
      \tdplottransformmainscreen{2}{0}{2}
      \shade[tdplot_screen_coords, ball color=blue, opacity=0.9] (\tdplotresx,\tdplotresy) circle (0.2);
    
      \tdplottransformmainscreen{0}{0}{2}
      \shade[tdplot_screen_coords, ball color=red, opacity=0.9] (\tdplotresx,\tdplotresy) circle (0.2);
    \end{scope}
  \end{tikzpicture}
  \quad
  {\Large =}
  \quad
  \begin{tikzpicture}[tdplot_main_coords, scale=0.5]
    \draw[fill=yellow, opacity=0.1] (2,-2,-2) -- (2,2,-2)-- (2,2,2) -- (2,-2,2)
    -- cycle;
    \draw[black!50, thick] (2,-2,-2) -- (2,2,-2) -- (2,2,2) -- (2,-2,2) -- cycle;
    
    \draw[fill=yellow, opacity=0.1] (-2,2,-2) -- (-2,2,2) -- (2,2,2) -- (2,2,-2) -- cycle;
    \draw[black!50, thick] (-2,2,-2) -- (-2,2,2) -- (2,2,2) -- (2,2,-2) -- cycle;
    
    \draw[fill=yellow, opacity=0.1] (-2,-2,2) -- (2,-2,2) -- (2,2,2) -- (-2,2,2)
    -- cycle;
    \draw[black!50, thick] (-2,-2,2) -- (2,-2,2) -- (2,2,2) -- (-2,2,2) -- cycle;

    \tdplottransformmainscreen{0}{0}{2}
    \shade[tdplot_screen_coords, ball color=red, opacity=0.9] (\tdplotresx,\tdplotresy) circle (0.25);
    \tdplottransformmainscreen{0}{2}{0}
    \shade[tdplot_screen_coords, ball color=red, opacity=0.9] (\tdplotresx,\tdplotresy) circle (0.25);
    \tdplottransformmainscreen{2}{0}{0}
    \shade[tdplot_screen_coords, ball color=red, opacity=0.9] (\tdplotresx,\tdplotresy) circle (0.25);
  \end{tikzpicture}

  \caption{Reformulation of the spin model arising from the hexahedral
    decomposition as a vertex model.  After the spin variables on the
    edges are doubled and those on the corners are tripled, the nine
    spin variables on and around each face are combined into one spin
    variable placed on the face.}
  \label{fig:quadrilaterals}
\end{figure}

We can reformulate this spin model as a model that has spin variables
only on the faces of the hexahedra.  Imagine making a hexahedron from
six quadrilaterals.  On each quadrilateral there are nine spin
variables, one on the face, four on the edges and four on the
vertices.  We can think of these nine spin variables as collectively
specifying a single ``big'' spin variable, which we can conveniently
place on the face.  A hexahedron in the reformulated model therefore
has six spin variables that are worth 54 original spin variables, but
for most configurations of these 54 spin variables we set the local
Boltzmann weight to zero.  The only nonvanishing configurations are
those for which the spin variables match on the glued edges and
vertices.  For these configuration, the local Boltzmann weight is
simply set to the corresponding value in the original model.  See
figure \ref{fig:quadrilaterals} for illustration.

In the reformulated model the spin variables live on the faces of the
hexahedra, or equivalently, on the edges of the lattice formed by
the surface defects.  The local Boltzmann weights are assigned to the
vertices of the lattice, located at the centers of the hexahedra.
Rephrased in this way, we see that the correlation function of surface
defects in the TQFT coincides with the partition function of a vertex
model on the lattice formed by the surface defects.

Applying this argument to the TQFT $\mathring{Z}$, we deduce that
$\mathring{Z}$ computes the partition function of a vertex model, with
local Boltzmann weights valued in $\Lambda = \Z[[q, q^{-1}]]$.

\subsection{Commuting transfer matrices}
\label{sec:[T0,T0]=0}

Because of its M-theory origin, the lattice model arising from the 3D
TQFT $\mathring{Z}$ has a special property which lattice models produced by
other 3D TQFTs do not possess in general: its transfer matrices
commute with each other.  As we will explain, this property is closely
related to the integrability of the model.

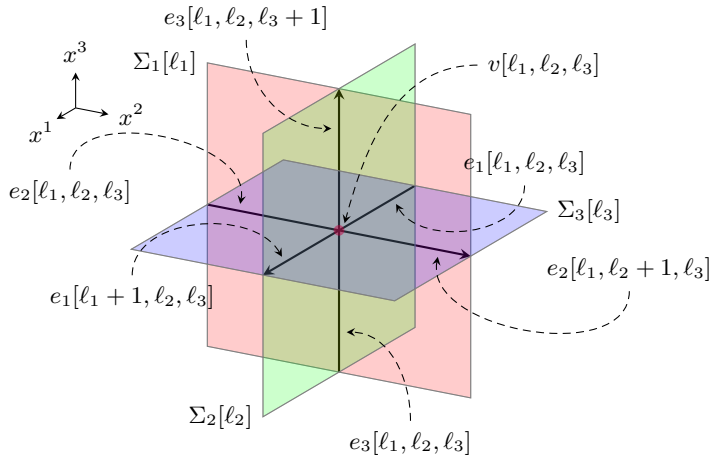
\begin{figure}
  \centering
  \tdplotsetmaincoords{70}{120}
  \begin{tikzpicture}[tdplot_main_coords]    
    \begin{scope}[scale=1/4]
      \draw[->, >=stealth] (0,-16,4)-- ++(2,0,0) node[below left=-4pt] {$x^1$};
      \draw[->, >=stealth] (0,-16,4) -- ++(0,2,0) node[right] {$x^2$};
      \draw[->, >=stealth] (0,-16,4) -- ++(0,0,2) node[above] {$x^3$};
    \end{scope}

    \draw[thick, -stealth] (-2,0,0) -- (2,0,0);
    \draw[thick, -stealth] (0,-2,0) -- (0,2,0);
    \draw[thick, -stealth] (0,0,-2) -- (0,0,2);

    \draw[fill=red, opacity=0.2] (0,-2,-2) -- (0,2,-2)-- (0,2,2) -- (0,-2,2)
    node[black, opacity=1, left] {$\Sigma_1[\ell_1]$} -- cycle;
    \draw[black!50] (0,-2,-2) -- (0,2,-2) -- (0,2,2) -- (0,-2,2) -- cycle;

    \draw[fill=green, opacity=0.2] (-2,0,-2) -- (-2,0,2) -- (2,0,2) -- (2,0,-2)
    node[black, opacity=1, left] {$\Sigma_2[\ell_2]$} -- cycle;
    \draw[black!50] (-2,0,-2) -- (-2,0,2) -- (2,0,2) -- (2,0,-2) -- cycle;

    \draw[fill=blue, opacity=0.2] (-2,-2,0) -- (2,-2,0) -- (2,2,0) -- (-2,2,0)
    node[black, opacity=1, right] {$\Sigma_3[\ell_3]$} -- cycle;
    \draw[black!50] (-2,-2,0) -- (2,-2,0) -- (2,2,0) -- (-2,2,0) -- cycle;

    \node[anchor=west] (n1) at (-3,0,0) {$e_1[\ell_1,\ell_2,\ell_3]$};
    \draw[-stealth, densely dashed] (n1) to[out=-90, in=-60] (-1.5,0,-0.1);
    \node[anchor=east] (n1') at (3,0,0) {$e_1[\ell_1+1,\ell_2,\ell_3]$};
    \draw[-stealth, densely dashed] (n1') to[out=90, in=120] (1.5,0,0.1);

    \node[anchor=east] (n2) at (0,-3,0) {$e_2[\ell_1,\ell_2,\ell_3]$};
    \draw[-stealth, densely dashed] (n2) to[out=90, in=80] (0,-1.5,0.1);
    \node[anchor=west] (n2') at (0,3,0) {$e_2[\ell_1,\ell_2+1,\ell_3]$};
    \draw[-stealth, densely dashed] (n2') to[out=-90, in=-100] (0,1.5,-0.1);
    
    \node[anchor=west] (n3) at (0,0,-3) {$e_3[\ell_1,\ell_2,\ell_3]$};
    \draw[-stealth, densely dashed] (n3) to[out=90, in=0] (-0.2,0,-1.5);
    \node[anchor=east] (n3') at (0,0,3) {$e_3[\ell_1,\ell_2,\ell_3+1]$};
    \draw[-stealth, densely dashed] (n3') to[out=-90, in=180] (0.2,0,1.5);

    \fill[purple, opacity=0.5] (0,0,0) circle (2pt);
    \node (line) at (5,6,5) 
    {$v[\ell_1,\ell_2,\ell_3]$};
    \draw[-stealth, densely dashed] (line) to[out=180,in=75] (0.2,0.2,0.2);
  \end{tikzpicture}
  \caption{A neighborhood of the vertex $v[\ell_1,\ell_2,\ell_3]$ of the
    lattice.  The M5-brane $\Mfive_i[\ell_i]$ lies on the surface
    $\Sigma_i[\ell_i]$.  M2-branes of type $i$ are placed along the edges
    $e_i[\ell_1,\ell_2,\ell_3]$ and
    $e_i[\ell_1+\delta_{i1},\ell_2+\delta_{i1},\ell_3+\delta_{i1}]$.}
  \label{fig:lattice-names}
\end{figure}

To facilitate discussions let us introduce various notations.  There
are $L_i$ M5-branes of type $i$, which we call $\Mfive_i[\ell_i]$,
$\ell_i = 1$, $\dotsc$, $L_i$.  The M5-brane $\Mfive_i[\ell_i]$ is
located at $x^i(\Mfive_i[\ell_i]) \in \circle_i$ and
$x^\mu(\Mfive_i[\ell_i]) \in \R_\mu$ for
$\mu \in \{4,5,6\} \setminus \{i+3\}$.  We choose the cyclic ordering
of the M5-branes such that
\begin{equation}
  \label{eq:xi-order}
  x^i(\Mfive_i[1])
  \leq
  x^i(\Mfive_i[2])
  \leq
  \dotsb
  \leq
  x^i(\Mfive_i[L_i])
  \leq
  x^i(\Mfive_i[1]) \,.
\end{equation}
The index $i$ is understood modulo $3$ when appearing as a label of a
brane.  The variable $\ell_i$ is defined modulo $L_i$, reflecting the
periodic boundary condition.  The worldvolume of $\Mfive_i[\ell_i]$
makes a surface $\Sigma_i[\ell_i]$ inside $\T^3_{123}$, which is a
$2$-torus.  The three surfaces $\Sigma_1[\ell_1]$, $\Sigma_2[\ell_2]$,
$\Sigma_3[\ell_3]$ intersect at a vertex $v[\ell_1, \ell_2, \ell_3]$
of the $L_1 \times L_2 \times L_3$ cubic lattice made by the $2$-tori.
The adjacent vertices
$v[\ell_1 - \delta_{i1}, \ell_2 - \delta_{i2}, \ell_3 - \delta_{i3}]$
and $v[\ell_1, \ell_2, \ell_3]$ are connected by an edge
$e_i[\ell_1, \ell_2, \ell_3]$.  See figure~\ref{fig:lattice-names}.

Given a configuration $\config{n}$ of the spin variables, the lattice
model assigns a local Boltzmann weight to every vertex.  As in the
discussion in section~\ref{sec:TQFT-lattice}, let us assume that all
M5-branes have the same sign, say $+$.  Then we have only one type of
surface defect and one type of local Boltzmann weight.  Writing
$n_i[\ell_1, \ell_2, \ell_3]$ for the value of the spin variable on
$e_i[\ell_1,\ell_2,\ell_3]$, let
\begin{equation}
  \label{eq:R-matrix-element}
  \TDR[\ell_1, \ell_2, \ell_3]
  _{n_1[\ell_1, \ell_2, \ell_3] n_2[\ell_1, \ell_2, \ell_3] n_3[\ell_1, \ell_2, \ell_3]}
  ^{n_1[\ell_1+1, \ell_2, \ell_3] n_2[\ell_1, \ell_2+1, \ell_3] n_3[\ell_1, \ell_2, \ell_3+1]}
\end{equation}
be the local Boltzmann weight at $v[\ell_1,\ell_2,\ell_3]$, determined
by the spin variables on the six edges connected to
$v[\ell_1,\ell_2,\ell_3]$.  The partition function of the lattice
model is defined by
\begin{equation}
  \label{eq:Z0-R}
  \mathring{Z}
  =
  \sum_{\config{n} \in \config{N}}
  \prod_{\ell_1, \ell_2, \ell_3}
  \TDR[\ell_1, \ell_2, \ell_3]
  _{n_1[\ell_1, \ell_2, \ell_3] n_2[\ell_1, \ell_2, \ell_3] n_3[\ell_1, \ell_2, \ell_3]}
  ^{n_1[\ell_1+1, \ell_2, \ell_3] n_2[\ell_1, \ell_2+1, \ell_3] n_3[\ell_1, \ell_2, \ell_3+1]} \,,
\end{equation}
where $\config{N}$ is the set of all spin configurations.

The local Boltzmann weight \eqref{eq:R-matrix-element} can be regarded
as a matrix element with six indices.  Let $\CV_i[\ell_1,\ell_2,\ell_3]$
be a complex vector space spanned by the set of all values the spin
variable on $e_i[\ell_1,\ell_2,\ell_3]$ can take.  The collection of
all possible values of the local Boltzmann weight at
$v[\ell_1,\ell_2,\ell_3]$ define a linear operator
\begin{multline}
  \TDR[\ell_1,\ell_2,\ell_3]\colon
  \CV_1[\ell_1, \ell_2, \ell_3] \otimes \CV_2[\ell_1, \ell_2, \ell_3] \otimes \CV_3[\ell_1, \ell_2, \ell_3]
  \\
  \to
  \CV_1[\ell_1+1, \ell_2, \ell_3] \otimes \CV_2[\ell_1, \ell_2+1, \ell_3] \otimes \CV_3[\ell_1, \ell_2, \ell_3+1] 
  \,,
\end{multline}
called the \emph{R-matrix} at $v[\ell_1,\ell_2,\ell_3]$.

With the help of the R-matrix we can map our 3D classical spin model
to a 2D quantum spin model.  The idea is to pick one of the periodic
directions of the lattice, say the ``vertical'' direction $\circle_3$,
and think of it as a time direction with a discrete time coordinate
$\ell_3$.  A time slice at time $\ell_3$ is a ``horizontal'' $2$-torus
intersecting the vertical edges $e_3[\ell_1,\ell_2,\ell_3]$,
$\ell_1=1$, $\dotsc$, $L_1$, $\ell_2=1$, $\dotsc$, $L_2$.  The vector
space $\CV_3[\ell_1,\ell_2,\ell_3]$ is interpreted as the Hilbert
space for a quantum mechanical degree of freedom attached to
$e_3[\ell_1,\ell_2,\ell_3]$.  The total Hilbert space on the time
slice is the tensor product
\begin{equation}
  \CV_3[\ell_3]
  =
  \bigotimes_{\ell_1, \ell_2} \CV_3[\ell_1,\ell_2,\ell_3] \,.
\end{equation}
Across the horizontal torus $\Sigma_3[\ell_3]$, a state in
$\CV_3[\ell_3]$ changes to another state in $\CV_3[\ell_3+1]$.  An
initial state at $\ell_3 = 1$ evolves in $L_3$ steps to a final state
at $\ell_3 = L_3+1$, which is required by the periodic boundary
condition to be the same as the initial state.

Discrete time evolution in the $x^3$-direction is generated by linear
operators called transfer matrices.  The \emph{layer-to-layer transfer
  matrix}
\begin{equation}
  \mathring{\TDT}_3[\ell_3]\colon \CV_3[\ell_3] \to \CV_3[\ell_3+1]
\end{equation}
at time $\ell_3$ is the composition of all R-matrices assigned to the
vertices lying on $\Sigma_3[\ell_3]$, with an appropriate trace taken
according to the periodicity of the torus:
\begin{equation}
  \mathring{\TDT}_3[\ell_3]
  =
  \Tr_{\bigotimes_{\ell_2} \CV_1[1,\ell_2,\ell_3] \otimes \bigotimes_{\ell_1}
    \CV_2[\ell_1,1,\ell_3]} \Biggl(
  \prod_{\ell_1, \ell_2} \TDR[\ell_1,\ell_2,\ell_3]
  \Biggr)
  \,.
\end{equation}
In this formula the operator ordering is chosen in accordance with the
orientations of the relevant edges.  Using the transfer matrix we can
write the partition function as
\begin{equation}
  \mathring{Z} = \Tr_{\CV_3[1]}\bigl(\mathring{\TDT}_3[L_3] \dotsm \mathring{\TDT}_3[2] \mathring{\TDT}_3[1]\bigr) \,.
\end{equation}

From the point of view of the TQFT, the vector space
$\CV_i[\ell_1,\ell_2,\ell_3]$ is, roughly speaking, the Hilbert space
on the faces of hexahedra that intersect $e_i[\ell_1,\ell_2,\ell_3]$.
It is the Hilbert space the theory assigns to a quadrilateral
intersected by the surface defects $\Sigma_{i+1}[\ell_{i+1}]$ and
$\Sigma_{i+2}[\ell_{i+2}]$.  (We use the same symbol
$\Sigma_i[\ell_i]$ to refer to the surface defect placed on the
surface $\Sigma_i[\ell_i]$.)  With this description it is clear that
the spaces $\CV_i[\ell_1,\ell_2,\ell_3]$ for different values of
$\ell_i$ are all isomorphic since they are determined solely by the
types of the intersecting surface defects, which we have been assuming
to be all $+$.  The space $\CV_3[\ell_3]$ is likewise independent of
$\ell_3$; it is the Hilbert space of the theory on a $2$-torus
intersected by the surface defects $\Sigma_1[\ell_1]$, $\ell_1 = 1$,
$\dotsc$, $L_1$, and $\Sigma_2[\ell_2]$, $\ell_2 = 1$, $\dotsc$,
$L_2$.  Since the theory is topological, a state evolves trivially
except when it encounters some objects.  In the present case, those
objects are the horizontal surface defects, and the transfer matrix
$\TDT_3[\ell_3]$ implements the action of $\Sigma_3[\ell_3]$ on the
Hilbert space.

Now, take a horizontal surface defect $\Sigma_3[\ell_3]$ and slide it
vertically upward.  The topological invariance of the theory implies
that continuous deformations of the geometry of the surface defects do
not affect $\mathring{Z}$ as long as the topology of the configuration
remains the same.  Therefore, $\mathring{Z}$ is invariant as
$\Sigma_3[\ell_3]$ slides upward until the point when it hits the next
surface defect $\Sigma_3[\ell_3+1]$.  If we further move
$\Sigma_3[\ell_3]$ past $\Sigma_3[\ell_3+1]$ and interchange their
vertical positions, $\mathring{Z}$ might change since the topology is
altered.  At least the structure of a TQFT does not prohibit such a
change.

This is, however, not what happens because our theory is more than
just a 3D TQFT: it has extra eight dimensions in the 11D spacetime of
M-theory.  Recall that the surface defect $\Sigma_i[\ell_i]$ is
created by the M5-brane $\Mfive_i[\ell_i]$ which intersects the origin
of $\R^3_{456}$.  Imagine that when we try to move $\Sigma_3[\ell_3]$
past $\Sigma_3[\ell_3+1]$, we displace $\Mfive_3[\ell_3]$ in
$\R^2_{45}$ to avoid collision with $\Mfive_3[\ell_3+1]$.  During this
``detour'' the Witten index of $\CH^j$, calculating the integer
coefficient of $q^j$ in $\mathring{Z}$, remains unchanged since
nothing is discontinuous about this process.  It follows that when
$\Mfive_3[\ell_3]$ comes back to the origin of $\R^2_{45}$ and is
placed above $\Mfive_3[\ell_3+1]$ in $\circle_3$, the value of
$\mathring{Z}$ is the same as the beginning of the process.

Translated to the quantum mechanical language, what we have just found
means that the transfer matrices $\mathring{\TDT}_3[\ell_3]$ and
$\mathring{\TDT}_3[\ell_3+1]$ commute:
\begin{equation}
  \label{eq:[T0,T0]=0}
  \mathring{\TDT}_3[\ell_3+1] \mathring{\TDT}_3[\ell_3]
  = \mathring{\TDT}_3[\ell_3] \mathring{\TDT}_3[\ell_3+1] \,.
\end{equation}
Under the current assumption that the signs of the M5-branes are all
$+$, this equation is vacuous since $\mathring{\TDT}_3[\ell_3]$ and
$\mathring{\TDT}_3[\ell_3+1]$ are actually the same operator.  As soon
as this assumption is dropped the equation becomes nontrivial.

In section \ref{sec:[T,T]=0} we will see that the transfer matrices of
the lattice model defined by the full index $Z$ also commute with each
other.  In this case the commutativity is meaningful even for transfer
matrices coming from surface defects of the same sign because they
depend on additional continuous parameters.  In fact, the
commutativity and the existence of continuous parameters imply the
integrability of the model.

\subsection{Tetrahedron equation}
\label{sec:tetrahedron}

We have seen that $\mathring{Z}$ is invariant under vertical movement
of horizontal surface defects.  The partition function also has an
invariance under local deformations of surface defects.  This local
invariance leads to a highly overdetermined system of equations
satisfied by the R-matrix, known as Zamolodchikov's tetrahedron
equation \cite{MR611994}.

\begin{figure}
  \centering
  \tdplotsetmaincoords{135}{0}
  \begin{tikzpicture}[tdplot_main_coords, xscale=1, yscale=1.2]
    \draw[thick, stealth-, shorten >=-12pt, shorten <=-12pt]
    ({-sqrt(3)},-1,0) -- node[below left=-4pt] {$(12)$}  (0,{sqrt(3)},0);

    \draw[thick, stealth-, shorten >=-12pt, shorten <=-12pt]
    ({sqrt(3)},-1,0) -- node[below right=-4pt] {$(23)$}  (0,{sqrt(3)},0);

    \draw[thick, -stealth, shorten >=-12pt, shorten <=-12pt]
    (0,0,{sqrt(3)}) -- node[above right=-4pt] {$(34)$}  ({sqrt(3)},-1,0);

    \draw[thick, stealth-, shorten >=-12pt, shorten <=-12pt]
    (0,0,{sqrt(3)}) -- node[above] {$(14)$}  ({-sqrt(3)},-1,0);

    \draw[thick, stealth-, shorten >=-12pt, shorten <=-12pt]
    (0,0,{sqrt(3)}) -- node[right=-2pt] {$(13)$}  (0,{sqrt(3)},0);

    \draw[line width=4pt, -, white]
    ({-sqrt(3)/2},-1,0) -- ({sqrt(3)/2},-1,0);
    \draw[thick, -stealth, shorten >=-12pt, shorten <=-12pt]
    ({-sqrt(3)},-1,0) -- node[below left, shift={(-8pt,2pt)}] {$(24)$} ({sqrt(3)},-1,0);

    \node[right=2pt] at (0,{sqrt(3)},0) {$(123)$};
    \node[above right, yshift=4pt] at (0,0,{sqrt(3)}) {$(134)$};
    \node[above left, xshift=-4pt] at ({-sqrt(3)},-1,0) {$(124)$};
    \node[above right, xshift=4pt] at ({sqrt(3)},-1,0) {$(234)$};
  \end{tikzpicture}
  {\Large =}
  \tdplotsetmaincoords{-45}{0}
  \begin{tikzpicture}[tdplot_main_coords, xscale=1, yscale=1.2, baseline=0pt]
    \draw[thick, -stealth, shorten >=-12pt, shorten <=-12pt]
    ({-sqrt(3)},-1,0) -- node[above left, shift={(-8pt,-2pt)}] {$(24)$} ({sqrt(3)},-1,0);

    \draw[thick, -stealth, shorten >=-12pt, shorten <=-12pt]
    ({-sqrt(3)},-1,0) -- node[above left=-4pt] {$(23)$}  (0,{sqrt(3)},0);

    \draw[thick, -stealth, shorten >=-12pt, shorten <=-12pt]
    ({sqrt(3)},-1,0) -- node[above right=-4pt] {$(12)$}  (0,{sqrt(3)},0);

    \draw[thick, -stealth, shorten >=-12pt, shorten <=-12pt]
    (0,0,{sqrt(3)}) -- node[below right=-4pt] {$(14)$}  ({sqrt(3)},-1,0);

    \draw[thick, stealth-, shorten >=-12pt, shorten <=-12pt]
    (0,0,{sqrt(3)}) -- node[below left=-4pt] {$(34)$}  ({-sqrt(3)},-1,0);

    \draw[line width=4pt, -, white]
    (0,0,{sqrt(3/2)}) -- (0,{sqrt(3)/2},0);
    \draw[thick, -stealth, shorten >=-12pt, shorten <=-12pt]
    (0,0,{sqrt(3)}) -- node[right=-2pt] {$(13)$}  (0,{sqrt(3)},0);

    \node[right=2pt] at (0,{sqrt(3)},0) {$(123)$};
    \node[below right, yshift=-4pt] at (0,0,{sqrt(3)}) {$(134)$};
    \node[below left, xshift=-4pt] at ({-sqrt(3)},-1,0) {$(234)$};
    \node[below right, xshift=4pt] at ({sqrt(3)},-1,0) {$(124)$};
  \end{tikzpicture}

  \caption{A graphical representation of the tetrahedron equation.}
  \label{fig:tetrahedron}
\end{figure}
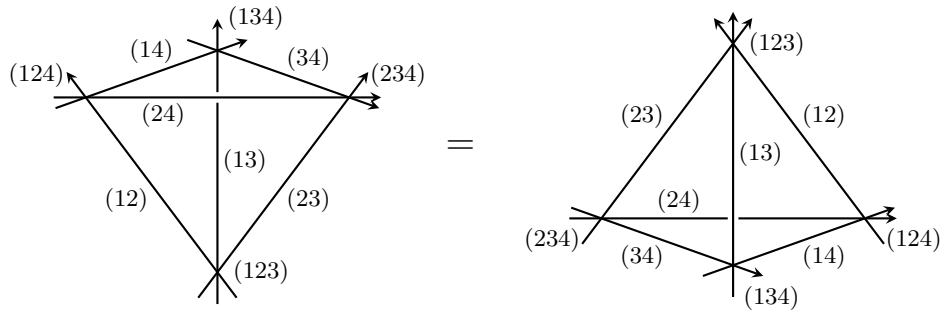

In the 3D TQFT $\mathring{Z}$, consider a configuration of four
surface defects that form a tetrahedron.  Introducing an ordering
among the four surfaces, we call them $\Sigma_1$, $\Sigma_2$,
$\Sigma_3$, $\Sigma_4$ from the smallest to the largest:
\begin{equation}
  \Sigma_1 < \Sigma_2 < \Sigma_3 < \Sigma_4 \,.
\end{equation}
For $a < b < c$, let
\begin{equation}
  (ab) = \Sigma_a \cap \Sigma_b \,,
  \quad
  (abc) = \Sigma_a \cap \Sigma_b \cap \Sigma_c \,.
\end{equation}
The ordering of the surfaces induces a lexicographic order of the
vertices of the tetrahedron: $(abc) < (def)$ if $abc < def$, read as
3-digit numbers.  We orient the edges so that they point from a
smaller vertex to a larger vertex.  See the left-hand side of
figure~\ref{fig:tetrahedron}

The R-matrix at the vertex $(abc)$
\begin{equation}
  \TDR_{(abc)} \colon
  \CV_{(bc)} \otimes \CV_{(ac)} \otimes \CV_{(ab)}
  \to
  \CV_{(bc)} \otimes \CV_{(ac)} \otimes \CV_{(ab)}
\end{equation}
is an endomorphism of the tensor product of the vector spaces
$\CV_{(bc)}$, $\CV_{(ac)}$, $\CV_{(ab)}$ assigned to the three lines
$(bc)$, $(ac)$, $(ab)$.  The topological invariance and the existence
of extra dimensions $\R^3_{456}$ imply that $\mathring{Z}$ remains
invariant under a local deformation of the configuration that
transforms the tetrahedron into another tetrahedron.  The equality of
the Boltzmann weights associated with the two tetrahedra leads to the
\emph{tetrahedron equation}:
\begin{equation}
  \label{eq:tetrahedron}
  \TDR_{(234)} \TDR_{(134)} \TDR_{(124)} \TDR_{(123)}
  =
  \TDR_{(123)} \TDR_{(124)} \TDR_{(134)} \TDR_{(234)}
  \,.
\end{equation}
The commutativity of transfer matrices \eqref{eq:[T0,T0]=0} follows
from the tetrahedron equation and the invertibility of the R-matrix,
as explained in section \ref{sec:[T,T]=0}.

\section{Properties of the model}
\label{sec:properties}

We have explained how a 3D lattice model originates from the brane
system in the special case in which the positions of the M5-branes are
adjusted so that all M2-branes become massless.  Now we consider the
case in which all M2-branes are massive: for distinct $i$, $j$,
$k \in \{1,2,3\}$, we assume
\begin{equation}
  x^{i+3}(\Mfive_j[\ell_j]) \neq x^{i+3}(\Mfive_k[\ell_k])
\end{equation}
for all $\ell_j = 1$, $\dotsc$, $L_j$ and $\ell_k = 1$, $\dotsc$,
$L_k$.  The massless case is recovered as a limit of the massive case.
In this section we discuss various properties that the model in this
general case is expected to possess.  In particular, we will explain
why the model is integrable.

\subsection{Spin variables}

The spin variables of the model are nonnegative integers that count
the numbers of M2-branes placed along the edges of the lattice.  More
precisely, $n_i[\ell_1, \ell_2, \ell_3]$ is the number of M2-branes
suspended between the M5-branes $\Mfive_{i+1}[\ell_{i+1}]$ and
$\Mfive_{i+2}[\ell_{i+2}]$ along $e_i[\ell_1,\ell_2,\ell_3]$.  (Recall
that the index $i$ labeling a brane is understood modulo $3$.)

The range in which the spin variable $n_i[\ell_1, \ell_2, \ell_3]$
takes values depends on the number of directions shared by the
worldvolumes of the two M5-branes.  Let
$\sigma(\Mfive_i[\ell_i]) \in \{+,-\}$ be the sign of
$\Mfive_i[\ell_i]$.  Then,
\begin{equation}
  n_i[\ell_1, \ell_2, \ell_3]
  \in
  \begin{cases}
    \{0,1,2,\dotsc\} &
    (\sigma(\Mfive_{i+1}[\ell_{i+1}]) = \sigma(\Mfive_{i+2}[\ell_{i+2}])) \,;
    \\
    \{0,1\} &
    (\sigma(\Mfive_{i+1}[\ell_{i+1}]) \neq \sigma(\Mfive_{i+2}[\ell_{i+2}])) \,.
  \end{cases}
\end{equation}

The reason the number of M2-branes is restricted if
$\sigma(\Mfive_{i+1}[\ell_{i+1}]) \neq
\sigma(\Mfive_{i+2}[\ell_{i+2}])$ is that in this case the ground
state of an M2-brane stretched between the two M5-branes is fermionic
\cite{Bachas:1997kn}, hence no more than one such M2-brane is allowed
to exist without becoming non-BPS.  This fact is related to the
``s-rule'' for the D3--D5--NS5 brane system of Hanany and Witten
\cite{Hanany:1996ie} by reduction on $\circle_{i+1}$ to type IIA
string theory and subsequent T-duality on $\circle_{i+2}$.

\subsection{Charge conservation}
\label{sec:charge}

There are also constraints on the numbers of M2-branes that can be
placed on a set of six edges meeting at a vertex.  To see why such
constraints exist, consider reduction on $\circle_0$ to type IIA
string theory.  Then, the M5-brane $\Mfive_i[\ell_i]$ becomes a
D4-brane $\Dfour_i[\ell_i]$ whose low-energy physics is described by
5D abelian $\TDN = 2$ super Yang--Mills theory, and M2-branes ending
on $\Dfour_i[\ell_i]$ become fundamental strings whose ends behave as
electrically charged particles in this theory.  The conservation of
electric charge leads to relations among the numbers of M2-branes.

Consider the fundamental strings coming from the M2-branes living on
the four edges $e_1[\ell_1,\ell_2,\ell_3]$,
$e_2[\ell_1,\ell_2,\ell_3]$, $e_1[\ell_1+1,\ell_2,\ell_3]$ and
$e_2[\ell_1,\ell_2+1,\ell_3]$.  They create charged particles in the
theory on $\Dfour_3[\ell_3]$.  We can view the edges
$e_1[\ell_1,\ell_2,\ell_3]$ and $e_2[\ell_1,\ell_2,\ell_3]$ as the
worldlines of two incoming charged particles.  They scatter at the
vertex $v[\ell_1,\ell_2,\ell_3]$ and fly away as two outgoing charged
particles whose worldlines are $e_1[\ell_1+1,\ell_2,\ell_3]$ and
$e_2[\ell_1,\ell_2+1,\ell_3]$.  Each string creates a charge of unit
magnitude, but the sign of the charge depends on the position of the
string relative to $\Dfour_3[\ell_3]$.  For strings along
$e_1[\ell_1,\ell_2,\ell_3]$ and $e_1[\ell_1+1,\ell_2,\ell_3]$, the
charges created by them are positive if
$x^4(\Mfive_2[\ell_2]) < x^4(\Mfive_3[\ell_3])$ and negative if
$x^4(\Mfive_2[\ell_2]) > x^4(\Mfive_3[\ell_3])$.  Similarly, for
strings along $e_2[\ell_1,\ell_2,\ell_3]$ and
$e_2[\ell_1,\ell_2+1,\ell_3]$, the charges are positive if
$x^5(\Mfive_1[\ell_1]) < x^5(\Mfive_3[\ell_3])$ and negative if
$x^5(\Mfive_1[\ell_1]) > x^5(\Mfive_3[\ell_3])$.%
\footnote{The charge assignments are symmetric under the cyclic
  permutation $(1,4) \to (2,5) \to (3,6) \to (1,4)$: in
  $\Dfour_1[\ell_1]$, the other ends of these strings are charged
  positively if $x^5(\Mfive_3[\ell_3]) < x^5(\Mfive_1[\ell_1])$ and
  negatively if $x^5(\Mfive_3[\ell_3]) > x^5(\Mfive_1[\ell_1])$.}

Equating the charges before and after the scattering, we obtain the
relation
\begin{equation}
  \charge{n}_1[\ell_1,\ell_2,\ell_3] - \charge{n}_2[\ell_1,\ell_2,\ell_3]
  =
  \charge{n}_1[\ell_1+1,\ell_2,\ell_3] - \charge{n}_2[\ell_1,\ell_2+1,\ell_3] \,,
\end{equation}
where we have defined
\begin{equation}
  \charge{n}_i[\ell_1,\ell_2,\ell_3]
  =
  \begin{cases}
    +n_i[\ell_1,\ell_2,\ell_3]
    & (x^{i+3}(\Mfive_{i+1}[\ell_{i+1}]) < x^{i+3}(\Mfive_{i+2}[\ell_{i+2}])) \,;
    \\
    -n_i[\ell_1,\ell_2,\ell_3]
    & (x^{i+3}(\Mfive_{i+1}[\ell_{i+1}]) > x^{i+3}(\Mfive_{i+2}[\ell_{i+2}])) \,.
  \end{cases}
\end{equation}
More generally, the charge conservation implies
\begin{equation}
  \charge{n}_i[\ell_1,\ell_2,\ell_3] - \charge{n}_j[\ell_1,\ell_2,\ell_3]
  =
  \charge{n}_i[\ell_1+\delta_{1i},\ell_2+\delta_{2i},\ell_3+\delta_{3i}]
  - \charge{n}_j[\ell_1+\delta_{1j},\ell_2+\delta_{2j},\ell_3+\delta_{3j}]
  \,.
\end{equation}

Summing this equation over $\ell_j$, we find that for distinct $i$,
$j$, $k \in \{1,2,3\}$, the quantity
\begin{equation}
  \charge{n}_i(\Sigma_k[\ell_k]) = \sum_{\ell_j} \charge{n}_i[\ell_1,\ell_2,\ell_3]
\end{equation}
is independent of $\ell_i$.  In other words, the charge flowing in the
$x^i$-direction on $\Sigma_k[\ell_k]$ is conserved.  Further summing
over $\ell_k$, we obtain a constant
\begin{equation}
  \charge{n}_i
  = \sum_{\ell_k} \charge{n}_i(\Sigma_k[\ell_k])
  = \sum_{\ell_j, \ell_k} \charge{n}_i[\ell_1,\ell_2,\ell_3] \,,
\end{equation}
which is the total charge flowing in the $x^i$-direction in the
system.

\subsection{Partition function}
\label{sec:pf}

For a spin configuration $\config{n}$, let $\CH_{\BPS}^{\config{n}}$ be the
subspace of $\CH_{\BPS}$ that contain $n_i[\ell_1,\ell_2,\ell_3]$
M2-branes on $e_i[\ell_1,\ell_2,\ell_3]$.  The partition function of
the model is given by the sum
\begin{equation}
  Z = \sum_{\config{n} \in \config{N}} Z_{\config{n}}
\end{equation}
over all spin configurations of the index $Z_{\config{n}}$ of
$\CH_{\BPS}^{\config{n}}$:
\begin{equation}
  Z_{\config{n}}
  =
  \Tr_{\CH_{\BPS}^{\config{n}}}\bigl((-1)^F q^{J_{78} - J_{9\ten}} e^{-\beta(H - E_0)}\bigr) \,.
\end{equation}
For each $\config{n}$, the factor $e^{-\beta(H - E_0)}$ in the trace
is constant and can be calculated as follows.

Each BPS M2-brane has a mass equal to the product of its area and the
unit M2-brane charge $Q_\Mtwo$.  The mass of the
$n_i[\ell_1, \ell_2, \ell_3]$ M2-branes placed along the edge
$e_i[\ell_1, \ell_2, \ell_3]$ is
\begin{multline}
  E_i[\ell_1,\ell_2,\ell_3]
  =
  Q_\Mtwo
  \charge{n}_i[\ell_1, \ell_2, \ell_3]
  \bigl(x^i(\Mfive_i[\ell_i]) - x^i(\Mfive_i[\ell_i-1])\bigr)
  \\
  \times
  \bigl(x^{i+3}(\Mfive_{i+2}[\ell_{i+2}]) - x^{i+3}(\Mfive_{i+1}[\ell_{i+1}])\bigr)
  \,.
\end{multline}
For a state in $\CH_\BPS^{\config{n}}$, the total mass of the M2-branes on the
edges is
\begin{multline}
  \label{eq:E}
  \sum_{\ell_1, \ell_2, \ell_3} \sum_{i=1}^3
  E_i[\ell_1,\ell_2,\ell_3]
  =
  \sum_{i=1}^3 Q_\Mtwo c_i \Biggl(
  \sum_{\ell_{i+2}}
  \charge{n}_i(\Sigma_{i+2}[\ell_{i+2}])
  x^{i+3}(\Mfive_{i+2}[{\ell_{i+2}}])
  \\
  -
  \sum_{\ell_{i+1}}
  \charge{n}_i(\Sigma_{i+1}[\ell_{i+1}])
  x^{i+3}(\Mfive_{i+1}[{\ell_{i+1}}])
  \Biggr) \,,
\end{multline}
where
\begin{equation}
  c_i
  = \sum_{\ell_i} \bigl(x^i(\Mfive_i[\ell_i]) - x^i(\Mfive_i[\ell_i-1])\bigr)
\end{equation}
is the circumference of $\circle_i$.  The configuration of M2-branes
close to the vertices may be complicated, but this part, being
localized, does not contribute to the energy.

Hence, $Z_{\config{n}}$ takes the form
\begin{equation}
  Z_{\config{n}}
  =
  \mathring{Z}_{\config{n}}
  \prod_{\ell_1}
  \frac{\lambda_2(\Sigma_1[\ell_1])^{\charge{n}_2(\Sigma_1[\ell_1])}} 
       {\lambda_3(\Sigma_1[\ell_1])^{\charge{n}_3(\Sigma_1[\ell_1])}}
  \prod_{\ell_2}
  \frac{\lambda_3(\Sigma_2[\ell_2])^{\charge{n}_3(\Sigma_2[\ell_2])}} 
       {\lambda_1(\Sigma_2[\ell_2])^{\charge{n}_1(\Sigma_2[\ell_2])}}
  \prod_{\ell_3}
  \frac{\lambda_1(\Sigma_3[\ell_3])^{\charge{n}_1(\Sigma_3[\ell_3])}}
       {\lambda_2(\Sigma_3[\ell_3])^{\charge{n}_2(\Sigma_3[\ell_3])}}
  \,,
\end{equation}
with
\begin{equation}
  \lambda_j(\Sigma_i[\ell_i])
  = \exp\bigl(-\beta Q_\Mtwo c_j x^{j+3}(\Mfive_i[\ell_i])\bigr)
\end{equation}
and
\begin{equation}
  \mathring{Z}_{\config{n}}
  =
  \Tr_{\CH_{\BPS}^{\config{n}}}\bigl((-1)^F q^{J_{78} - J_{9\ten}}\bigr) \,.
\end{equation}
The factor $\mathring{Z}_{\config{n}}$ is the summand for
$\config{n} \in \config{N}$ in the definition \eqref{eq:Z0-R} of
$\mathring{Z}$ (when the signs of all surfaces are $+$).

The factor
$\lambda_j(\Sigma_i[\ell_i])^{\charge{n}_j(\Sigma_i[\ell_i])}$ is
independent of the $x^j$-coordinate of the point on $\Sigma_i[\ell_i]$
at which it is evaluated.  If it is evaluated between, say
$x^j(\Sigma_j[L_j])$ and $x^j(\Sigma_j[1])$, then it can be
interpreted as the contribution from twisting of the periodic boundary
condition for $\Sigma_i[\ell_i]$ in the $x^j$-direction.  The TQFT
interpretation of this factor is a Wilson loop of a background gauge
field on $\Sigma_i[\ell_i]$ wrapped around $\circle_j$; the
$x^j$-component of the gauge field is
$-\epsilon_{ijk} \beta Q_\Mtwo x^{j+3}(\Mfive_i[\ell_i])$, where
$k \in \{1,2,3\} \setminus \{i,j\}$ and $\epsilon_{ijk}$ is the
component of the completely antisymmetric tensor with
$\epsilon_{123} = 1$.  Since the Wilson loop is gauge invariant, it
can be evaluated in any gauge.  In particular, we can choose a gauge
such that the gauge field is zero everywhere except between
$x^j(\Sigma_j[L_j])$ and $x^j(\Sigma_j[1])$, thereby connecting to the
interpretation as a twisting parameter for the periodic boundary
condition.

The Wilson loop interpretation is not just an analogy.  Consider the
type IIA setup described in section~\ref{sec:charge}, which is
obtained by reduction on $\circle_0$.  Each D4-brane in that setup
supports a $\U(1)$ gauge field as well as five scalar fields
parametrizing its transverse positions.  Take $\Dfour_3[\ell_3]$ for
definiteness.  The components $A_1$, $A_2$, $A_6$ of the gauge field
$A$ on $\Dfour_3[\ell_3]$ combines with the scalar fields $\phi_4$,
$\phi_5$, $\phi_3$ parametrizing the positions in $\R_4$, $\R_5$,
$\circle_3$ to form a partial complex $\GL(1)$ gauge field
\begin{equation}
  \CA
  =
  (A_1 + \iu\phi_4) \rmd x^1 + (A_2 + \iu\phi_5) \rmd x^2
  + (A_6 + \iu\phi_3) \rmd x^6 \,.
\end{equation}
(In fact, the BPS sector the worldvolume theory on $\Dfour_3[\ell_3]$
is described by Chern--Simons theory on $\T_{12} \times \R_6$ with
gauge field $\CA$ and gauge group $\GL(1)$ \cite{Luo:2014sva,
  Yagi:2013fda, Lee:2013ida}.)  At low energies $\CA$ can be
considered as nondynamical.  The value of $A$ can be taken to be zero
here because we did not turn on background fields on the M5-branes,
whereas the value of $\phi^j$ sets the $x^j$-coordinate of
$\Dfour_3[\ell_3]$ (and hence of $\Mfive_3[\ell_3]$).  The coupling of
$\CA$ to open strings ending on $\Dfour_3[\ell_3]$ produces the Wilson
loop.

\subsection{R-matrices}
\label{sec:R-matrices}

We have thus reduced the computation of $Z_{\config{n}}$ to that of
$\mathring{Z}_{\config{n}}$.  As explained in section
\ref{sec:TQFT-lattice}, the latter is given by the product of the
local Boltzmann weights for the configuration $\config{n}$.  The local
Boltzmann weights are described by the R-matrices assigned to the
vertices, defined via the TQFT that emerges in the limit in which all
M2-branes become massless.  Let us deduce some properties these
R-matrices should have.

First of all, we need to address one important subtlety which we did
not discuss in section~\ref{sec:TQFT}: the massless limit is realized
by going to a special point in the configuration space of the
M5-branes, but the brane system may not behave continuously at this
point.  For example, M2-branes may be created or annihilated when two
M5-branes pass through each other.  In order to describe the lattice
model using the R-matrices, we need to specify the relative positions
of the M5-branes from which we start to take the massless limit.  We
will not try to present a completely satisfactory treatment of this
point.  Instead, in what follows we will content ourselves with the
following particular choice of the relative positions.

To the surface defects in the TQFT forming the cubic lattice, we
introduce the ordering
\begin{equation}
  \label{eq:ordering}
  \Sigma_1[1] < \Sigma_1[2] < \dotsb < \Sigma_1[L_1]
  < \Sigma_2[1] < \dotsb < \Sigma_2[L_2]
  < \Sigma_3[1] < \dotsb < \Sigma_3[L_3] \,.
\end{equation}
For any two surfaces $\Sigma_a$, $\Sigma_b$ such that
$\Sigma_a < \Sigma_b$ and $\Sigma_a \cap \Sigma_b \neq \emptyset$, we
require that the corresponding M5-branes $\Mfive_a$ and $\Mfive_b$,
which intersect along $\Sigma_a \cap \Sigma_b \subset M \subset T^*M$,
are infinitesimally separated in the fiber direction so that
$\Mfive_a$ is ``closer'' to the base than $\Mfive_b$.  In other words,
we take the massless limit from a configuration such that
\begin{equation}
  \label{eq:M5-order}
  x^{k+3}(\Mfive_i[\ell_i]) < x^{k+3}(\Mfive_j[\ell_j]) \,,
  \quad
  i < j \,,
  \quad
  k \in \{1,2,3\} \setminus \{i,j\} \,.
\end{equation}
With this convention, all factors of R-matrices involved in the
computation of $\mathring{Z}$ are specified unambiguously.

Let $\Sigma_a$, $\Sigma_b$, $\Sigma_c$ be three surfaces such that
$\Sigma_a < \Sigma_b < \Sigma_c$ and
$\Sigma_a \cap \Sigma_b \cap \Sigma_c \neq \emptyset$.  The space
$\CV_{(ab)}$ assigned to $\Sigma_a \cap \Sigma_b$ is isomorphic to a
vector space $\CV^{\sigma_a\sigma_b}$ determined by the signs $\sigma_a$
of $\Sigma_a$ and $\sigma_b$ of $\Sigma_b$.  It is either the bosonic
Fock space $\CF^{(0)}$ or the fermionic Fock space $\CF^{(1)}$:
\begin{equation}
  \CV^{\sigma_a\sigma_b}
  =
  \begin{cases}
    \CF^{(0)} = \bigoplus_{n \in \Z_{\geq0}} \C \ket{n}
    & (\sigma_a = \sigma_b) \,;
    \\
    \CF^{(1)} = \C\ket{0} \oplus \C\ket{1}
    & (\sigma_a \neq \sigma_b) \,.
  \end{cases}
\end{equation}
The state $\ket{1} \in \CF^{(1)}$ is fermionic, whereas
$\ket{0} \in \CF^{(1)}$ and all states in $\CF^{(0)}$ are bosonic.
The R-matrix $R_{(abc)}$ assigned to the intersection
$\Sigma_a \cap \Sigma_b \cap \Sigma_c$ is a linear operator
\begin{equation}
  R^{\sigma_a\sigma_b\sigma_c}
  \colon
  \CV^{\sigma_b\sigma_c} \otimes \CV^{\sigma_a\sigma_c} \otimes \CV^{\sigma_a\sigma_b}
  \to
  \CV^{\sigma_b\sigma_c} \otimes \CV^{\sigma_a\sigma_c} \otimes \CV^{\sigma_a\sigma_b}
  \,.
\end{equation}

The matrix elements of $R^{\sigma_a\sigma_b\sigma_c}$ are defined by
\begin{equation}
  R^{\sigma_a\sigma_b\sigma_c} (\ket{l} \otimes \ket{m} \otimes \ket{n})
  \\
  =
  \sum_{l', m', n'}
  \ket{l'} \otimes \ket{m'} \otimes \ket{n'}
  (R^{\sigma_a\sigma_b\sigma_c})_{lmn}^{l'm'n'}
  \,.
\end{equation}
Some of them are easily determined.  The matrix element
$(R^{\sigma_a\sigma_b\sigma_c})_{000}^{000}$ corresponds to the unique
situation in which there is no M2-brane, so we have
\begin{equation}
  \label{eq:R000000}
  (R^{\sigma_a\sigma_b\sigma_c})_{000}^{000} = 1 \,.
\end{equation}
With the convention~\eqref{eq:M5-order},
$\ket{l} \otimes \ket{m} \otimes \ket{n} \in \CV^{\sigma_b\sigma_c}
\otimes \CV^{\sigma_a\sigma_c} \otimes \CV^{\sigma_a\sigma_b}$ has the
charges $\charge{l} = l$, $\charge{m} = -m$, $\charge{n} = n$.  Charge
conservation then implies that
$(R^{\sigma_a\sigma_b\sigma_c})_{lmn}^{l'm'n'} = 0$ unless
\begin{equation}
  \label{eq:charge-consv}
  l + m = l' + m' \,,
  \quad
  m + n = m' + n' \,.
\end{equation}
Note that $l + n \equiv l' + n'$ mod $2$.  Since either none of
$\CV_{(ab)}$, $\CV_{(ac)}$, $\CV_{(bc)}$ or exactly two of them are
the fermionic Fock space, $R^{\sigma_a\sigma_b\sigma_c}$ is a bosonic
operator.

The R-matrix also has properties that reflect symmetries of the brane
system.  Consider a rotation that exchanges $\R^{78}$ and
$\R^{9\ten}$.  This is a symmetry of the brane system if the signs of
all M5-branes are reversed at the same time.  The definition of $Z$,
however, involves the operator $q^{J_{78} - J_{9\ten}}$ whose exponent
changes sign under the rotation.  It follows that
\begin{equation}
  R^{-\sigma_a \, -\sigma_b \, -\sigma_c}
  = \underline{R}^{\sigma_a\sigma_b\sigma_c} \,,
\end{equation}
where $\underline{R}^{\sigma_a\sigma_b\sigma_c}$ is
$R^{\sigma_a\sigma_b\sigma_c}$ with $q$ replaced by $q^{-1}$.  The
reflection $x^\mu \to -x^\mu$, $\mu = 1$, $2$, $\dotsc$, $6$, leaving
the symplectic structure of $T^*M$ invariant, is likewise a symmetry.
This reverses the ordering of surfaces \eqref{eq:ordering}, so we find
\begin{equation}
  \label{eq:RR-symmetry}
  (R^{\sigma_a\sigma_b\sigma_c})_{lmn}^{l'm'n'}
  =
  (R^{\sigma_c\sigma_b\sigma_a})_{nml}^{n'm'l'}
  \,.
\end{equation}
(Imagine the parity reversal of the brane picture in
figure~\ref{fig:lattice-names}.)

\begin{figure}
  \centering
  \tdplotsetmaincoords{12}{80}
  \begin{tikzpicture}[tdplot_main_coords, scale=0.3]
    \draw[thick, -stealth, rounded corners=8pt]
    ({-sqrt(3)},-1,-1) node[left] {$(ac)$} --
    ({-sqrt(3)},0,-1) -- ({sqrt(3)},5,1)
    -- ({-sqrt(3)},10,-1) -- ({-sqrt(3)},11,-1);

    \draw[thick, -stealth, rounded corners=8pt]
    ({sqrt(3)},-1,-1) node[left] {$(bc)$} -- 
    ({sqrt(3)},0,-1) -- ({-sqrt(3)},5,1)
    -- ({sqrt(3)},10,-1) -- ({sqrt(3)},11,-1);

    \draw[thick, -stealth, rounded corners=8pt]
    (0,-1,{sqrt(3)}) node[left] {$(ab)$} -- 
    (0,0,{sqrt(3)}) -- (0,5,{-sqrt(3)})
    -- (0,10,{sqrt(3)}) -- (0,11,{sqrt(3)});
  \end{tikzpicture}
  \quad
  {\Large =}
  \tdplotsetmaincoords{12}{80}
  \begin{tikzpicture}[tdplot_main_coords, scale=0.3]
    \draw[thick, -stealth, rounded corners=8pt]
    ({-sqrt(3)},-1,-1) node[left] {$(ac)$} -- ({-sqrt(3)},11,-1);

    \draw[thick, -stealth, rounded corners=8pt]
    ({sqrt(3)},-1,-1) node[left] {$(bc)$} -- ({sqrt(3)},11,-1);

    \draw[thick, -stealth, rounded corners=8pt]
    (0,-1,{sqrt(3)}) node[left] {$(ab)$} -- (0,11,{sqrt(3)});
  \end{tikzpicture}

  \caption{A graphical representation of the relation
    $(R^{\sigma_a\sigma_b\sigma_c})^2 = 1$.}

  \label{fig:R^2=1}
\end{figure}
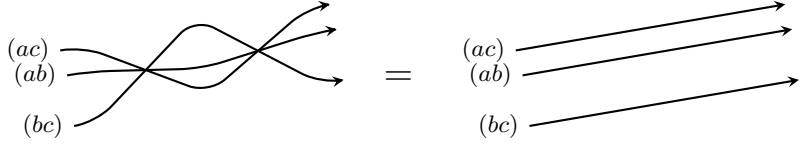

Lastly, there are properties that follow from the protected nature of
$\mathring{Z}$ in the same way as how the commutativity of transfer
matrices was deduced in section \ref{sec:[T,T]=0}.  One such property is that
the R-matrix is an involution:
\begin{equation}
  \label{eq:R^2=1}
  R^{\sigma_a\sigma_b\sigma_c} = (R^{\sigma_a\sigma_b\sigma_c})^{-1} \,.
\end{equation}
This property
expresses the equality between the two configurations of surfaces
shown in figure \ref{fig:R^2=1}.  Another property of this kind is the
supertetrahedron equation, to which we now turn.

\subsection{Supertetrahedron equation}

The most remarkable property of the R-matrices is that they solve the
supertetrahedron equation.  We have introduced this equation in
section \ref{sec:tetrahedron} in the case in which all surfaces have
the same sign and hence all states are bosonic.  In this case the
supertetrahedron equation reduces to the ordinary tetrahedron
equation~\eqref{eq:tetrahedron}.  In general, the surfaces can have
either sign and states can be bosonic or fermionic.  These extra
degrees of freedom make the equation more complex.

The supertetrahedron equation can be written most concisely in the
language of Fock spaces.  Consider four surfaces $\Sigma_1$,
$\Sigma_2$, $\Sigma_3$, $\Sigma_4$ as in figure~\ref{fig:tetrahedron}.
On the intersections of these surfaces live the six spaces $\CV_{(12)}$,
$\CV_{(13)}$, $\CV_{(14)}$, $\CV_{(23)}$, $\CV_{(24)}$, $\CV_{(34)}$.  The space
$\CV_{(ab)}$ is a Fock space generated from a vacuum state
$\ket{0}_{(ab)}$ by the action of a creation operator $c_{(ab)}$.  We
introduce the Fock space $\CV_{1234}$ generated from a vacuum state
$\ket{0}_{1234}$ by the action of the creation operators $c_{(12)}$,
$c_{(13)}$, $c_{(14)}$, $c_{(23)}$, $c_{(24)}$, $c_{(34)}$; it is
spanned by the states of the form
\begin{equation}
  c^{n_{(12)}}_{(12)}
  c^{n_{(13)}}_{(13)}
  c^{n_{(14)}}_{(14)}
  c^{n_{(23)}}_{(23)}
  c^{n_{(24)}}_{(24)}
  c^{n_{(34)}}_{(34)}
  \ket{0}_{1234}
  \,,
\end{equation}
where $n_{(ab)} \in \Z_{\geq0}$ or $\{0,1\}$ depending on whether
$c_{(ab)}$ is bosonic or fermionic.  For distinct $a$, $b$, $c$,
$d \in \{1,2,3,4\}$, we have an isomorphism
\begin{equation}
  \label{eq:V1234-tensor}
  \CV_{1234}
  \iso
  \CV_{(ab)} \otimes \CV_{(ac)} \otimes \CV_{(ad)}
  \otimes \CV_{(bc)} \otimes \CV_{(bd)} \otimes \CV_{(cd)}
\end{equation}
as vector spaces.  The notation
\begin{equation}
  \ket{l}_{(ab)}
  \ket{m}_{(ac)}
  \ket{n}_{(ad)}
  \ket{p}_{(bc)}
  \ket{q}_{(bd)}
  \ket{r}_{(cd)}
  =
  c^{l}_{(ab)}
  c^{m}_{(ac)}
  c^{n}_{(ad)}
  c^{p}_{(bc)}
  c^{q}_{(bd)}
  c^{r}_{(cd)}
  \ket{0}_{1234}
\end{equation}
makes the isomorphism manifest.

In this notation the R-matrix at $(abc)$ is represented in
$\CV_{1234}$ by the operator
\begin{equation}
  \CR_{(abc)}
  =
  \sum_{l,m,n,l',m',n'}
  (R^{\sigma_a\sigma_b\sigma_c})_{lmn}^{l'm'n'}
  \ket{l'}_{(bc)}
  \ket{m'}_{(ac)}
  \ket{n'}_{(ab)}
  {}_{(ab)}\bra{n}
  {}_{(ac)}\bra{m}
  {}_{(bc)}\bra{l}
  \,.
\end{equation}
The action of $\CR_{(abc)}$ is defined by
\begin{multline}
  \CR_{(abc)}
  \ket{l}_{(bc)}
  \ket{m}_{(ac)}
  \ket{n}_{(ab)}
  \ket{p}_{(ad)}
  \ket{q}_{(bd)}
  \ket{r}_{(cd)}
  \\
  =
  \sum_{l',m',n'}
  (R^{\sigma_a\sigma_b\sigma_c})_{lmn}^{l'm'n'} 
  \ket{l'}_{(bc)}
  \ket{m'}_{(ac)}
  \ket{n'}_{(ab)}
  \ket{p}_{(ad)}
  \ket{q}_{(bd)}
  \ket{r}_{(cd)}
  \,.
\end{multline}
Translating the two brane configurations in
figure~\ref{fig:tetrahedron} to the operator language we obtain the
\emph{supertetrahedron equation}
\begin{equation}
  \label{eq:supertetrahedron}
  \CR_{(234)} \CR_{(134)} \CR_{(124)} \CR_{(123)}
  =
  \CR_{(123)} \CR_{(124)} \CR_{(134)} \CR_{(234)}
  \,.
\end{equation}

The tetrahedron equation is usually expressed as an equality between
operators acting on the right-hand side of the
isomorphism~\eqref{eq:V1234-tensor}.  Let us recast the
supertetrahedron equation~\eqref{eq:supertetrahedron} in this form.
Doing so introduces nontrivial sign factors.

Let us define the fermion number operator $F$ by
\begin{equation}
  F \ket{n}
  =
  \begin{cases}
    0 & (\ket{n} \in \CF^{(0)}) \,;
    \\
    n \ket{n} & (\ket{n} \in \CF^{(1)}) \,,
  \end{cases}
\end{equation}
and let $P$ be the operator that swaps factors in the tensor product
of two spaces with an appropriate sign:
\begin{equation}
  P (\ket{m} \otimes \ket{n})
  =
  (-1)^{F \otimes F} \ket{n} \otimes \ket{m} \,.
\end{equation}
We define the operator
\begin{equation}
  \breve R^{\sigma_a\sigma_b\sigma_c}
  = 
  R^{\sigma_a\sigma_b\sigma_c} (P \otimes 1) (1 \otimes P) (P \otimes 1) \,.
\end{equation}
The matrix elements of
$\breve R^{\sigma_a\sigma_b\sigma_c}\colon \CV^{\sigma_a\sigma_b}
\otimes \CV^{\sigma_a\sigma_c} \otimes \CV^{\sigma_b\sigma_c} \to
\CV^{\sigma_b\sigma_c} \otimes \CV^{\sigma_a\sigma_c} \otimes
\CV^{\sigma_a\sigma_b}$ are given by
\begin{equation}
  (\breve R^{\sigma_a\sigma_b\sigma_c})_{nml}^{l'm'n'}
  =
  (-1)^{lm \deltab_{\sigma_b\sigma_c} \deltab_{\sigma_c\sigma_a}
    +  mn \deltab_{\sigma_c\sigma_a} \deltab_{\sigma_a\sigma_b}
    + nl \deltab_{\sigma_a\sigma_b} \deltab_{\sigma_b\sigma_c}}
    (R^{\sigma_a\sigma_b\sigma_c})_{lmn}^{l'm'n'} \,,
\end{equation}
where $\deltab_{\sigma_a\sigma_b} = 1 - \delta_{\sigma_a \sigma_b}$.
Finally, we rename the vector spaces involved in the supertetrahedron
equation as
\begin{equation}
  \CV_1 = \CV^{\sigma_3\sigma_4} \,,
  \
  \CV_2 = \CV^{\sigma_2\sigma_4} \,,
  \
  \CV_3 = \CV^{\sigma_2\sigma_3} \,,
  \
  \CV_4 = \CV^{\sigma_1\sigma_4} \,,
  \
  \CV_5 = \CV^{\sigma_1\sigma_3} \,,
  \
  \CV_6 = \CV^{\sigma_1\sigma_2} \,.
\end{equation}

In terms of the operators and spaces introduced above, the
supertetrahedron equation can be expressed as the following equality
between homomorphisms from
$\CV_6 \otimes \CV_5 \otimes \CV_4 \otimes \CV_3 \otimes \CV_2 \otimes
\CV_1$ to
$\CV_1 \otimes \CV_2 \otimes \CV_3 \otimes \CV_4 \otimes \CV_5 \otimes
\CV_6$:
\begin{multline}
  \label{eq:supertetrahedron-breve-R}
  \breve R^{\sigma_2\sigma_3\sigma_4}_{123}
  \breve R^{\sigma_1\sigma_3\sigma_4}_{345}
  P_{23} P_{56}
  \breve R^{\sigma_1\sigma_2\sigma_4}_{345}
  \breve R^{\sigma_1\sigma_2\sigma_3}_{123}
  P_{34}
  \\
  =
  P_{34}
  \breve R^{\sigma_1\sigma_2\sigma_3}_{456}
  \breve R^{\sigma_1\sigma_2\sigma_4}_{234}
  P_{12} P_{45}
  \breve R^{\sigma_1\sigma_3\sigma_4}_{234}
  \breve R^{\sigma_2\sigma_3\sigma_4}_{456}
  \,.
\end{multline}
The subscripts on an operator indicate the positions in the relevant
tensor product space at which the operator acts.
In components,
\begin{equation}
  \begin{split}
    &
    \sum_{n'_1, n'_2, n'_3, n'_4, n'_5, n'_6}
    (-1)^
    {n_1 n''_6 \deltab_{\sigma_3\sigma_4} \deltab_{\sigma_1\sigma_2}
      + n'_2 n'_5 \deltab_{\sigma_2\sigma_4} \deltab_{\sigma_1\sigma_3}
      + n_3 n_4 \deltab_{\sigma_2\sigma_3} \deltab_{\sigma_1\sigma_4}}
    \\
    & \qquad
    \times
    (\breve R^{\sigma_2\sigma_3\sigma_4})^{n''_1 n''_2 n''_3}_{n'_3 n'_2 n'_1}
    (\breve R^{\sigma_1\sigma_3\sigma_4})^{n'_1 n''_4 n''_5}_{n'_5 n'_4 n_1}
    (\breve R^{\sigma_1\sigma_2\sigma_4})^{n'_2 n'_4 n''_6}_{n'_6 n_4 n_2}
    (\breve R^{\sigma_1\sigma_2\sigma_3})^{n'_3 n'_5 n'_6}_{n_6 n_5 n_3}
    \\
    & \qquad
    =
    \sum_{n'_1, n'_2, n'_3, n'_4, n'_5, n'_6}
    (-1)^
    {n''_1 n_6 \deltab_{\sigma_3\sigma_4} \deltab_{\sigma_1\sigma_2}
      + n'_2 n'_5 \deltab_{\sigma_2\sigma_4} \deltab_{\sigma_1\sigma_3}
      + n''_3 n''_4 \deltab_{\sigma_2\sigma_3} \deltab_{\sigma_1\sigma_4}}
    \\
    & \qquad\qquad\quad
    \times (\breve R^{\sigma_1\sigma_2\sigma_3})^{n''_3 n''_5 n''_6}_{n'_6 n'_5 n'_3}
    (\breve R^{\sigma_1\sigma_2\sigma_4})^{n''_2 n''_4 n'_6}_{n_6 n'_4 n'_2}
    (\breve R^{\sigma_1\sigma_3\sigma_4})^{n''_1 n'_4 n'_5}_{n_5 n_4 n'_1}
    (\breve R^{\sigma_2\sigma_3\sigma_4})^{n'_1 n'_2 n'_3}_{n_3 n_2 n_1}
    \,.
  \end{split}
\end{equation}
Defining
$\breve R^{\sigma_a\sigma_b\sigma_c}_{\alpha,\beta,\gamma}
\in \End(\CV_1 \otimes \CV_2 \otimes \CV_3 \otimes \CV_4 \otimes \CV_5 \otimes
\CV_6)$ by
\begin{multline}
  \label{eq:breve-R-End}
  \breve R^{\sigma_a\sigma_b\sigma_c}_{\alpha,\beta,\gamma}
  (\dotsb \otimes
  \ket{n_\alpha} \otimes \dotsb \otimes
  \ket{n_\beta} \otimes \dotsb \otimes
  \ket{n_\gamma} \otimes \dotsb)
  \\
  =
  \sum_{n'_\alpha, n'_\beta, n'_\gamma}
  (\breve R^{\sigma_a\sigma_b\sigma_c})_{n_\gamma n_\beta n_\alpha}^{n'_\alpha n'_\beta n'_\gamma}
  (\dotsb \otimes
  \ket{n'_\alpha} \otimes \dotsb \otimes
  \ket{n'_\beta} \otimes \dotsb \otimes
  \ket{n'_\gamma} \otimes \dotsb) \,,
\end{multline}
we can also write the supertetrahedron equation as
\begin{multline}
  \label{eq:supertetrahedron-2}
  \breve R^{\sigma_2\sigma_3\sigma_4}_{1,2,3}
  \breve R^{\sigma_1\sigma_3\sigma_4}_{1,4,5}
  (-1)^{F_1 F_6 + F_2 F_5}
  \breve R^{\sigma_1\sigma_2\sigma_4}_{2,4,6}
  \breve R^{\sigma_1\sigma_2\sigma_3}_{3,5,6}
  (-1)^{F_3 F_4}
  \\
  =
  (-1)^{F_3 F_4}
  \breve R^{\sigma_1\sigma_2\sigma_3}_{3,5,6}
  \breve R^{\sigma_1\sigma_2\sigma_4}_{2,4,6}
  (-1)^{F_1 F_6 + F_2 F_5}
  \breve R^{\sigma_1\sigma_3\sigma_4}_{1,4,5}
  \breve R^{\sigma_2\sigma_3\sigma_4}_{1,2,3}
  \,,
\end{multline}
The sign factors are transparent in this expression.

The supertetrahedron equation combines $16$ quartic relations
corresponding to the $16$ choices of the quadruple of signs
$(\sigma_1, \sigma_2, \sigma_3, \sigma_4)$.  To unpack these
relations, let us define
\begin{alignat}{2}
  \TDR &= R^{+++} &\in \End(\CF^{(0)} \otimes \CF^{(0)} \otimes \CF^{(0)}) \,,
  \\
  \TDL &= R^{++-} &\in \End(\CF^{(1)} \otimes \CF^{(1)} \otimes \CF^{(0)}) \,,
  \\
  \TDM &= R^{-++} &\in \End(\CF^{(0)} \otimes \CF^{(1)} \otimes \CF^{(1)}) \,,
  \\
  \TDN &= R^{-+-} &\in \End(\CF^{(1)} \otimes \CF^{(0)} \otimes \CF^{(1)}) \,.
\end{alignat}
In view of the properties of R-matrices deduced in section
\ref{sec:R-matrices}, these operators satisfy
\begin{align}
  \label{eq:RR}
  \TDR_{lmn}^{l'm'n'} &= \TDR_{nml}^{n'm'l'} \,,
  \\
  \label{eq:LM}
  \TDL_{lmn}^{l'm'n'} &= \TDM_{nml}^{n'm'l'} \,,
  \\
  \label{eq:NN}
  \TDN_{lmn}^{l'm'n'} &= \TDN_{nml}^{n'm'l'}
\end{align}
and
\begin{alignat}{2}
  \underline{\TDR} &= R^{---} &\in \End(\CF^{(0)} \otimes \CF^{(0)} \otimes \CF^{(0)}) \,,
  \\
  \underline{\TDL} &= R^{--+} &\in \End(\CF^{(1)} \otimes \CF^{(1)} \otimes \CF^{(0)}) \,,
  \\
  \underline{\TDM} &= R^{+--} &\in \End(\CF^{(0)} \otimes \CF^{(1)} \otimes \CF^{(1)}) \,,
  \\
  \underline{\TDN} &= R^{+-+} &\in \End(\CF^{(1)} \otimes \CF^{(0)} \otimes \CF^{(1)}) \,.
\end{alignat}
Eight of the $16$ relations are
\begin{alignat}{2}
  \label{eq:RRRR}
  ({+}{+}{+}{+})
  & \qquad & &
  \breve\TDR_{1,2,3} \, \breve\TDR_{1,4,5} \,
  \breve\TDR_{2,4,6} \, \breve\TDR_{3,5,6}
  = \breve\TDR_{3,5,6} \, \breve\TDR_{2,4,6} \,
  \breve\TDR_{1,4,5} \, \breve\TDR_{1,2,3}
  \,,
  \\
  \label{eq:LLLR}
  ({+}{+}{+}{-})
  & \qquad & &
  \breve\TDL_{1,2,3} \, \breve\TDL_{1,4,5} \,
  \breve\TDL_{2,4,6} \, \breve\TDR_{3,5,6}
  = \breve\TDR_{3,5,6} \, \breve\TDL_{2,4,6} \,
  \breve\TDL_{1,4,5} \, \breve\TDL_{1,2,3}
  \,,
  \\
  \label{eq:NNRL}
  ({+}{+}{-}{+})
  & \qquad & &
  \breve{\underline\TDN}_{1,2,3} \, \breve{\underline\TDN}_{1,4,5} \,
  \breve\TDR_{2,4,6} \, \breve\TDL_{3,5,6}
  =
  \breve\TDL_{3,5,6} \, \breve\TDR_{2,4,6} \,
  \breve{\underline\TDN}_{1,4,5} \, \breve{\underline\TDN}_{1,2,3}
  \,,
  \\
  \label{eq:MRNN}
  ({+}{-}{+}{+})
  & \qquad & &
  \breve\TDM_{1,2,3} \, \breve\TDR_{1,4,5} \,
  \breve{\underline\TDN}_{2,4,6} \, \breve{\underline\TDN}_{3,5,6}
  =
  \breve{\underline\TDN}_{3,5,6} \, \breve{\underline\TDN}_{2,4,6} \,
  \breve\TDR_{1,4,5} \, \breve\TDM_{1,2,3}
  \,,
  \\
  \label{eq:MMLL}
  ({+}{+}{-}{-})
  & \qquad & &
  \breve{\underline\TDM}_{1,2,3}\, \breve{\underline\TDM}_{1,4,5}
  (-1)^{\hat{n}_2 \hat{n}_5}
  \breve\TDL_{2,4,6} \, \breve\TDL_{3,5,6}
  (-1)^{\hat{n}_{3} \hat{n}_{4}}
  \nonumber
  \\
  & & & \qquad
  =
  (-1)^{\hat{n}_{3} \hat{n}_{4}}
  \breve\TDL_{3,5,6} \, \breve\TDL_{2,4,6}
  (-1)^{\hat{n}_2 \hat{n}_5}
  \breve{\underline\TDM}_{1,4,5} \, \breve{\underline\TDM}_{1,2,3}
  \,,
  \\
  \label{eq:NLMN}
  ({+}{-}{+}{-})
  & \qquad & &
  \breve\TDN_{1,2,3} \, \breve\TDL_{1,4,5}
  (-1)^{\hat{n}_1 \hat{n}_6}
  \breve{\underline\TDM}_{2,4,6} \, \breve{\underline\TDN}_{3,5,6}
  (-1)^{\hat{n}_{3} \hat{n}_{4}}
  \nonumber
  \\
  & & & \qquad
  =
  (-1)^{\hat{n}_{3} \hat{n}_{4}}
  \breve{\underline\TDN}_{3,5,6} \, \breve{\underline\TDM}_{2,4,6}
  (-1)^{\hat{n}_1 \hat{n}_6}
   \breve\TDL_{1,4,5} \,  \breve\TDN_{1,2,3}
  \,,
  \\
  \label{eq:LNNM}
  ({+}{-}{-}{+})
  & \qquad & &
  \breve{\underline\TDL}_{1,2,3} \, \breve{\underline\TDN}_{1,4,5}
  (-1)^{\hat{n}_1 \hat{n}_6 + \hat{n}_2 \hat{n}_5}
  \breve{\underline\TDN}_{2,4,6} \,\breve{\underline\TDM}_{3,5,6}
  \nonumber
  \\
  & & & \qquad
  =
  \breve{\underline\TDM}_{3,5,6} \, \breve{\underline\TDN}_{2,4,6}
  (-1)^{\hat{n}_1 \hat{n}_6 + \hat{n}_2 \hat{n}_5}
  \breve{\underline\TDN}_{1,4,5} \,\breve{\underline\TDL}_{1,2,3}
  \,,
  \\
  \label{eq:RMMM}
  ({+}{-}{-}{-})
  & \qquad & &
  \breve{\underline\TDR}_{1,2,3} \, \breve{\underline\TDM}_{1,4,5} \,
  \breve{\underline\TDM}_{2,4,6} \, \breve{\underline\TDM}_{3,5,6}
  = \breve{\underline\TDM}_{3,5,6} \, \breve{\underline\TDM}_{2,4,6} \,
  \breve{\underline\TDM}_{1,4,5} \, \breve{\underline\TDR}_{1,2,3}
  \,,
\end{alignat}
where $\hat{n}$ is the number operator defined by
\begin{equation}
  \hat{n} \ket{n} = n \ket{n} \,.
\end{equation}
The remaining eight are obtained from the above relations by the
replacement $q \to q^{-1}$.  The exchange of indices
$(1,2) \leftrightarrow (6,5)$ interchanges relations \eqref{eq:NNRL}
and \eqref{eq:MRNN} and relations~\eqref{eq:LLLR} and \eqref{eq:RMMM}
(without the underlines).

\subsection{Integrability}
\label{sec:[T,T]=0}

In section \ref{sec:[T0,T0]=0} we deduced that the transfer matrices
commute in the special case in which all M2-branes are massless.  A
surface defect $\Sigma$ in the emergent TQFT comes with a single
discrete parameter, the sign $\sigma(\Sigma)$ specifying which of the
planes $\R^2_{78}$ and $\R^2_{9\ten}$ the corresponding M5-brane
covers.  Accordingly, the transfer matrix $\mathring{T}_3[\ell_3]$ for
$\mathring{Z}$ has a discrete parameter $\sigma(\Mfive_3[\ell_3])$.
In our discussion in section \ref{sec:[T0,T0]=0}, we chose this sign
to be $+$ and wrote $\mathring{T}_3[\ell_3]$ as
$\mathring{\TDT}_3[\ell_3]$.

The transfer matrix $T_3[\ell_3]$ of the model defined by the index
$Z$, as opposed to its massless limit $\mathring{Z}$, has additional
continuous parameters, namely the coordinates of $\Mfive_3[\ell_3]$ in
$\R^2_{45}$.  Expanding $T_3[\ell_3]$ in these parameters, we obtain a
series of operators acting on the Hilbert space of the equivalent 2D
quantum spin model, and we can declare that one of them is the
Hamiltonian.  If the commutativity of transfer matrices continues to
hold, then all of these operators commute with each other.  The
commutativity therefore implies that the model is quantum integrable:
there is a series of commuting conserved charges generated by the
transfer matrix.

The commutativity of transfer matrices can be argued as follows.  Fix
the size of the lattice and the positions of the M5-branes in
$\R^3_{456}$.  Then, the energy of a BPS state is determined solely by
the charges flowing on the surface defects according to formula
\eqref{eq:E}.  Let $\CE$ be the spectrum of $H - E_0$ in $\CH_\BPS$,
which is discrete, and let $\CH^{E,j}$ be the subspace of $\CH$ in
which $H - E_0 = E$ and $J_{78} - J_{9\ten} = j$.  We can write the
partition function as
\begin{equation}
  Z
  =
  \sum_{E \in \CE} \sum_{j=-\infty}^\infty
  \Tr_{\CH_\BPS^{E,j}}(-1)^F e^{-\beta E} q^j \,,
\end{equation}
where $\CH_\BPS^{E,j} = \CH^{E,j} \cap \CH_\BPS$.  The coefficient of
$e^{-\beta E} q^j$ is the Witten index of $\CH^{E,j}$ and invariant
under continuous movement of the M5-branes in $\T_{123}$.  In
particular, we can switch the positions of $\Mfive_3[\ell_3]$ and
$\Mfive_3[\ell_3+1]$ in $\circle_3$ continuously.  Throughout the
process the two M5-branes do not intersect in the 11D spacetime, so
$Z$ remains the same.  Hence, $T_3[\ell_3]$ and $T_3[\ell_3+1]$
commute.

Since the above argument is a little abstract, we provide below a more
direct derivation of the commutativity of transfer matrices using the
supertetrahedron equation and charge conservation.

In figure~\ref{fig:tetrahedron}, replace the surface $\Sigma_1$ with a
stack of surfaces $\Sigma_1[\ell_1]$, $\ell_1 = 1$, $\dotsc$, $L_1$,
that are all parallel to $\Sigma_1$.  Similarly, replace $\Sigma_2$
with a stack of parallel surfaces $\Sigma_2[\ell_2]$, $\ell_2 = 1$,
$\dotsc$, $L_2$.  These surfaces should be thought of as (parts of)
the $2$-tori with the same names in the lattice of the 3D lattice
model.  The surfaces $\Sigma_3$ and $\Sigma_4$ will be identified with
$\Sigma_3[\ell_3]$ and $\Sigma_3[\ell_3+1]$.  We place these surfaces
in such a way that their intersections look like the corresponding
horizontal layers of the 3D lattice, and introduce the following
ordering of the surfaces:
\begin{equation}
  \Sigma_1[1] < \Sigma_1[2] < \dotsb < \Sigma_1[L_1]
  < \Sigma_2[1] < \Sigma_2[2] < \dotsb < \Sigma_2[L_2]
  < \Sigma_3 < \Sigma_4 \,.
\end{equation}

Let $\CV_{(1[\ell_1]2[\ell_2])}$ and $\CV_{(a[\ell_a]b)}$, $a = 1$, $2$, \
$b = 3$, $4$, be the Fock spaces assigned to
$\Sigma_1[\ell_1] \cap \Sigma_2[\ell_2]$ and
$\Sigma_a[\ell_a] \cap \Sigma_b$.  We use $\CV_{(\mathbf{12})}$ and
$\CV_{(\mathbf{a}b)}$ to denote the Fock spaces such that
\begin{equation}
  \CV_{(\mathbf{12})}
  \iso \bigotimes_{\ell_1, \ell_2} \CV_{(1[\ell_1]2[\ell_2])} \,,
  \quad
  \CV_{(\mathbf{a}b)}
  \iso \bigotimes_{\ell_a} \CV_{(a[\ell_a]b)} \,.
\end{equation}
Also, let $\CR_{(1[\ell_1]2[\ell_2]b)}$ and $\CR_{(a[\ell_a]34)}$,
$a = 1$, $2$, denote the R-matrices at
$\Sigma_1[\ell_1] \cap \Sigma_2[\ell_2] \cap \Sigma_b$ and
$\Sigma_a[\ell_a] \cap \Sigma_3 \cap \Sigma_4$, and define
\begin{equation}
  \CR_{(\mathbf{1}34)}
  = \prod_{\ell_1}^{\longleftarrow} \CR_{(1[\ell_1]34)} \,,
  \quad
  \CR_{(\mathbf{2}34)}
  = \prod_{\ell_2}^{\longrightarrow} \CR_{(2[\ell_2]34)} \,,
  \quad
  \CR_{(\mathbf{1}\mathbf{2}b)}
  =
  \prod_{\ell_1, \ell_2}^{\longleftarrow} \CR_{(1[\ell_1]2[\ell_2]b)} \,.
\end{equation}
The symbols $\overset{\leftarrow}{\prod}$ and
$\overset{\rightarrow}{\prod}$ mean that the operator product is
arranged in the lexicographic order of the vertices from left to right
and from right to left, respectively.  By repeated application of the
supertetrahedron equation \eqref{eq:supertetrahedron} we obtain the
equality
\begin{equation}
  \label{eq:enhanced-tetrahedron}
  \CR_{(\mathbf{2}34)}
  \CR_{(\mathbf{1}34)}
  \CR_{(\mathbf{1}\mathbf{2}4)}
  \CR_{(\mathbf{1}\mathbf{2}3)}
  =
  \CR_{(\mathbf{1}\mathbf{2}3)}
  \CR_{(\mathbf{1}\mathbf{2}4)}
  \CR_{(\mathbf{1}34)}
  \CR_{(\mathbf{2}34)}
\end{equation}
between endomorphisms of
$\CV_{\mathbf{1}\mathbf{2}34} \iso \CV_{(\mathbf{1}\mathbf{2})}
\otimes \CV_{(\mathbf{1}3)} \otimes \CV_{(\mathbf{1}4)} \otimes
\CV_{(\mathbf{2}3)} \otimes \CV_{(\mathbf{2}4)} \otimes \CV_{(34)}$.

By charge conservation,
$\hat{n}_{(2[\ell_2]3)} + \hat{n}_{(2[\ell_2]4)}$,
$\sum_{\ell_2} \hat{n}_{(2[\ell_2]3)} - \hat{n}_{(34)}$ and
$\sum_{\ell_2} \hat{n}_{(2[\ell_2]4)} + \hat{n}_{(34)}$ commute with
$\CR_{(\mathbf{2}34)}$.  Multiplying both sides of
\eqref{eq:enhanced-tetrahedron} by
\begin{equation}
  \Biggl(
  \frac{
    \nu_3^{\sum_{\ell_2} \hat{n}_{(2[\ell_2]3)} - \hat{n}_{(34)}}
    \nu_4^{\sum_{\ell_2} \hat{n}_{(2[\ell_2]4)} - \hat{n}_{(34)}}
  }
  {
    \prod_{\ell_2} \mu_1[\ell_2]^{\hat{n}_{(2[\ell_2]3)} + \hat{n}_{(2[\ell_2]4)}}
  }
  \Biggr)
  \CR_{(\mathbf{2}34)}^{-1}
\end{equation}
and taking the supertrace over
$\CV_{(\mathbf{2}3)} \otimes \CV_{(\mathbf{2}4)} \otimes \CV_{(34)}$,%
\footnote{By taking the supertrace over
  $\CV_{(34)} \otimes \CV_{(\mathbf{2}4)} \otimes \CV_{(\mathbf{2}3)}$, we
  mean that we multiply by
  $(-1)^{F_{(34)} + \sum_{\ell_2} F_{(2[\ell_2]4)} + \sum_{\ell_2}
    F_{(2[\ell_2]3)}}$,
  sandwich by
  $\ket{l}_{(34)} \overset{\leftarrow}{\prod}_{\ell_2}
  \ket{m[\ell_2]}_{(2[\ell_2]4)} \overset{\leftarrow}{\prod}_{\ell_2}
  \ket{n[\ell_2]}_{(2[\ell_2]3)}$ and its dual, and sum over $l$,
  $m[\ell_2]$, $n[\ell_2]$, $\ell_2 = 1$, $\dotsc$, $L_2$.}
we obtain
\begin{equation}
  \label{eq:enhanced-YBE}
  \CS_{\bullet34} \CS_{\bullet\mathbf{2}4} \CS_{\bullet\mathbf{2}3}
  = \CS_{\bullet\mathbf{2}3} \CS_{\bullet\mathbf{2}4} \CS_{\bullet34}
  \,,
\end{equation}
where
\begin{equation}
  \CS_{\bullet\mathbf{2}b}
  =
  \Str_{\CV_{(\mathbf{2}b)}}
  \Biggl(
  \prod_{\ell_2}
  \biggl(\frac{\nu_b}{\mu_1[\ell_2]}\biggr)^{\hat{n}_{(2[\ell_2]b)}}
  \CR_{(\mathbf{1}\mathbf{2}b)}
  \Biggr)
  \,,
  \quad
  \CS_{\bullet34}
  =
  \Str_{\CV_{(34)}}
  \bigl(
  (\nu_3\nu_4)^{-\hat{n}_{(34)}}
  \CR_{(\mathbf{1}34)}
  \bigr) \,.
\end{equation}

Charge conservation implies that
$\hat{n}_{(1[\ell_1]3)} + \hat{n}_{(1[\ell_1]4)}$,
$\sum_{\ell_1} \hat{n}_{(1[\ell_1]3)}$ and
$\sum_{\ell_1} \hat{n}_{(1[\ell_1]4)}$ commute with $\CS_{\bullet34}$.
Multiplying both sides of \eqref{eq:enhanced-YBE} by
\begin{equation}
  \Biggl(\prod_{\ell_1}
  \frac{\rho_3^{\hat{n}_{(1[\ell_1]3)}} \rho_4^{\hat{n}_{(1[\ell_1]4)}}}
  {\mu_2[\ell_1]^{\hat{n}_{(1[\ell_1]3)} + \hat{n}_{(1[\ell_1]4)}}}
  \Biggr)
  \CS_{\bullet34}^{-1}
\end{equation}
and taking the supertrace over
$\CV_{(\mathbf{1}3)} \otimes \CV_{(\mathbf{1}4)}$, we find
$\CT_4 \CT_3 = \CT_3 \CT_4$ with
\begin{equation}
  \begin{split}
    \CT_b
    &=
    \Str_{\CV_{(\mathbf{1}b)}}
    \Biggl(
    \prod_{\ell_1}
    \biggl(\frac{\rho_b}{\mu_2[\ell_1]}\biggr)^{\hat{n}_{(1[\ell_1]b)}}
    \CS_{\bullet\mathbf{2}b}
    \Biggr)
    \\
    &=
    \Str_{\CV_{(\mathbf{2}b)} \otimes \CV_{(\mathbf{1}b)}}
    \Biggl(
    \prod_{\ell_1}
    \biggl(\frac{\rho_b}{\mu_2[\ell_1]}\biggr)^{\hat{n}_{(1[\ell_1]b)}}
    \prod_{\ell_2}
    \biggl(\frac{\nu_b}{\mu_1[\ell_2]}\biggr)^{\hat{n}_{(2[\ell_2]b)}}
    \CR_{(\mathbf{1}\mathbf{2}b)}
    \Biggr)
    \,.
  \end{split}
\end{equation}

Finally, take $\Sigma_{3} = \Sigma_3[\ell_3]$,
$\Sigma_{4} = \Sigma_3[\ell_3+1]$ and set
\begin{equation}
  \frac{\nu_b}{\mu_1[\ell_2]}
  = \frac{\lambda_1(\Sigma_3[\ell_3+\delta_{b4}])}{\lambda_1(\Sigma_2[\ell_2])} \,,
  \quad
  \frac{\rho_b}{\mu_2[\ell_1]}
  = \frac{\lambda_2(\Sigma_3[\ell_3+\delta_{b4}])}{\lambda_2(\Sigma_1[\ell_1])} \,.
\end{equation}
Then we obtain
\begin{equation}
  \label{eq:[T,T]=0}
  T_3[\ell_3+1] T_3[\ell_3] = T_3[\ell_3] T_3[\ell_3+1] \,,
\end{equation}
where $T_3[\ell_3]$ can be written, in the notations of
section~\ref{sec:[T0,T0]=0}, as
\begin{multline}
  T_3[\ell_3]
  =
  \Str_{\bigotimes_{\ell_2} \CV_1[1,\ell_2,\ell_3] \otimes \bigotimes_{\ell_1} \CV_2[\ell_1,1,\ell_3]}
  \Biggl(
  \prod_{\ell_1, \ell_2}^{\longleftarrow} \CR[\ell_1,\ell_2,\ell_3]
  \\
  \times
  \frac{\lambda_1(\Sigma_3[\ell_3])^{\hat{n}_1(\Sigma_3[\ell_3])}}
       {\prod_{\ell_2} \lambda_1(\Sigma_2[\ell_2])^{\hat{n}_1[1,\ell_2,\ell_3]}}
  \frac{\lambda_2(\Sigma_3[\ell_3])^{\hat{n}_2(\Sigma_3[\ell_3])}}{\prod_{\ell_1} \lambda_2(\Sigma_1[\ell_1])^{\hat{n}_2[\ell_1,1,\ell_3]}}
  \Biggr)
  \,.
\end{multline}
(Here we have suppressed complicated sign factors by employing the
Fock space notation for the R-matrices.)  The operator $T_3[\ell_3]$
is the transfer matrix with twisted periodic boundary conditions, with
the twisting in the $x^i$-direction controlled by the parameter
$\lambda_i(\Sigma_3[\ell_3])$: the partition function can be expressed
as
\begin{equation}
  Z
  =
  \Str_{\CV_3[1]}\Biggl(
  T_3[L_3] \dotsm T_3[2] T_3[1]
  \frac{\prod_{\ell_2} \lambda_3(\Sigma_2[\ell_2])^{\hat{n}_3(\Sigma_2[\ell_2])}}{\prod_{\ell_1} \lambda_3(\Sigma_1[\ell_1])^{\hat{n}_3(\Sigma_1[\ell_1])}}
  \Biggr) \,.
\end{equation}
Thus we have demonstrated the integrability of the model.

\subsection{Reduction to 2D lattice models}

Given any 3D lattice model on a periodic cubic lattice, one can always
view it as a 2D lattice model by treating one of the directions of the
lattice as internal degrees of freedom.  If the 3D lattice model is
integrable, then the resulting 2D lattice model is also integrable, in
the sense that the equivalent 1D quantum spin model has commuting
transfer matrices with continuous parameters.

\begin{figure}
  \centering
  \tdplotsetmaincoords{135}{0}
  \begin{tikzpicture}[tdplot_main_coords, xscale=1, yscale=1.2]
    \draw[thick, stealth-, shorten >=0pt, shorten <=-12pt]
    ({-sqrt(3)},-1,0) -- ({-sqrt(3)},0.45,0);
    \draw[thick, densely dotted] ({-sqrt(3)},0.5,0) -- ({-sqrt(3)},1,0);
    \draw[thick] ({-sqrt(3)},0.95,0) -- ({-sqrt(3)},{sqrt(3)},0);

    \draw[thick, stealth-, shorten >=0pt, shorten <=-12pt]
    ({sqrt(3)},-1,0) -- ({sqrt(3)},0.45,0);
    \draw[thick, densely dotted] ({sqrt(3)},0.5,0) -- ({sqrt(3)},1,0);
    \draw[thick] ({sqrt(3)},0.95,0) -- ({sqrt(3)},{sqrt(3)},0);

    \draw[thick, stealth-, shorten >=0pt, shorten <=-12pt]
    (0,0,{sqrt(3)}) -- (0,1,0);
    \draw[line width=4pt, white, shorten >=19pt, shorten <=35pt]
    (0,0,{sqrt(3)}) -- (0,1,0);
    \draw[thick, densely dotted, shorten >=20pt, shorten <=36pt]
    (0,0,{sqrt(3)}) -- (0,1,0);

    \begin{scope}
      \draw[thick, -stealth, shorten >=-12pt, shorten <=-12pt]
      (0,0,{sqrt(3)}) -- ({sqrt(3)},-1,0);

      \draw[thick, stealth-, shorten >=-12pt, shorten <=-12pt]
      (0,0,{sqrt(3)}) -- ({-sqrt(3)},-1,0);
    \end{scope}

    \begin{scope}[yshift=-12pt]
      \draw[thick, -stealth, shorten >=-12pt, shorten <=-12pt]
      (0,0,{sqrt(3)}) -- ({sqrt(3)},-1,0);

      \draw[thick, stealth-, shorten >=-12pt, shorten <=-12pt]
      (0,0,{sqrt(3)}) -- ({-sqrt(3)},-1,0);
    \end{scope}
      
    \begin{scope}[yshift=-24pt]
      \draw[thick, -stealth, shorten >=-12pt, shorten <=-12pt]
      (0,0,{sqrt(3)}) -- ({sqrt(3)},-1,0);

      \draw[thick, stealth-, shorten >=-12pt, shorten <=-12pt]
      (0,0,{sqrt(3)}) -- ({-sqrt(3)},-1,0);
    \end{scope}
      
    \begin{scope}[yshift=-44pt]
      \draw[thick, -stealth, shorten >=-12pt, shorten <=-12pt]
      (0,0,{sqrt(3)}) -- ({sqrt(3)},-1,0);

      \draw[thick, stealth-, shorten >=-12pt, shorten <=-12pt]
      (0,0,{sqrt(3)}) -- ({-sqrt(3)},-1,0);
    \end{scope}

    \begin{scope}
      \draw[line width=4pt, shorten >=12pt, shorten <=12pt, white]
      ({-sqrt(3)},-1,0) -- ({sqrt(3)},-1,0);
      
      \draw[thick, -stealth, shorten >=-12pt, shorten <=-12pt]
      ({-sqrt(3)},-1,0) -- ({sqrt(3)},-1,0);
    \end{scope}

    \begin{scope}[yshift=-12pt]
      \draw[line width=4pt, shorten >=12pt, shorten <=12pt, white]
      ({-sqrt(3)},-1,0) -- ({sqrt(3)},-1,0);
      
      \draw[thick, -stealth, shorten >=-12pt, shorten <=-12pt]
      ({-sqrt(3)},-1,0) -- ({sqrt(3)},-1,0);
    \end{scope}

    \begin{scope}[yshift=-24pt]
      \draw[line width=4pt, shorten >=12pt, shorten <=12pt, white]
      ({-sqrt(3)},-1,0) -- ({sqrt(3)},-1,0);
      
      \draw[thick, -stealth, shorten >=-12pt, shorten <=-12pt]
      ({-sqrt(3)},-1,0) -- ({sqrt(3)},-1,0);
    \end{scope}

    \begin{scope}[yshift=-44pt]
      \draw[line width=4pt, shorten >=12pt, shorten <=12pt, white]
      ({-sqrt(3)},-1,0) -- ({sqrt(3)},-1,0);
      
      \draw[thick, -stealth, shorten >=-12pt, shorten <=-12pt]
      ({-sqrt(3)},-1,0) -- ({sqrt(3)},-1,0);
    \end{scope}
  \end{tikzpicture}
  \quad\ \
  {\Large =}
  \quad
  \tdplotsetmaincoords{-45}{0}
  \begin{tikzpicture}[tdplot_main_coords, xscale=1, yscale=1.2]
    \begin{scope}
      \draw[thick, -stealth, shorten >=-12pt, shorten <=-12pt]
      (0,0,{sqrt(3)}) -- ({sqrt(3)},-1,0);

      \draw[thick, -stealth, shorten >=-12pt, shorten <=-12pt]
      ({-sqrt(3)},-1,0) -- ({sqrt(3)},-1,0);

      \draw[thick, stealth-, shorten >=-12pt, shorten <=-12pt]
      (0,0,{sqrt(3)}) -- ({-sqrt(3)},-1,0);
    \end{scope}

    \begin{scope}[yshift=12pt]
      \draw[line width=4pt, shorten >=30pt, shorten <=2pt, white]
      (0,0,{sqrt(3)}) -- ({sqrt(3)},-1,0);

      \draw[line width=4pt, shorten >=30pt, shorten <=2pt, white]
      (0,0,{sqrt(3)}) -- ({-sqrt(3)},-1,0);

      \draw[thick, -stealth, shorten >=-12pt, shorten <=-12pt]
      (0,0,{sqrt(3)}) -- ({sqrt(3)},-1,0);

      \draw[thick, -stealth, shorten >=-12pt, shorten <=-12pt]
      ({-sqrt(3)},-1,0) -- ({sqrt(3)},-1,0);

      \draw[thick, stealth-, shorten >=-12pt, shorten <=-12pt]
      (0,0,{sqrt(3)}) -- ({-sqrt(3)},-1,0);
    \end{scope}

    \begin{scope}[yshift=24pt]
      \draw[line width=4pt, shorten >=30pt, shorten <=2pt, white]
      (0,0,{sqrt(3)}) -- ({sqrt(3)},-1,0);

      \draw[line width=4pt, shorten >=30pt, shorten <=2pt, white]
      (0,0,{sqrt(3)}) -- ({-sqrt(3)},-1,0);

      \draw[thick, -stealth, shorten >=-12pt, shorten <=-12pt]
      (0,0,{sqrt(3)}) -- ({sqrt(3)},-1,0);

      \draw[thick, -stealth, shorten >=-12pt, shorten <=-12pt]
      ({-sqrt(3)},-1,0) -- ({sqrt(3)},-1,0);

      \draw[thick, stealth-, shorten >=-12pt, shorten <=-12pt]
      (0,0,{sqrt(3)}) -- ({-sqrt(3)},-1,0);
    \end{scope}

    \begin{scope}[yshift=44pt]
      \draw[line width=4pt, shorten >=30pt, shorten <=2pt, white]
      (0,0,{sqrt(3)}) -- ({sqrt(3)},-1,0);

      \draw[line width=4pt, shorten >=30pt, shorten <=2pt, white]
      (0,0,{sqrt(3)}) -- ({-sqrt(3)},-1,0);

      \draw[thick, -stealth, shorten >=-12pt, shorten <=-12pt]
      (0,0,{sqrt(3)}) -- ({sqrt(3)},-1,0);

      \draw[thick, -stealth, shorten >=-12pt, shorten <=-12pt]
      ({-sqrt(3)},-1,0) -- ({sqrt(3)},-1,0);

      \draw[thick, stealth-, shorten >=-12pt, shorten <=-12pt]
      (0,0,{sqrt(3)}) -- ({-sqrt(3)},-1,0);
    \end{scope}

    \draw[thick, shorten >=0pt, shorten <=-12pt]
    ({-sqrt(3)},-1,0) -- ({-sqrt(3)},0.45,0);
    \draw[thick, densely dotted] ({-sqrt(3)},0.5,0) -- ({-sqrt(3)},1,0);
    \draw[thick, -stealth] ({-sqrt(3)},0.95,0) -- ({-sqrt(3)},{sqrt(3)},0);

    \draw[thick, shorten >=0pt, shorten <=-12pt]
    ({sqrt(3)},-1,0) -- ({sqrt(3)},0.45,0);
    \draw[thick, densely dotted] ({sqrt(3)},0.5,0) -- ({sqrt(3)},1,0);
    \draw[thick, -stealth] ({sqrt(3)},0.95,0) -- ({sqrt(3)},{sqrt(3)},0);

    \draw[thick, -stealth, shorten >=0pt, shorten <=-12pt]
    (0,0,{sqrt(3)}) -- (0,1,0);
    \draw[line width=4pt, white, shorten >=19pt, shorten <=35pt]
    (0,0,{sqrt(3)}) -- (0,1,0);
    \draw[thick, densely dotted, shorten >=20pt, shorten <=36pt]
    (0,0,{sqrt(3)}) -- (0,1,0);
  \end{tikzpicture}

  \caption{A graphical representation of the Yang--Baxter equation
    that follows from the tetrahedron equation.  The vertical
    direction is periodic.}
  \label{fig:YBE}
\end{figure}
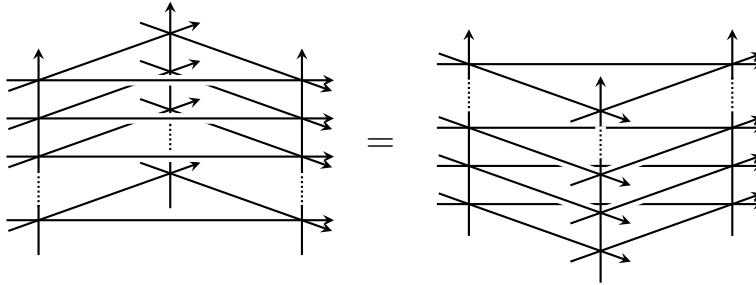

For the 3D lattice model constructed from branes, we can make a
stronger statement since its R-matrices satisfy the supertetrahedron
equation.  Replace $\Sigma_4$ in figure~\ref{fig:tetrahedron} with a
stack of surfaces $\Sigma_4[\ell]$, $\ell = 1$, $\dotsc$, $L$.  The
product R-matrix
\begin{equation}
  \CR_{(ab\mathbf{4})}
  =
  \CR_{(ab4[L])} \CR_{(ab4[L-1])} \dotsm \CR_{(ab4[1])}
\end{equation}
satisfies the equation
\begin{equation}
  \label{eq:ml-tetrahedron}
  \CR_{(23\mathbf{4)}}
  \CR_{(13\mathbf{4)}}
  \CR_{(12\mathbf{4)}}
  \CR_{(123)}
  =
  \CR_{(123)}
  \CR_{(12\mathbf{4)}}
  \CR_{(13\mathbf{4)}}
  \CR_{(23\mathbf{4)}} \,.
\end{equation}
Multiplying both sides by
\begin{equation}
  \CR_{(123)}^{-1}
  \biggl(\frac{z_1}{z_2}\biggr)^{\hat{n}_{(12)} - \hat{n}_{(23)}}
  \biggl(\frac{z_1}{z_3}\biggr)^{\hat{n}_{(13)} + \hat{n}_{(23)}}
  =
  \biggl(\frac{z_1}{z_2}\biggr)^{\hat{n}_{(12)} - \hat{n}_{(23)}}
  \biggl(\frac{z_1}{z_3}\biggr)^{\hat{n}_{(13)} + \hat{n}_{(23)}}
  \CR_{(123)}^{-1}
\end{equation}
and taking the supertrace over
$\CV_{(23)} \otimes \CV_{(13)} \otimes \CV_{(12)}$, we obtain the
\emph{Yang--Baxter equation}
\begin{equation}
  \label{eq:YBE}
  \CS_{23}\biggl(\frac{z_2}{z_3}\biggr)
  \CS_{13}\biggl(\frac{z_1}{z_3}\biggr)
  \CS_{12}\biggl(\frac{z_1}{z_2}\biggr)
  =
  \CS_{12}\biggl(\frac{z_2}{z_3}\biggr)
  \CS_{13}\biggl(\frac{z_1}{z_3}\biggr)
  \CS_{23}\biggl(\frac{z_1}{z_2}\biggr)
  \,,
\end{equation}
where
\begin{equation}
  \CS_{ab}(z)
  =
  \Str_{\CV_{(ab)}}(z^{\hat{n}_{ab}} \CR_{(ab4[L])} \CR_{(ab4[L-1])} \dotsm \CR_{(ab4[1])}) \,.
\end{equation}
See figure~\ref{fig:YBE} for a graphical representation of the
Yang--Baxter equation.  The operator $\CS_{ab}(z)$ is the R-matrix
describing the local Boltzmann weights of a 2D lattice model, and in
this context the parameter $z$ is called the \emph{spectral
  parameter}.  The integrability of the 2D lattice model is a
consequence of the Yang--Baxter equation with spectral parameter.

For comparison with known results in the literature, let us take
\begin{equation}
  \sigma_1 = \sigma_2 = \sigma_3 = + \,.
\end{equation}
Then, the sign factors in the supertetrahedron equation
\eqref{eq:supertetrahedron-2} drop out.  Since the sign of
$\Sigma_4[\ell]$ can be either $+$ or $-$, we obtain a set of $2^{L}$
Yang--Baxter equations.  Choose an $L$-tuple
\begin{equation}
  \boldsymbol{\epsilon} = (\epsilon_1, \dotsc, \epsilon_{L}) \in \{0,1\}^{L}
\end{equation}
and take
\begin{equation}
  \sigma(\Sigma_4[\ell]) = (-1)^{\epsilon_{\ell}} \,.
\end{equation}
The corresponding solution of the Yang--Baxter equation is
\begin{equation}
  \label{eq:S^epsilon}
  S^{(\boldsymbol{\epsilon})}(z)
  =
  \Tr_{\CF^{(0)}}(z^{\hat{n}} \breve\TDS^{(\epsilon_{L})} \dotsm \breve\TDS^{(\epsilon_1)})
  \in
  \End(\CF^{(\boldsymbol{\epsilon})} \otimes \CF^{(\boldsymbol{\epsilon})}) \,,
\end{equation}
where we have defined
\begin{equation}
  \TDS^{(0)} = \TDR \,,
  \quad
  \TDS^{(1)} = \TDL
\end{equation}
and
\begin{equation}
  \CF^{(\boldsymbol{\epsilon})}
  =
  \CF^{(\epsilon_L)} \otimes \dotsb \otimes \CF^{(\epsilon_1)}
  \,.
\end{equation}

The vector space $\CF^{(\boldsymbol{\epsilon})}$ is the Fock space with $L_+$
bosonic creation operators and $L_-$ fermionic creation operators,
where $L_+$ and $L_-$ are the numbers of $0$ and $1$ in $\boldsymbol{\epsilon}$,
respectively.  Let $\CF^{(\boldsymbol{\epsilon})}_n$ be the level-$n$ subspace of
$\CF^{(\boldsymbol{\epsilon})}$, spanned by the vectors of the form
\begin{equation}
  \ket{n_L} \otimes \dotsb \otimes \ket{n_1}
  \in \CF^{(\epsilon_L)} \otimes \dotsb \otimes \CF^{(\epsilon_1)} \,,
  \quad
  n_1 + \dotsb + n_L = n \,.
\end{equation}
By charge conservation, $S^{(\boldsymbol{\epsilon})}$ leaves
$\CF^{(\boldsymbol{\epsilon})}_l \otimes
\CF^{(\boldsymbol{\epsilon})}_m$ invariant.  Thus,
$S^{(\boldsymbol{\epsilon})}$ decomposes as
\begin{equation}
  S^{(\boldsymbol{\epsilon})}
  =
  \bigoplus_{l,m=0}^\infty S^{(\boldsymbol{\epsilon})}_{l,m} \,,
  \quad
  S^{(\boldsymbol{\epsilon})}_{l,m} \in \End(\CF^{(\boldsymbol{\epsilon})}_l \otimes
  \CF^{(\boldsymbol{\epsilon})}_m)
  \,,
\end{equation}
and each summand satisfies the Yang--Baxter equation.  The Lie
supergroup $\GL(L_+|L_-)$ acts on $\CF^{(\boldsymbol{\epsilon})}_n$ by identifying
the creation operators with the standard basis vectors of
$\C^{L_+|L_-}$.  As a $\GL(L_+|L_-)$-module, $\CF^{(\boldsymbol{\epsilon})}_n$ is
the $n$th supersymmetric power of the natural $L$-dimensional
representation:
\begin{equation}
  \CF^{(\boldsymbol{\epsilon})}_n
  \iso
  \bigoplus_{\substack{l, m \in \Z_{\geq0} \\ l+m=n}}
  S^l(\C^{L_+}) \otimes \Lambda^m(\C^{L_-}) \,.
\end{equation}
Here $S^n(V)$ and $\Lambda^n(V)$ denote the $n$th symmetric and
antisymmetric powers of $V$.

From the point of view of the brane system, the reduction of the 3D
lattice model to a 2D lattice model amounts to reducing M-theory on
$\circle_3$ to type IIA string theory.  Let us further apply T-duality
along the periodic Euclidean time direction $\circle_0$ and S-duality.
Equivalently, we can reduce the M-theory setup on $\circle_0$ and then
take T-duality on $\circle_3$.  We will adopt the second presentation
and let $\check\circle_3$ denote the circle dual to $\circle_3$.  If
we do so, the M5-branes become D3-branes and D5-branes, and the
M2-branes become fundamental strings.  The resulting configuration is
summarized in table \ref{tab:D5-D3-F1}.  The situation considered
above is when the signs of the D3-branes are all $+$ and the sign of
the $\ell$th D5-brane is $(-1)^{\ell}$.

\begin{table}
  \centering
  \begin{tabular}{l|cc|cccc|cc|cc}
    \hline \hline
    & 1 & 2 & 3 & 4 & 5 & 6 & 7 & 8 & 9 & $\ten$
    \\
    \hline
    $\Dthree_{1+}$
    & $\point$  & $\rline$
    & $\point$
    & $\rline$ & $\point$ & $\point$
    & $\rline$ & $\rline$
    & $\atzero$ & $\atzero$
    \\
    $\Dthree_{2+}$
    & $\rline$ & $\point$
    & $\point$
    & $\point$ & $\rline$ & $\point$
    & $\rline$ & $\rline$
    & $\atzero$ & $\atzero$
    \\
    $\Dfive_{3+}$
    & $\rline$ & $\rline$
    & $\rline$
    & $\point$ & $\point$ & $\rline$
    & $\rline$ & $\rline$
    & $\atzero$ & $\atzero$
    \\
    \hline
    $\Dthree_{1-}$
    & $\point$  & $\rline$
    & $\point$
    & $\rline$ & $\point$ & $\point$
    & $\atzero$ & $\atzero$
    & $\rline$ & $\rline$
    \\
    $\Dthree_{2-}$
    & $\rline$ & $\point$
    & $\point$
    & $\point$ & $\rline$ & $\point$
    & $\atzero$ & $\atzero$
    & $\rline$ & $\rline$
    \\
    $\Dfive_{3-}$
    & $\rline$ & $\rline$
    & $\rline$
    & $\point$ & $\point$ & $\rline$
    & $\atzero$ & $\atzero$
    & $\rline$ & $\rline$
    \\
    \hline
    $\Fone_1$
    & $\rline$
    & $\point$ & $\point$
    & $\interval$ & $\point$ & $\point$
    & $\atzero$ & $\atzero$
    & $\atzero$ & $\atzero$
    \\
    $\Fone_2$
    & $\point$
    & $\rline$ & $\point$
    & $\point$ & $\interval$ & $\point$
    & $\atzero$ & $\atzero$
    & $\atzero$ & $\atzero$
    \\
    $\Fone_3$
    & $\point$
    & $\point$ & $\rline$
    & $\point$ & $\point$ & $\interval$
    & $\atzero$ & $\atzero$
    & $\atzero$ & $\atzero$
    \\
    \hline \hline
  \end{tabular}

  \caption{A configuration of D5-branes, D3-branes and fundamental
    strings.}
  \label{tab:D5-D3-F1}
\end{table}

This D3--D5 brane system was studied in \cite{Ishtiaque:2021jan}.  (In
fact, a more general setup with D3-branes of both signs was considered
there.)  Let us summarize the results of~\cite{Ishtiaque:2021jan}
pertinent to the present discussion.

The $L_+$ D5-branes of sign $+$ produce 6D $\CN = 2$ super Yang--Mills
theories with gauge group $\U(L_+)$.  The twisting of the periodic
boundary condition in the $x^0$-direction induces a Ramond--Ramond
$2$-form background field deforming this theory.  The deformation has
the effect of reducing the relevant BPS sector of the 6D theory to 4D
Chern--Simons theory \cite{Costello:2013zra, Costello:2013sla,
  Costello:2017dso} with gauge group $\GL(L_+)$ and coupling
$\hbar \propto \theta$ \cite{Yagi:2014toa, Luo:2014sva,
  Costello:2018txb}.  The $L_-$ D5-branes of sign $-$ likewise produce
4D Chern--Simons theory with gauge group $\GL(L_-)$ and coupling
$-\hbar$.  Finally, open strings connecting D5-branes of opposite
signs produce 4D $\CN = 2$ hypermultiplet in the bifundamental
representation of $\GL(L_+) \times \GL(L_-)$, the BPS sector of which
is described by a fermionic $1$-form field valued in the direct sum of
a pair of bifundamental representations of $\GL(L_+) \times \GL(L_-)$.
Together, these ingredients combine into 4D Chern--Simons theory on
$\T^2_{12} \times \check\circle_3 \times \R_6$ with gauge group
$\GL(L_+|L_-)$ and coupling constant $\hbar$.

The appearance of 4D Chern--Simons theory naturally explains the
origin of the integrable 2D lattice model \cite{Costello:2013zra,
  Costello:2013sla}.  The cylinder $\check\circle_3 \times \R_6$,
equipped with the standard complex structure, is conformally
equivalent to a punctured plane $\check\C^\times_{36}$.  4D
Chern--Simons theory formulated on
$\T^2_{12} \times \check\C^\times_{36}$ is topological on $\T^2_{12}$
and holomorphic on $\check\C^\times_{36}$.  It has Wilson line
operators lying in $\T^2_{12}$ and supported at points in
$\check\C^\times_{36}$.  We can wrap these Wilson lines around
$\T^2_{12}$ to form a lattice.  The correlation function of such a
configuration of Wilson lines computes the partition function of a 2D
lattice model, defined by a trigonometric solution of the Yang--Baxter
equation associated with the Lie algebra $\gf$ of the gauge group of
the theory.  Crossing of two Wilson lines yields a factor of the
R-matrix, evaluated in the representations of the Wilson lines and
with the spectral parameter given by the ratio of the coordinates of
the two lines in $\check\C^\times_{36}$.

The D3-branes create line defects in the 4D Chern--Simons theory,
extending in the $x^1$-direction and the $x^2$-direction.  In the
setup studied in \cite{Ishtiaque:2021jan}, where the holomorphic
directions were taken to be $\C$, it was shown that a D3-brane
intersecting the D5-branes realizes a quantum mechanical system with
global symmetry $\GL(L_+|L_-)$, coupled to the 4D Chern--Simons theory
via gauging.  Open strings localized at the intersection of the
D3-brane and a D5-brane give rise to an oscillator algebra, which is
bosonic if the two branes have the same sign and fermionic otherwise.
If we fix the number of open strings attached to the D3-brane to $n$,
the Hilbert space of the quantum mechanical system is projected to
$\CF^{(\boldsymbol{\epsilon})}_n$.  Integrating out this quantum
mechanical system leaves a Wilson line in the representation
$\CF^{(\boldsymbol{\epsilon})}_n$.  When two such Wilson lines
intersect in $\T^2_{12}$, they exchange gluons.  These gluons are open
strings stretched between two D3-branes.  The interaction produces a
rational $\glf(L_+|L_-)$ R-matrix acting on the tensor products of the
representations of the Wilson lines.

The setup is slightly different in the case at hand in that the
holomorphic directions are $\check\C^\times$ here.  Correspondingly,
we expect that the R-matrix $S^{(\boldsymbol{\epsilon})}$ is a trigonometric
version of this rational $\glf(L_+|L_-)$ R-matrix.

\section{Identification of the model}
\label{sec:identification}

A solution of the supertetrahedron equation
\eqref{eq:supertetrahedron-breve-R} that has all of the properties
deduced in section \ref{sec:properties} is known
\cite{Yoneyama:2020duw}.  I propose that this solution describes the
3D lattice model constructed from branes.

The R-matrices for the solution in question have the following nonzero
matrix elements:%
\footnote{In the notation of \cite{Yoneyama:2020duw},
  $\breve\TDR_{nml}^{l'm'n'}$, $\breve\TDL_{nml}^{l'm'n'}$,
  $\breve\TDM_{nml}^{l'm'n'}$, $\breve\TDN_{nml}^{l'm'n'}$ are written
  as $\TDR_{lmn}^{l'm'n'}$, $\TDL_{lmn}^{l'm'n'}$,
  $\TDM_{lmn}^{l'm'n'}$, $(-1)^m \TDN_{lmn}^{l' m' n'}$,
  respectively.}
\begin{gather}
  \breve\TDR_{nml}^{l'm'n'}
  =
  \delta^{l'+m'}_{l+m}
  \delta^{m'+n'}_{m+n}
  \sum_{\substack{\lambda, \mu \in \Z_{\geq0} \\ \lambda+\mu=m'}}
  (-1)^\lambda
  q^{l(n'-m) + (n+1)\lambda + \mu(\mu - n)}
  \frac{(q^2)_{n'+\mu}}{(q^2)_{n'}}
  {l \choose \mu}_{q^2}
  {m \choose \lambda}_{q^2} \,,
  \\
  \begin{gathered}
    \breve\TDL_{n00}^{00n'}
    = \breve\TDL_{n11}^{11n'}
    = \delta_n^{n'} \,,
    \quad
    \breve\TDL_{n10}^{01n'} = -\delta_n^{n'} q^{n+1} \,,
    \quad
    \breve\TDL_{n01}^{10n'} = \delta_n^{n'} q^{n} \,,
    \\
    \breve\TDL_{n01}^{01n'} = \delta_{n-1}^{n'} (1-q^{2n}) \,,
    \quad
    \breve\TDL_{n10}^{10n'} = \delta_{n+1}^{n'} \,,
  \end{gathered}
  \\
  \begin{gathered}
    \breve\TDN_{0n0}^{0n'0} = (-1)^n \delta_n^{n'} q^n \,,
    \quad
    \breve\TDN_{1n1}^{1n'1} = (-1)^{n+1} \delta_n^{n'} q^{n+1} \,,
    \quad
    \breve\TDN_{0n1}^{1n'0} = \breve\TDN_{1n0}^{0n'1} = (-1)^n \delta_n^{n'} \,,
    \\
    \breve\TDN_{1n1}^{0n'0} = (-1)^n \delta_{n+1}^{n'} q^n (1-q^2) \,,
    \quad
    \breve\TDN_{0n0}^{1n'1} = (-1)^n \delta_{n-1}^{n'} [n]_q \,,
  \end{gathered}
\end{gather}
and $\breve\TDM_{lmn}^{n'm'l'} = \breve\TDL_{nml}^{l'm'n'}$.  Here
\begin{equation}
  (q)_n = \prod_{k=1}^n (1 - q^k) \,,
  \quad
  {m \choose n}_q
  =
  \frac{(q)_m}{(q)_{m-n} (q)_n} \,,
  \quad
  [n]_q = \frac{q^n - q^{-n}}{q - q^{-1}}
\end{equation}
are the $q$-factorial, the $q$-binomial and the $q$-number, and the
$q$-binomial is understood to be zero unless $0 \leq n \leq m$.

The operator $\TDR$ is a solution of the RRRR relation~\eqref{eq:RRRR}
that was constructed by Kapranov and Voevodsky \cite{MR1278735} from
the intertwiner of two irreducible representations of the quantum
coordinate ring $A_q(\slf(3))$.  (The relevant representations are
reducible if $q$ is a root of unity.  When $q$ is a primitive root of
unity of odd order, they contain finite-dimensional irreducible
quotient representations and the corresponding R-matrix was obtained
in \cite{MR1247288}.)  Later, this R-matrix was rediscovered by
Bazhanov and Sergeev \cite{Bazhanov:2005as} in a different but
equivalent \cite{Kuniba:2012ei} form.  The operator $\TDL$ which
solves the RLLL relation~\eqref{eq:LLLR} was discovered in
\cite{Bazhanov:2005as}.  Finally, Yoneyama~\cite{Yoneyama:2020duw},
building on earlier works by Sergeev \cite{MR2379711} and Kuniba,
Okado and Yamada \cite{MR3116185}, provided a uniform characterization
of $\TDR$, $\TDL$, $\TDM$ and $\TDN$ as transition matrices between
the Poincar\'e--Birkhoff--Witt bases for nilpotent subalgebras of
$U_q(\slf(L_+|L_-))$ with $L_+ + L_- = 3$.

The matrix elements of the above R-matrices are polynomials in $q$ and
$q^{-1}$ with integer coefficients, in agreement with the integrality
of the reduced index $\mathring Z$.  The R-matrices satisfy the
normalization condition~\eqref{eq:R000000}, the charge conservation
rule~\eqref{eq:charge-consv} and the symmetry~\eqref{eq:RR-symmetry}.
Moreover, they have the involutivity~\eqref{eq:R^2=1}.  The relation
$\TDR^{-1} = \TDR$ is proved in \cite{Kuniba:2012ei}.  In
\cite{Yoneyama:2020duw} it is shown that
\begin{equation}
  \sum_{l',m',n'} \breve\TDL_{n'm'l'}^{l''m''n''} \breve\TDL_{nml}^{l'm'n'}
  = \sum_{l',m',n'}
    (-1)^{m'} \breve\TDN_{n'm'l'}^{l''m''n''} (-1)^{m} \breve\TDN_{nml}^{l'm'n'}
  = \delta_l^{l''} \delta_m^{m''} \delta_n^{n''} \,.
\end{equation}
Since $(-1)^{lm+l'm'} \TDL_{lmn}^{l'm'n'} = \TDL_{lmn}^{l'm'n'}$ and
$(-1)^{ln+l'n'+m+m'} \TDN_{lmn}^{l'm'n'} = \TDN_{lmn}^{l'm'n'}$, we
have $\TDL^{-1} = \TDL$ and $\TDN^{-1} = \TDN$.

This solution also has the expected relation to solutions of the
Yang--Baxter equation.  In \cite{Bazhanov:2005as} it was found that
the trace of the product of $L$ copies of $\TDR$ and the trace of the
product of $L$ copies of $\TDL$ produce trigonometric $\slf(L)$
R-matrices valued in the direct sum of all symmetric tensor
representations and in the direct sum of all antisymmetric tensor
representations, respectively.  This result was extended in
\cite{MR2554447, Kuniba:2015sca} to more general combinations of
R-matrices.  In \cite{Kuniba:2015sca}, the trace of the product of
$L_+$ copies $\TDR$ and $L_-$ copies $\TDL$ was studied.  It was shown
that $S^{(\boldsymbol{\epsilon})}_{l,m}$ is an R-matrix acting on the
tensor product of two irreducible modules
$\CW^{(\boldsymbol{\epsilon})}_l$, $\CW^{(\boldsymbol{\epsilon})}_m$
of a Hopf algebra $\CU(\boldsymbol{\epsilon})$ called the generalized
quantum group of type $A$, which is an affine analog of the quantized
enveloping algebra $U_q(\glf(L_+|L_-))$ of $\glf(L_+|L_-)$.  We
identify $\CW^{(\boldsymbol{\epsilon})}_n$ with
$\CF^{(\boldsymbol{\epsilon})}_n$.

One of the interesting features of this solution is that the
R-matrices are constant, and the spectral parameters of the solutions
of the Yang--Baxter equations and transfer matrices originate from
twist parameters for the periodic boundary conditions.  As we saw in
section~\ref{sec:pf}, this feature is nicely explained in the brane
picture: the only continuous parameters of the model, the positions of
the M5-branes in $\R^3_{456}$, can be identified with Wilson loops of
complex gauge fields in a dual frame and, as such, show up in the
partition function as twist parameters.

As a last piece of evidence, we point out that there is another way of
reducing the 3D lattice model to a 2D lattice model.  Let us consider
the circle $\circle_3$ as the interval $[-c_3/2,+c_3/2]$ with the two
ends identified.  Instead of simply reducing the lattice along
$\circle_3$, we can first take the orbifold $\circle_3/\Z_2$ by the
$\Z_2$-action $x^3 \mapsto -x^3$ and then perform reduction.  For the
orbifolding to be possible, an M5-brane located at $x^3 \in \circle_3$
must have its image at $-x^3$, unless it is located at one of the
fixed points at $x^3 = 0$ and $x^3 = \pm c_3/2$.  The orbifold
$\circle_3/\Z_2$ can also be regarded as the interval $[0,c_3/2]$
sandwiched by a pair of M9-branes \cite{Horava:1995qa}.  The M5-branes
extending along the interval end on the
M9-branes~\cite{Bergshoeff:2006bs}.  Each M9-brane provides a boundary
state of the lattice model, and we expect that a product of R-matrices
evaluated between the two boundary states is a solution of the
Yang--Baxter equation.  There is indeed such a construction in the
lattice model~\cite{MR2525474, MR3123535}.  In fact, two different
boundary states were introduced in~\cite{MR3123535}.  The difference
between the two is probably whether a ``half'' M5-brane is stuck on
the fixed point or not.

It may be possible to reproduce the partition function and the
R-matrices of the 3D lattice model as quantities calculated in QFTs
describing the brane system.  An interesting problem in this regard is
to identify the quantum mechanical system supported on the
intersection of three M5-branes, one of type $1\sigma_1$, one of type
$2\sigma_2$ and one of type $3\sigma_3$, and some numbers of M2-branes
of any type.  The supersymmetric index of this system is equal to a
matrix element of the R-matrix $R^{\sigma_1\sigma_2\sigma_3}$.  The
system can be thought of as a junction of six theories, each living on
a stack of M2-branes suspended between a pair of M5-branes, and this
viewpoint may be helpful in approaching the problem.  Another
interesting problem is to concretely describe the 3D TQFT discussed in
section~\ref{sec:branes-TQFT}, using the 7D theory on the D6-brane or
its dimensional reduction on $\circle_0$.  The R-matrices are obtaned
from correlation functions of intersecting surface defects in this
theory, which we can try to compute in perturbation theory in $q$ or
$q^{-1}$.  A quantitative verification of the proposal of this paper,
along these lines or with any other methods, is left for future
research.

\appendix

\section{Analysis of supersymmetry}
\label{sec:susy}

In this appendix we analyze the supersymmetry preserved by the
M5-brane configuration introduced in section \ref{sec:branes}.  We
will mostly follow the notations of \cite{West:1998ey}.

\subsection{Spinors in ten and eleven dimensions}

Let $\Gamma_\mu$, $\mu = 0$, $1$, $\dotsc$, $9$, $\ten$, be the 11D
gamma matrices.  They generate the Clifford algebra, defined by the
anticommutation relation
\begin{equation}
  \{\Gamma_\mu, \Gamma_\nu\} = 2\eta_{\mu\nu} \,,
\end{equation}
where $\eta = \diag(-1, +1, +1, \dotsc, +1)$ is the 11D Minkowski
metric.  We write $\Gamma_{\mu_1 \mu_2 \dotsc \mu_k}$ for the
antisymmetrized product of $\Gamma_{\mu_1}$, $\Gamma_{\mu_2}$,
$\dotsc$, $\Gamma_{\mu_k}$; if the indices are all distinct,
$\Gamma_{\mu_1 \mu_2 \dotsc \mu_k} = \Gamma_{\mu_1} \Gamma_{\mu_2}
\dotsm \Gamma_{\mu_k}$.

There are two inequivalent irreducible representations of the Clifford
algebra with dimension greater than one.  We pick one by requiring
\begin{equation}
  \label{eq:0123456789ten}
  \Gamma_{0123456789\ten} = +1 \,.
\end{equation}
(The other irreducible representation has
$\Gamma_{0123456789\ten} = -1$.)  This representation is
$32$-dimensional.  The gamma matrices are represented by matrices
acting on a complex $32$-dimensional vector space, an element of which
is called a Dirac spinor.

The minus of the transpose of the gamma matrices $\{-\Gamma_\mu^T\}$
also satisfy the defining relations of this irreducible
representation, so they are related to $\{\Gamma_\mu\}$ by a change of
basis.  It follows that there exists a charge conjugation matrix $C$
such that
\begin{equation}
  C\Gamma_\mu C^{-1} = -\Gamma_\mu^T \,.
\end{equation}
Since $C^{-1} C^T$ commutes with the gamma matrices, it is
proportional to the identity matrix.

The quadratic elements $\{\iu \Gamma_{\mu\nu}\}$ generate the action
of the Lorentz group $\SO(10,1)$ on the Clifford algebra and Dirac
spinors.  Since $C\Gamma_{\mu\nu}C^{-1} = -\Gamma_{\mu\nu}^T$, Dirac
spinors $\zeta$ and $C\zeta$ transform in the dual representations of
$\SO(10,1)$.  If $\zeta$, $\chi$ are two Dirac spinors, then
$\zetab \Gamma_{\mu_1 \mu_2 \dotsc \mu_k} \chi$ transforms in the same
way as a component of an antisymmetric tensor of type $(0,k)$, where
the Majorana conjugate $\zetab$ of $\zeta$ is defined by
\begin{equation}
  \zetab = \zeta^T C \,.
\end{equation}

There is a basis, called a Majorana basis, such that all gamma
matrices are real $32 \times 32$ matrices, $\Gamma_0$ is antisymmetric
and the others are symmetric.  In such a basis we can take
$C = \Gamma^0$.  A Dirac spinor $\eps$ that is real in a Majorana
basis is called a Majorana spinor.

\subsection{Supersymmetry preserved by the M5-brane system}

11D supersymmetry is generated by $32$ hermitian fermionic conserved
charges $Q_\alpha$, $\alpha = 1$, $\dotsc$, $32$, satisfying the
anticommutation relations
\begin{equation}
  \{Q_\alpha, Q_\beta\}
  =
  -(\Gamma^\mu C^{-1})_{\alpha\beta} P_\mu
  - \frac12
    (\Gamma^{\mu\nu} C^{-1})_{\alpha\beta} Z^{(2)}_{\mu\nu}
  - \frac{1}{5!}
    (\Gamma^{\mu\nu\rho\sigma\tau} C^{-1})_{\alpha\beta} Z^{(5)}_{\mu\nu\rho\sigma\tau}
  \,,
\end{equation}
where $P$ is the momentum, $Z^{(2)}$ is a 2-form charge and $Z^{(5)}$
is a 5-form charge, all commuting with $Q_\alpha$ and with each other.
Under the Lorentz transformations the supercharges transform as the
components of a spinor $Q$.  In what follows we choose a Majorana
basis so that the gamma matrices and the parameters of supersymmetry
transformations are all real.

The M5-branes of type $i+$, $i = 1$, $2$, $3$, are invariant under
supersymmetry transformations generated by $\epsb Q$ if the constant
Majorana spinor $\eps$ satisfies
\begin{equation}
  \label{eq:012378}
  \eps
  = \Gamma^{i \, i+3} \Gamma^{012378} \eps \,.
\end{equation}
Similarly, the M5-branes of type $i-$ are invariant for $\eps$
satisfying
\begin{equation}
  \eps
  = \Gamma^{i \, i+3} \Gamma^{01239\ten} \eps \,.
\end{equation}
The signs of the right-hand sides of the above equations are
determined by the orientations of the M5-branes, which we have chosen
in such a way that the brane system is invariant under the cyclic
permutation $(1,4) \to (2,5) \to (3,6) \to (1,4)$ and the exchange
$(7,8) \leftrightarrow (9,\ten)$ of directions.

The presence of the M5-branes leads to six constraint equations for
$\eps$, but not all of them are independent.  Imposing these six
constraints is equivalent to requiring that $\eps$ satisfies one of
them, say
\begin{equation}
  \eps = \Gamma^{14} \Gamma^{012378} \eps  = -\Gamma^{023478} \eps \,,
\end{equation}
as well as the conditions
\begin{equation}
  \label{eq:ii+3=jj+3}
  \Gamma^{i \, i+3} \eps
  = \Gamma^{j \, j+3} \eps \,,
  \quad
  i, \, j = 1, \, 2, \, 3 \,,
\end{equation}
and
\begin{equation}
  \label{eq:78=9ten}
  \Gamma^{78} \eps = \Gamma^{9\ten} \eps \,.
\end{equation}

Equivalently, $\eps$ must satisfy the following four independent
equations:
\begin{align}
  \Gamma^{023478} \eps &= -\eps \,,
  \label{eq:0233478} \\
  \Gamma^{1245} \eps &= +\eps \,,
  \label{eq:1278} \\
  \Gamma^{2356} \eps &= +\eps \,,
  \label{eq:2356} \\
  \Gamma^{789\ten} \eps &= -\eps \,.
  \label{eq:789ten}
\end{align}
The products of gamma matrices in these four equations are symmetric
matrices commuting with each other, hence simultaneously
diagonalizable.  Moreover, they square to the identity matrix and are
traceless (by the cyclicity of trace and the anticommutation relations
of the gamma matrices), so their eigenvalues are $+1$ and $-1$ and the
corresponding eigenspaces have the same dimension.  Therefore,
imposing each of the above condition reduces the dimension of the
space of possible choices of $\eps$ by half.  In total, there is a
two-dimensional space of solutions to the constraint equations.

The complex matrices $\iu\Gamma^{78}$ and $\iu\Gamma^{9\ten}$ are
hermitian, traceless, square to the identity matrix and commute with
each other and with those products of gamma matrices that appear in
the above conditions.  Let $\zeta$ be a simultaneous eigenvector with
eigenvalue $+1$:
\begin{equation}
  \label{eq:Gamma78-Gamma9ten}
  \iu\Gamma^{78} \zeta = \iu\Gamma^{9\ten} \zeta = \zeta \,.
\end{equation}
We normalize $\zeta$ so that
\begin{equation}
  \zeta^\dagger \zeta = 1
\end{equation}
and define the complex supercharge
\begin{equation}
  Q_\zeta = \zetab Q
\end{equation}
corresponding to $\zeta$.  Using two hermitian supercharges $Q_+$,
$Q_-$ we can write
\begin{equation}
  Q_\zeta = Q_+ + \iu Q_- \,.
\end{equation}
The complex conjugate $\zeta^*$ of $\zeta$ satisfies
$\iu\Gamma^{78} \zeta^* = \iu\Gamma^{9\ten} \zeta^* = -\zeta^*$ and
$Q_\zeta^\dagger = \overline{\zeta^*} Q$.

Let us calculate the anticommutation relation
\begin{equation}
  \{Q_\zeta^\dagger, Q_\zeta\}
  =
  (\zeta^\dagger \Gamma^0 \Gamma^\mu \zeta) P_\mu
  + \frac{1}{2}
    (\zeta^\dagger \Gamma^0 \Gamma^{\mu\nu} \zeta) Z^{(2)}_{\mu\nu}
  + \frac{1}{5!}
    (\zeta^\dagger \Gamma^0 \Gamma^{\mu\nu\rho\sigma\tau} \zeta) Z^{(5)}_{\mu\nu\rho\sigma\tau}
  \,.
\end{equation}
The choice of $\zeta$ is unique up to an overall phase factor but the
anticommutator is independent of this factor, so the right-hand side
is completely determined by the conditions that fix $\zeta$ up to a
phase.  In other words, the coefficients in front of the charges on
the right-hand side can be reduced to multiples of
$\zeta^\dagger \zeta$ by the constraint equations for $\epsilon$
written above.  We need not use \eqref{eq:Gamma78-Gamma9ten} since
$\{Q_\zeta^\dagger, Q_\zeta\}$ is symmetric with respect to $\zeta$
and $\zeta^*$.

We can reduce $\zeta^\dagger \Gamma^0 \Gamma^\mu \zeta$ to a multiple
of $\zeta^\dagger \zeta$ only if $\mu = 0$:
\begin{equation}
  \zeta^\dagger \Gamma^0 \Gamma^0 \zeta = -\zeta^\dagger \zeta = -1 \,;
\end{equation}
and reduce $\zeta^\dagger \Gamma^0 \Gamma^{\mu\nu} \zeta$ only if
$\{\mu, \nu\} = \{i, i+3\}$ with $i = 1$, $2$, $3$:
\begin{equation}
  \zeta^\dagger \Gamma^0 \Gamma^{i \, i+3} \zeta
  = -\zeta^\dagger \Gamma^0 \Gamma^{i \, i+3}
     \Gamma^{023478} \Gamma^{2356}  \Gamma^{0123456789\ten} \zeta
  = -\zeta^\dagger \Gamma^{i \, i+3} \Gamma^{01239\ten} \zeta
  = -1 \,.
\end{equation}
As for $\zeta^\dagger \Gamma^0 \Gamma^{\mu\nu\rho\sigma\tau} \zeta$,
we can reduce it if
$\{\mu, \nu, \rho, \sigma, \tau\} \in \{1,2,3,i+3,7,8\} \setminus
\{i\}$ for $i = 1$, $2$, $3$ using \eqref{eq:012378}:
\begin{equation}
  \zeta^\dagger \Gamma^0 \Gamma^{i \, i+3} \Gamma^{12378} \zeta
  = 1 \,,
  \quad
  i = 1, \, 2, \, 3 \,;
\end{equation}
if $\{\mu, \nu, \rho, \sigma, \tau\} \in \{4,5,6,7,8\}$ using
\eqref{eq:0233478} and \eqref{eq:2356}:
\begin{equation}
  \zeta^\dagger \Gamma^0 \Gamma^{45678} \zeta
  = -\zeta^\dagger \Gamma^{023478} \Gamma^{2356} \zeta
  = 1 \,;
\end{equation}
 if $\{\mu, \nu, \rho, \sigma, \tau\} \in \{0, i, j, i+3, j+3\}$
for distinct $i$, $j \in \{1, 2, 3\}$ using \eqref{eq:ii+3=jj+3}:
\begin{equation}
  \zeta^\dagger \Gamma^0 \Gamma^{0ij \, i+3 \, j+3} \zeta
  = \zeta^\dagger \Gamma^{i \, i+3} \Gamma^{j \, j+3} \zeta
  = -1 \,;
\end{equation}
and if $\{\mu, \nu, \rho, \sigma, \tau\} \in \{0, 7,8,9,\ten\}$
 using \eqref{eq:789ten}:
\begin{equation}
  \zeta^\dagger \Gamma^0 \Gamma^{0789\ten} \zeta
  = -\zeta^\dagger \Gamma^{789\ten} \zeta
  = 1 \,.
\end{equation}
Also, $\zeta^\dagger \Gamma^0 \Gamma^{\mu\nu\rho\sigma\tau} \zeta$ can
be reduced to a multiple of $\zeta^\dagger\zeta$ in those cases that
are related to the above cases by the exchange
$(7,8) \leftrightarrow (9,\ten)$.  In the other cases the coefficients
vanish.

Thus we find
\begin{equation}
  \{Q_\zeta^\dagger, Q_\zeta\}
  =
  -P_0 - Z^{(2)}_{14} - Z^{(2)}_{25} - Z^{(2)}_{36} - Y \,,
\end{equation}
where
\begin{multline}
  \label{eq:Y}
  Y
  =
  Z^{(5)}_{23478}
  - Z^{(5)}_{13578}
  + Z^{(5)}_{12678}
  + Z^{(5)}_{2349\ten}
  - Z^{(5)}_{1359\ten}
  + Z^{(5)}_{1269\ten}
  \\
  - Z^{(5)}_{45678} 
  - Z^{(5)}_{4569\ten} 
  + Z^{(5)}_{01245}
  + Z^{(5)}_{01346}
  + Z^{(5)}_{02356}
  - Z^{(5)}_{0789\ten} \,.
\end{multline}

Finally, we mention two properties of $Q_\zeta$ which are important in
connection with the brane setup in section \ref{sec:TQFT}.  The
conditions \eqref{eq:ii+3=jj+3} can be rewritten as
\begin{equation}
  (\Gamma^{ij} + \Gamma^{i+3 \, j+3}) \eps = 0 \,,
  \quad
  i, \, j = 1, \, 2, \, 3 \,.
\end{equation}
These equations say that $\eps$ is invariant under the action of the
diagonal subgroup of $\SO(3)_{123} \times \SO(3)_{456}$, hence so is
$Q_\zeta$.  By the same token, condition \eqref{eq:78=9ten} shows that
$Q_\zeta$ is invariant under the action of the antidiagonal subgroup
of $\SO(2)_{78} \times \SO(2)_{9\ten}$.  From these properties we
immediately see that the only component of $P$ that can appear in
$\{Q_\zeta^\dagger, Q_\zeta\}$ is $P_0$ since the other components are
not invariant under either the diagonal subgroup of
$\SO(3)_{123} \times \SO(3)_{456}$ or the antidiagonal subgroup of
$\SO(2)_{78} \times \SO(2)_{9\ten}$.

\providecommand{\href}[2]{#2}\begingroup\raggedright\endgroup
\end{document}